\documentclass[letterpaper,10pt,fleqn]{article}
\pdfoutput=1
\usepackage{jheppub_v1}

\usepackage{graphicx,epsfig}
\usepackage[figuresright]{rotating}
\usepackage{bm,amssymb}
\usepackage{amsmath}
\usepackage[mathscr]{eucal}
\usepackage{color}
\usepackage[utf8]{inputenc}

\long\def\comment#1{ }
\newcommand{\beq}{\begin{eqnarray}}
\newcommand{\eeq}{\end{eqnarray}}

\newcommand{\nn}{\nonumber\\}
\newcommand{\order}[1]{\mathcal{O}{\left(#1\right)}}

\newcommand{\abar}{\bar{\alpha}}
\newcommand{\mcal}{\mathcal}
\newcommand{\rmtr}{{\rm tr}}
\newcommand{\del}{\partial}
\newcommand{\lan}{\langle}
\newcommand{\ran}{\rangle}

\def\del{\partial}        
\newcommand{\rmd}{{\rm d}}
\newcommand{\dif}{{\rm d}}
\newcommand{\rme}{{\rm e}}
\newcommand{\rmi}{{\rm i}}

\newcommand{\bme}{\bm{\epsilon}}
\newcommand{\bk}{\bm{k}}
\newcommand{\bp}{\bm{p}}

\newcommand{\bx}{\bm{x}}
\newcommand{\bz}{\bm{z}}
\newcommand{\by}{\bm{y}}
\newcommand{\bb}{\bm{b}}

\newcommand{\bu}{\bm{u}}

\newcommand{\br}{\bm{r}}

\newcommand{\bbr}{\overline{\bm{r}}}
\newcommand{\bbb}{\overline{\bm{b}}}

\newcommand{\xbj}{x_{_{\rm Bj}}}

\newcommand{\bK}{\bm{K}}
\newcommand{\bP}{\bm{P}}

\title{The Colour Glass Condensate}

\author{Edmond Iancu}

\affiliation{Université Paris-Saclay, CNRS, CEA, Institut de physique théorique, F-91191, Gif-sur-Yvette, France}

\emailAdd{edmond.iancu@cea.fr}

\abstract{We present a comprehensive introduction to the Colour Glass Condensate (CGC) effective theory, a first-principles framework for hadronic interactions in high-energy Quantum Chromodynamics (QCD) at large parton densities. The focus is on the underlying physical picture and associated emergent phenomena, such as gluon saturation, transverse momentum broadening, and geometric scaling. We introduce the key components of the CGC formalism: the semiclassical approximation for dense gluon systems at saturation, the nonlinear JIMWLK equation governing the energy evolution of multi-gluon correlations, and its mean-field approximation, the BK equation. The formalism is illustrated through applications to selected "dilute-dense" processes, namely high-energy electron-nucleus and proton-nucleus collisions. Finally, we briefly discuss advanced topics including higher-order radiative corrections, the Langevin formulation of the JIMWLK equation, and the emergence of transverse-momentum-dependent (TMD) factorisation from the CGC effective theory. 
}

\keywords{\textit{Quantum Chromodynamics at High Energy; Deep Inelastic Scattering at small $x$; Proton-Nucleus Collisions; Parton Distributions; Gluon Saturation; Colour Glass Condensate Effective Theory; High Energy Evolution in QCD; Transverse Momentum Dependent  Factorisation}}
\begin{document}

\maketitle 


\bigskip

\section{Introduction} \label{sec:intro}
\setcounter{equation}{0} 

A fundamental objective of nuclear physics is to understand the structure of hadrons and nuclei (with nucleon number $A>1$) in terms of the fundamental degrees of freedom of the QCD Lagrangian—quarks and gluons. At high energies, gluons carrying a small fraction $x\ll 1$ of their parent nucleon longitudinal momentum are copiously produced via bremsstrahlung emissions from valence quarks or from higher-$x$ gluons. Because these small-$x$ gluons possess large longitudinal wavelengths, they overlap extensively, even when sourced by different nucleons within a nucleus. Consequently, the quantum evolution of QCD toward decreasing $x$ leads to an increasingly dense system of gluons. In heavy nuclei such as lead, gold, or uranium ($A\sim 200$), this gluon density is further amplified by the large number of sources ($3A$ valence quarks). For sufficiently small values of $x$ and large values of $A$, the gluon density increases to the point where gluons become closely packed and their mutual interactions dominate. Non-linear effects like gluon recombination are then expected to saturate the growth, capping the gluon occupation numbers to values of the order of $1/\alpha_s$, where $\alpha_s$ is the QCD running coupling~\cite{Gribov:1984tu,Mueller:1985wy,McLerran:1993ni,McLerran:1993ka}. This phenomenon, known as \textit{gluon saturation}, is expected to be universal—manifesting in all hadronic systems, from pions to protons and nuclei—though its onset is significantly enhanced in nuclear targets. 

In this high-density regime, multi-parton correlations dominate and the traditional description of hadrons based on independent parton distributions loses its validity. While highly non-linear, this regime is weakly coupled due to the combination of high gluon density and asymptotic freedom, thereby allowing for first-principles theoretical investigations. Such investigations are, however, challenging: even if the coupling is weak, ordinary perturbative techniques fail due to the large gluon occupation numbers. Furthermore, the relevant values of $\alpha_s$ are not asymptotically small: the characteristic momentum scale—the \textit{saturation momentum} $Q_s(x,A)$ which defines the boundary where gluon recombination balances bremsstrahlung—increases with $A$ and with decreasing $x$, roughly like
\begin{equation}
Q_s^2(x, A) \propto A^{1/3} \left(\frac{1}{x}\right)^{\lambda_s},
\end{equation}
with $\lambda _s\sim 0.2\text{--}0.3$, but for the current phenomenology it remains \textit{semi-hard}: for a lead nucleus at $x\sim 10^{-3}$, one expects $Q_s\sim 1\text{--}2$~GeV, which implies $\alpha_s(Q_s)\sim 0.3$.

These considerations demonstrate that a meaningful description of saturation physics requires all-order resummations of non-linear effects combined with accurate perturbative calculations. The \textit{colour Glass Condensate (CGC)} effective theory offers a systematic framework for such advanced calculations in perturbative QCD~\cite{Iancu:2002xk,Iancu:2003xm,Gelis:2010nm,Kovchegov:2012mbw}. Within this framework, the small-$x$ gluons in a hadron wavefunction are described in terms of strong, random classical colour fields that encode multi-gluon correlations. Central to the CGC framework is an infinite set of coupled non-linear equations—the B-JIMWLK hierarchy (Balitsky, Jalilian-Marian, Iancu, McLerran, Weigert, Leonidov, and Kovner)~\cite{Balitsky:1995ub,JalilianMarian:1997jx,JalilianMarian:1997gr,Kovner:2000pt,Weigert:2000gi,Iancu:2000hn,Iancu:2001ad,Ferreiro:2001qy}—which governs the evolution of these gluon correlations with decreasing $x$. In practice, a mean-field truncation of this hierarchy, known as the BK (Balitsky-Kovchegov) equation~\cite{Kovchegov:1999yj}, offers a highly tractable and widely used alternative. The predictions of the CGC theory have been extensively developed—in several instances up to next-to-leading order (NLO) in the perturbative expansion in powers of $\alpha_{s}$—and have demonstrated excellent agreement with a wealth of experimental data spanning electron-proton ($ep$) collisions at HERA, deuteron-gold collisions at RHIC, and proton-lead ($pA$) collisions at the LHC~\cite{GolecBiernat:1998js,GolecBiernat:1999qd,Stasto:2000er,Bartels:2002cj,Iancu:2002tr,Mueller:2002zm,Triantafyllopoulos:2002nz,Iancu:2003ge,Kharzeev:2003wz,Iancu:2004bx,Soyez:2007kg,Albacete:2003iq,Albacete:2010pg,Albacete:2010bs,Albacete:2010sy,Mantysaari:2016jaz,Ducloue:2019jmy,Beuf:2020dxl,Shi:2021hwx,Hanninen:2022gje,Mantysaari:2023xcu,Caucal:2025zkl,Korcyl:2026nrz,JalilianMarian:2005jf,Lappi:2010ek,Albacete:2014fwa,Klein:2019qfb,Morreale:2021pnn}. Furthermore, the CGC provides a compelling description of the early stages of ultrarelativistic heavy-ion collisions as probed at RHIC and the LHC~\cite{Baier:2000sb,Lappi:2006fp,Lappi:2010ek,Gelis:2010nm,Gelis:2012ri,Albacete:2013tpa,Berges:2020fwq}.

Yet, despite its robust theoretical foundations and phenomenological successes, a direct and unambiguous observation of gluon saturation is still elusive. This difficulty stems primarily from the fact that saturation is restricted to a limited region of phase space that is challenging to disentangle from linear QCD evolution, standard nuclear shadowing, or non-perturbative effects like confinement. Nevertheless, its imprint is felt over much wider kinematic ranges—as evidenced, for instance, by the observation of geometric scaling~\cite{Stasto:2000er} in electron-proton deep inelastic scattering (DIS) at HERA. Moreover, special observables, such as the diffractive structure function in DIS, the diffractive particle production in photon-nucleus interactions, and the semi-inclusive production of forward di-jets, or of heavy-flavour mesons, in photon-nucleus or proton-nucleus collisions, remain highly sensitive to gluon saturation even when the external scales (the photon virtuality, the transverse momenta of the jets,  the heavy quark mass) are considerably larger than $Q_s$. While the existence of such hard scales allows for controlled calculations in perturbative QCD (pQCD), it also complicates the search for gluon saturation by opening the phase space for large radiative corrections that dilute the saturation signals. To overcome these challenges, state-of-the-art CGC calculations must achieve the precision necessary to clearly distinguish radiative effects from the non-linear signatures of saturation.

To summarise, convincingly establishing the onset of gluon saturation requires a three-pronged approach: identifying observables tailored to the gluon dynamics at the scale $Q_s$, constructing dedicated experimental sub-systems capable of measuring these observables with high precision, and delivering theory calculations accurate enough to prevent ambiguous data interpretation.

On the experimental front, major breakthroughs are expected from the upcoming Run 4 at the LHC and the future Electron-Ion Collider (EIC) under construction at Brookhaven National Laboratory (BNL). For Run 4, the LHC will deploy major detector upgrades, most notably the Forward Calorimeter (FoCal)~\cite{ALICE:2023fov,ALICE:2024jtt} within the ALICE experiment. FoCal will enable pioneering measurements of forward particle production—including isolated direct photons, neutral mesons, photon-jet correlations, and dijets—in $pp$, $p\text{--}\text{Pb}$, and ultra-peripheral $\text{Pb}\text{--}\text{Pb}$ collisions at unprecedentedly small momentum fractions ($x \sim 10^{-6}$). In parallel, exploring gluon saturation stands as a primary scientific pillar of the future EIC~\cite{Boer:2011fh,Accardi:2012qut,Aschenauer:2017jsk,AbdulKhalek:2022hcn}. Although its center-of-mass energies will be more modest than those of HERA and the LHC, the EIC capability to collide electrons with heavy ions up to lead, coupled with its high luminosity and beam polarization, will permit exquisite, multi-dimensional scans of gluon saturation through diffractive processes, exclusive vector meson production, and di-hadron angular correlations.

On the theory side, there is an ongoing, community-wide effort to promote the CGC effective theory to next-to-leading order (NLO) accuracy and beyond. This programmatic shift encompasses both the aforementioned non-linear (B-JIMWLK and BK) evolution equations and the impact factors expressing partonic cross-sections in a dense gluon environment. Such advanced perturbative calculations are particularly challenging due to the need to account for strong non-linear effects; indeed, multiple scatterings must be resummed to all orders in the gluon density (or the associated strong colour fields) at each order in $\alpha_s$. Special techniques have been developed to this end, such as the colour dipole picture for photon-nucleus interactions or the eikonal approximation for a parton multiple scattering off a nuclear shockwave, which exploit the high-energy kinematics and the wide separation of spatio-temporal (or energy-momentum) scales. This review paper aims to provide a pedagogical introduction to these modern techniques, with a primary focus on the underlying physical picture.

\section{Gluon evolution towards saturation: an elementary discussion}
\label{sec:gluons}
\setcounter{equation}{0} 

At rest or low energies, a proton is a highly non-perturbative system of virtual quark and gluon fluctuations bound by confinement. These elementary degrees of freedom carry the colour charge of the non-Abelian $SU(N_c)$ gauge symmetry (with $N_c=3$), which gets neutralised ("bleached") over a characteristic distance defined by the confinement scale, $R_p \sim 1/\Lambda_{\rm QCD} \sim 1$~fm. 
This scale also controls the running of the strong coupling $\alpha_s(Q^2)$, which becomes $\mathcal{O}(1)$ at $Q^2 \sim \Lambda_{\rm QCD}^2$. Driven by strong interactions, these virtual fluctuations possess transverse momenta and virtualities of order $\Lambda_{\rm QCD}$ (hereafter denoted simply as $\Lambda$). By the uncertainty principle, they exhibit lifetimes $\Delta t \sim 1/\Lambda$, continuously emerging from and disappearing into the vacuum, meaning that the internal structure of a low-energy proton is inherently time-dependent on the interaction timescale.

This picture changes dramatically when the proton is accelerated to ultrarelativistic energies along the $z$-axis ($E = \gamma M_p\simeq P_z $, with $\gamma \gg 1$). Due to Lorentz time dilation, the lifetimes of the virtual fluctuations expand to $\Delta t \sim \gamma / \Lambda$, allowing them to behave as nearly on-shell, quasi-free, and quasi-collinear constituents known as \textit{partons}. Up to corrections of order $1/\gamma$, a parton 4-momentum can be written as $k^\mu \simeq (k_z, 0, 0, k_z) \simeq xP^\mu$, where $P^\mu$ is the proton 4-momentum and $x \equiv k_z/P_z < 1$ is the longitudinal momentum fraction. 

While parton interactions remain strong on these dilated timescales, Lorentz contraction alters their spatial distribution. The fast valence quarks ($x \sim 1/3$) are compressed into a thin "pancake" or shockwave with transverse radius $R_p$ and a narrow longitudinal width $\Delta z \sim 1/P_z$. This dense core is surrounded by a cloud of long-lived "wee" partons ($x \ll 1$) whose longitudinal extension increases with decreasing $x$.

Although the bulk parton distributions are non-perturbative, hard scattering processes involving a large transverse momentum transfer $q_\perp \gg \Lambda$ can be calculated using perturbative QCD (pQCD), since $\alpha_s(q_\perp^2) \ll 1$. Because the collision duration ($\Delta t_{\rm coll} \sim 1/q_\perp$) is much shorter than the partonic fluctuation lifetime ($\Delta t \sim \gamma/\Lambda$), the hard scattering factorises from the long-lived quantum fluctuations within the incoming hadrons. For a generic hard process $p+p \to 3+4+X$, the cross-section satisfies \textit{collinear factorisation}~\cite{Sterman:2026vuh}:
\begin{align}\label{coll}
\frac{\rmd \sigma}{\rmd x_1 \rmd x_2 \rmd q_{\perp}^2} (p+p\to 3+4 + X) = \sum_{ij} F_{i}(x_1,\mu^2)\, F_{j}(x_2,\mu^2)\, \frac{\rmd \hat\sigma_{ij}}{\rmd q^2_{\perp}}(1+2\to 3+4)\,,
\end{align}
where $F_{i}(x,\mu^2)$ are the non-perturbative \textit{parton distribution functions} (PDFs) for quarks, antiquarks, and gluons ($i=q,\bar q,g$), and the factorisation (or ``resolution'') scale $\mu^2$ can be understood as an upper limit on the transverse momenta of the partons that participate in the scattering. By the uncertainty principle, any parton with $k_\perp \lesssim q_\perp$ occupies a transverse area $S_\perp \sim 1/k_\perp^2 \gtrsim 1/q_\perp^2$, allowing it to overlap with the hard process and thus contribute to the scattering. This resolution argument shows that $\mu^2$ must be chosen of the order $q_\perp^2$. Because $\mu^2 \gg \Lambda^2$, the evolution of the PDFs with increasing the resolution scale can be computed in pQCD. In the leading logarithmic approximation with respect to (w.r.t) the large transverse logarithm $\ln(\mu^2/\Lambda^2)$, this evolution is governed by the \textit{DGLAP equations} (Dokshitzer, Gribov, Lipatov, Altarelli, and Parisi)~\cite{Gribov:1972ri,Altarelli:1977zs,Dokshitzer:1977sg}, which describe the proliferation of partons through successive hard splittings ($q \to qg$, $g \to gg$, $g\to q\bar q$).

The DGLAP  solutions resum perturbative corrections of the type $[\alpha_s(\mu^2)\ln (\mu^2/\Lambda^2)]^n$ to all orders $n\ge 1$ for generic values of $x$. However, the radiative processes in pQCD also lead to a rapid proliferation of partons with small values of  $x$. This growth is dominated by soft ($x\ll 1$) gluon emissions, eventually generating a high-density form of gluonic matter. To see this, consider the leading-order (LO) bremsstrahlung spectrum for a soft gluon with transverse momentum $k_\perp\gg \Lambda$ emitted by a valence quark:
\begin{equation}\label{bremR} 
\frac{\rmd N_g^{(0)}}{\rmd x\rmd k_\perp^2} \,\simeq\,\frac{\alpha_s C_F}{\pi}\,\frac{1}{x}\,\frac{1}{k_\perp^2}\,,
\end{equation}
with the colour factor $C_F\equiv (N_c^2-1)/2N_c$. The characteristic $1/x$ infrared pole is unique to gluon radiation and drives the QCD evolution towards small $x$. 
Integrating this bremsstrahlung spectrum up to a hard resolution scale $Q^2$ yields the LO gluon PDF sourced by the $N_c=3$ valence quarks:
\begin{equation}\label{xGxR}
xG^{(0)}(x, Q^2) =  N_c \int_{\Lambda^2}^{Q^2} \rmd k_\perp^2\,x \frac{\rmd N_g^{(0)}}{\rmd x\rmd k_\perp^2} = \frac{\alpha_s C_F N_c}{\pi}\,\ln\frac{Q^2}{\Lambda^2}\,.
\end{equation}
As shown by the notation, it is customary to multiply $G(x, Q^2)$ by $x$ and to refer to the product 
$x G(x, Q^2)$ as {\it the} gluon distribution function.  With this normalisation, the gluon PDF at small $x$ and LO is independent of $x$. The integral representation in \eqref{xGxR} motivates the interpretation of the integrand  as the {\it unintegrated gluon distribution} (UGD), which in general is defined as:
\begin{equation}\label{unint}
x f_g(x, \bk_\perp) \equiv x \frac{\rmd N_g}{\rmd x\rmd^2\bk_\perp}\,.
\end{equation}

\begin{figure}[t]
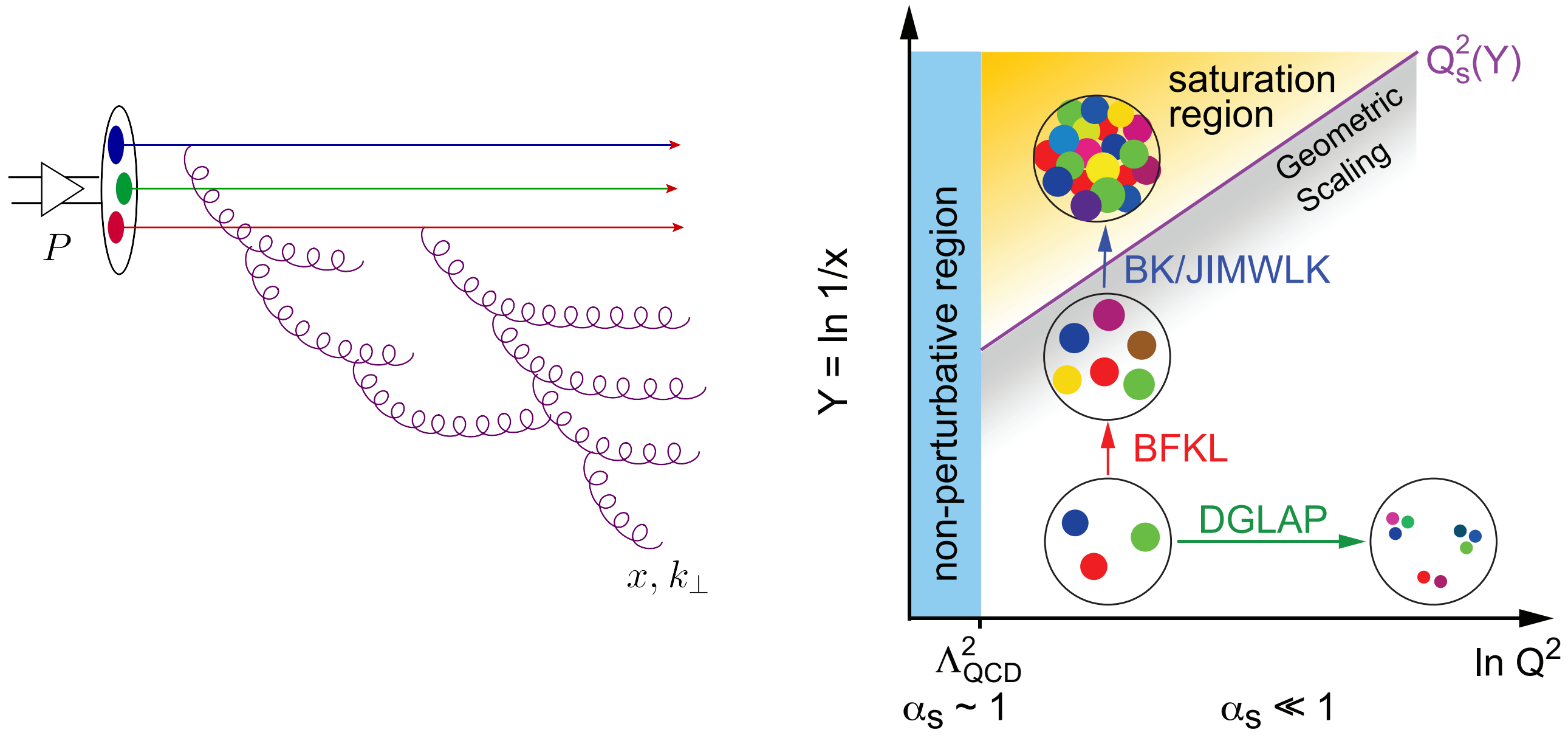
 
  \centering
  \begin{minipage}[t]{0.45\textwidth}
    \mbox{ }\\
    \vspace*{1.cm}
      \includegraphics[page=,width=\textwidth]{PDF/JIMWLK_Cascade.pdf}
  \end{minipage}
  \hfill%
  \begin{minipage}[t]{0.48\textwidth}
    \mbox{ }\\
      \includegraphics[page=,width=\textwidth]{PDF/CGC_phase-space.pdf}
        \end{minipage}
        \caption{\sl Left:  A gluon cascade in the high density regime at small $x$, including a recombination process ($gg\to g$) leading to gluon saturation.
           Right: the ``phase-diagram'' for gluon evolution
in pQCD; each coloured blob represents a gluon with transverse area
$S_\perp \sim 1/Q^2$ and longitudinal momentum $k_z=xP$. The
saturation line $\ln Q_s^2(x)=\lambda_s Y$, with $Y\equiv \ln(1/x)$,  separates between dense and dilute regimes.
  \label{XPhase}}
\end{figure}

Unlike photons, gluons carry colour charge and can source secondary gluons with even smaller momentum fractions ($x'\ll x$), initiating a gluon cascade (see the left panel in Fig.~\ref{XPhase} for an illustration). At sufficiently low $x$ and large $k_\perp^2$, the lowest-order approximation \eqref{bremR}  breaks down due to multiple emissions. For instance, a secondary emission where an intermediate unobserved gluon ($x_1, p_\perp$) splits into the final observed gluon ($x, k_\perp$) generates the following first-order correction to the UGD:
\begin{align}\label{xf1}
x f_g^{(1)}(x, \bk_\perp) &= \frac{\alpha_s C_F N_c}{\pi^2}\,\frac{1}{k_\perp^2} \,\frac{\alpha_s C_A}{\pi}\int_x^1\frac{\rmd x_1}{x_1} \int_{\Lambda^2}^{k_\perp^2} \frac{\rmd p_\perp^2}{p_\perp^2} \nonumber \\[0.2cm]
&= \abar\ln\frac{1}{x}\ln\frac{k_\perp^2}{\Lambda^2} \,x f_g^{(0)}(x, \bk_\perp)\,,
\end{align} 
where $C_A \equiv N_c$ is the colour factor for the $g\to gg$ vertex and
$\abar\equiv \alpha_s N_c/\pi$. Although suppressed by a power of $\alpha_s$, this contribution is amplified by the product of a large longitudinal logarithm, $\ln(1/x)$, and a large transverse logarithm, $\ln({k_\perp^2}/{\Lambda^2})$. In the small-$x$ and/or large-$k_\perp^2$ regime, such radiative corrections become large, necessitating an all-orders resummation of gluon configurations strongly ordered in either $x$ ($x_1\gg x_2 \gg \dots \gg x$), or $p_\perp^2$ ($p_{1\perp}^2\ll p_{2\perp}^2\ll \dots \ll k_\perp^2$), or both.

In the \textit{double-logarithmic approximation} (DLA), which retains only those radiative corrections where each power of $\alpha_s$ is accompanied by both a small-$x$ and a transverse logarithm, this resummation can be performed analytically. Iterating the multi-step emission process leading to Eq.~\eqref{xf1} and summing over $n$ intermediate gluon emissions yields:
\begin{align}\label{fgDLA}
x f_g(x, \bk_\perp) &= x f_g^{(0)}(x, \bk_\perp) \sum_{n\ge 0} \frac{1}{(n!)^2} \left[\abar \ln\frac{1}{x}\ln\frac{k_\perp^2}{\Lambda^2}\right]^n \nonumber \\[0.2cm]
&= \frac{\alpha_s C_F N_c}{\pi^2 k_\perp^2} \,{\rm I}_0\left(2\sqrt{\abar Y\rho}\right) \sim \frac{\alpha_s(N_c^2-1)}{k_\perp^2} \exp\left\{2\sqrt{\abar \ln\frac{1}{x} \ln\frac{k_\perp^2}{\Lambda^2}}\right\}\,,
\end{align} 
where $Y\equiv \ln(1/x)$ and $\rho\equiv \ln({k_\perp^2}/\Lambda^2)$ are assumed to be large, and ${\rm I}_0(z)$ is the modified Bessel function, that in the final approximation is evaluated for asymptotically large $z$. Eq.~\eqref{fgDLA} exhibits a rapid growth of the gluon distribution with decreasing $x$ and increasing $k_\perp^2$. However, DLA (and by extension the full DGLAP evolution) is not the optimal regime to study saturation. Indeed, in these evolutions, the transverse momentum $k_\perp$ is assumed to strongly increase from one generation to the next one, meaning that the newly emitted gluons occupy a progressively smaller transverse area $S_\perp \sim 1/k_\perp^2$, which reduces their probability to overlap and interact. Consequently, DGLAP and DLA describe an evolution towards a \textit{decreasing local gluon density}.  The above discussion leads to two important lessons:

\begin{description}
\item[\texttt{(i)} The phase-space occupation number:] To properly characterize high-density QCD effects, counting the spatial gluon density ($xG(x, Q^2)/\pi R_p^2$) is insufficient. One must account for the fact that gluons are extended objects whose interaction cross-section is governed by their transverse size. Two overlapping gluons with momenta $k_{1\perp}$ and $k_{2\perp}$ interact with a strength proportional to $\alpha_s N_c (k_<^2/k_>^2)$ [with $k_<$ ($k_>$) the smaller (the larger) among $k_{1\perp}$ and $k_{2\perp}$], effectively restricting strong interactions to partons of comparable transverse size. A rigorous measure of these non-linear effects is provided by the gluon \textit{occupation number} $n(x, \bk_\perp, \bx_\perp)$ in transverse phase-space. Assuming a homogeneous distribution in the transverse plane, this is estimated as (see Sect.~\ref{sec:sat} for a more precise definition)
\begin{equation}\label{nxQ}
n(x, Q) \sim \frac{1}{2(N_c^2-1)} \frac{xG(x, Q^2)}{\pi R_p^2 Q^2}\,.
\end{equation}
The factor ${1}/{[2(N_c^2-1)]}$ accounts for specific colour and polarisation states: gluons can exist in $N_c^2-1$ colour states and 2 transverse polarisation states. The second factor represents the ratio between the total transverse area $xG(x, Q^2)/Q^2$ that would be occupied by gluons with an individual  area $S_\perp=1/Q^2$ and the proton area $S_p = \pi R_p^2$. When this factor becomes larger than one, gluons with a given transverse area are bound to overlap with each other. Notice that no special constraint is required on the longitudinal phase-space: small-$x$ gluons are strongly delocalised along the beam axis ($\Delta z \sim 1/xP_z$), ensuring they naturally overlap in the longitudinal direction.

\item[\texttt{(ii)} The BFKL evolution:] To achieve large occupation numbers, one must maximize gluon multiplication at a fixed transverse momentum scale. This is achieved via the \textit{BFKL evolution} (Balitsky, Fadin, Kuraev, and Lipatov)~\cite{Lipatov:1976zz,Kuraev:1977fs,Balitsky:1978ic}, which resums successive soft gluon emissions strongly ordered in $x$ but featuring comparable transverse momenta. (The DLA evolution can be viewed as a common limit of the DGLAP and BFKL equations, where the evolution proceeds diagonally in the $(Y,\,\rho)$ space while remaining strictly within the dilute regime.)
This framework resums terms of order $[\alpha_s \ln(1/x)]^n$ to all orders, predicting a sharp power-law growth of the gluon density: $x f_g(x, \bk_\perp) \sim (1/x)^\omega$, where the leading-order BFKL Pomeron intercept is given by $\omega = 4\ln 2 \, \abar \approx 2.77 \, \abar$.
\end{description}

To estimate the effects of the linear BFKL evolution on the gluon occupation number, we substitute the asymptotic power-law growth $x G(x,Q^2)\sim \alpha_s(N_c^2-1)(1/x)^\omega$ into Eq.~\eqref{nxQ}, yielding:
\begin{equation}\label{nxQBFKL}
n(x, Q^2)\sim\,\frac{\alpha_s}{R_p^2 Q^2}\,\frac{1}{x^\omega}\,.
\end{equation}
While Eq.~\eqref{nxQBFKL} predicts that $n(x, Q^2)$ grows boundlessly as $x \to 0$, this linear framework is valid only in the dilute regime where $n\ll 1/\alpha_s$. When the occupation number reaches $n\sim 1/\alpha_s$, a dense system of $\sim 1/\alpha_s$ gluons piles up in phase space, generating a collective interaction strength of $\alpha_s n \sim \mathcal{O}(1)$. At this stage, the system becomes intrinsically non-linear, meaning that recombination effects ($gg\to g$) become important and the linear BFKL evolution is not appropriate anymore. A qualitative picture of the recombination effects within the gluon cascades at small $x$ is shown in the left panel in Fig.~\ref{XPhase}, while the right panel illustrates the regions for linear (DGLAP, BFKL, DLA) and respectively non-linear (BK/JIMWLK; see below) quantum evolution in the $(Y, \rho)$ phase-space, with $Y= \ln(1/x)$ and $\rho= \ln(Q^2/\Lambda^2)$.

Although not accurate anymore, the linear BFKL solution can be used to estimate the boundary of this dense regime, which defines the characteristic transverse momentum scale---the \textit{saturation momentum} $Q_s(x)$ where the gluon occupation number becomes of $\order{1/\alpha_s}$. Setting $n(x, Q_s(x))\simeq 1/\alpha_s$ in Eq.~\eqref{nxQBFKL} yields the parametric relation
\begin{equation}\label{Qs0}
Q_s^2(x) \,\sim \,\frac{\alpha_s^2}{R_p^2}\,\frac{1}{x^{\lambda_s}}\,,\qquad\text{with}\qquad \lambda_s = \omega \simeq 2.77\alpha_s.
\end{equation}
As we will explain in Sect.~\ref{sec:QsY}, this qualitative estimate is strongly modified by the effects of saturation, the sub-asymptotic kinematics, the non-locality of the transverse evolution, and the higher-order running coupling corrections. Incorporating these physical effects maintains the power-law scaling $Q_s^2(x)\sim (1/x)^{\lambda_s}$, but significantly suppresses the \textit{saturation exponent} $\lambda_s$ to a phenomenologically acceptable value of $\lambda_s\simeq 0.2\text{--}0.3$~\cite{Triantafyllopoulos:2002nz}.

Eq.~\eqref{Qs0} implies that, for sufficiently small $x$, $Q_s(x)\gg \Lambda$, which drives the running coupling into the perturbative regime, $\alpha_s(Q_s)\ll 1$. Thus, dense gluonic matter appears to be \textit{weakly coupled}, which legitimates perturbative investigations, like our above arguments. As we shall see via explicit calculations later on, the gluon occupation numbers \textit{saturate} at a value $n\sim 1/\alpha_s$, due to destructive non-linear interactions: further radiation of gluons with $k_\perp \lesssim Q_s(x)$ is dynamically blocked, forcing any additional emissions to populate larger transverse momenta $k_\perp > Q_s$~\cite{Gribov:1984tu,Mueller:1985wy,McLerran:1993ni,McLerran:1993ka,Mueller:1999wm}.

Eq.~\eqref{nxQ} suggests an alternative strategy to amplify the gluon density: utilizing heavy nuclei with mass number $A\approx 200$ (e.g., Pb or Au). A large nucleus contains $A$ times more colour sources than a proton, meaning $xG_A(x, Q^2)\approx A xG_p(x,Q^2)$ if the valence quarks act independently. Because these sources are distributed over a larger nuclear area $\pi R_A^2\simeq A^{2/3} \pi R_p^2$, Eq.~\eqref{nxQ} predicts that the gluon occupation number—and consequently the saturation scale—is enhanced by a nuclear "oomph" factor: $Q_s^2(x,A) \approx A^{1/3} Q_s^2(x,p)$. This nuclear enhancement is a major cornerstone for upcoming programs at the Electron-Ion Collider (EIC)~\cite{Boer:2011fh,Accardi:2012qut,Aschenauer:2017jsk,AbdulKhalek:2022hcn} and forward upgrades at the LHC like FoCal~\cite{ALICE:2024jtt}, where heavy-ion targets bring the saturation regime into a fully accessible kinematic window.

The study of the saturated gluon matter and of the QCD evolution towards saturation is the scope of the   \textit{colour Glass Condensate (CGC)} effective theory to be discussed in this review paper. This framework relies on two pillars: \texttt{(i)} a  factorisation scheme adapted to high energy collisions off dense hadrons which includes the effects of gluon saturation and multiple scattering, and  \texttt{(ii)}  non-linear evolution equations which describe the evolution of the QCD scattering amplitude with decreasing $x$ in a dense gluon environment. These tools generalise the traditional high-energy factorisation~\cite{Catani:1990eg,Catani:1994sq}, valid in the dilute regime and based on the BFKL equation. In a dense gluon environment, single-gluon distributions extend to multi-parton correlations governed by the infinite Balitsky hierarchy of coupled equation~\cite{Balitsky:1995ub} or the functional JIMWLK equation~\cite{JalilianMarian:1997jx,JalilianMarian:1997gr,Kovner:2000pt,Iancu:2000hn,Iancu:2001ad,Ferreiro:2001qy}. Under a mean-field truncation, this infinite hierarchy reduces to the non-linear BK equation~\cite{Kovchegov:1999yj}. Grounded in perturbative QCD, these tools are currently known up to next-to-leading order (NLO) accuracy in $\alpha_s$, enabling precision mapping of the high-density frontier. A fundamental consequence of the non-linear phenomena embedded in these tools is the fulfilment of the  \textit{unitarity constraints} by the QCD scattering amplitudes in the high-energy limit.

\section{Dilute-dense collisions in the CGC: two examples}
\label{sec:DD}

Throughout this review, we will be exclusively concerned with ``dilute-dense collisions,'' wherein a dilute projectile (such as a virtual photon, or a proton probed at large longitudinal fractions $x$) probes the high-density gluon distribution of a dense hadronic target (such as a heavy nucleus, or a proton probed at small $x$). These processes are particularly well suited for exploring gluon saturation, as collective high-density effects can be directly mapped from measured observables. This is at variance with the more intricate case of nucleus-nucleus (``dense-dense'') collisions, where initial-state signatures—including gluon saturation in the incoming nuclear wavefunctions—are often obscured or entirely washed out by strong final-state interactions, such as multiple parton scattering occurring within the hot and dense Quark-Gluon Plasma (QGP) created in the wake of the collision. 

Dilute-dense collisions are also significantly more tractable from a theoretical standpoint: for our purposes, multiple scatterings can be systematically resummed to all orders within the eikonal approximation. In this section, we introduce two benchmark dilute-dense processes—photon-nucleus ($\gamma A$) and proton-nucleus ($pA$)  collisions—that will allow us to explain the CGC factorisation and the eikonal approximation. These processes will subsequently serve as our baseline to illustrate the predictive power and physical consequences of the CGC approach. For extensions of the CGC framework to dense-dense environments, we refer the reader to~\cite{Lappi:2006fp,Gelis:2008rw,Gelis:2008ad,Gelis:2009wh,Lappi:2010ek,Gelis:2010nm,Gelis:2012ri,Albacete:2013tpa}, while developments going beyond the eikonal approximation can be found in~\cite{Altinoluk:2014oxa,Altinoluk:2015gia,Altinoluk:2021lvu,Altinoluk:2022jkk,Altinoluk:2023qfr,Altinoluk:2025ivn,Kovchegov:2015pbl,Kovchegov:2018znm,Cougoulic:2022gbk,Kovchegov:2021iyc,Chirilli:2018kkw,Chirilli:2021lif,Li:2026azt} and references therein.

\subsection{Photon-nucleus collisions at small $x$: the dipole picture}
\label{sec:DIS}

\begin{figure}[t]
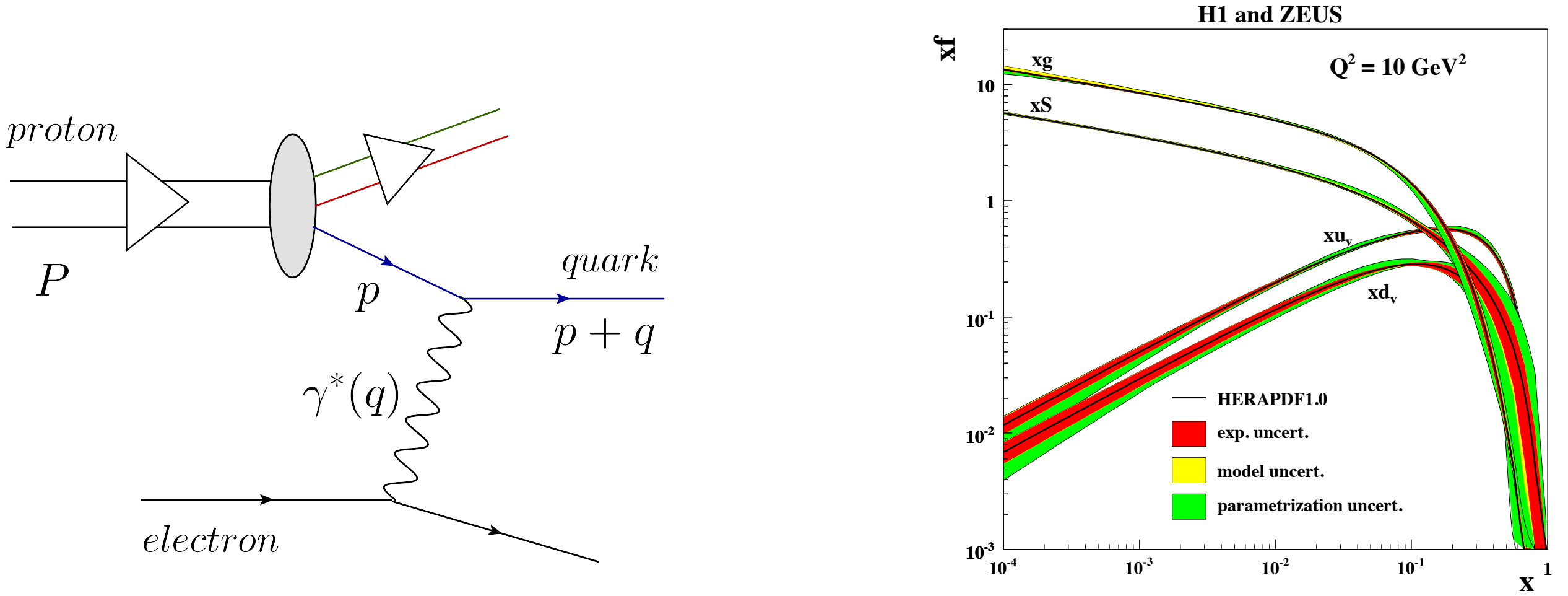
 
  \centering
  \begin{minipage}[t]{0.45\textwidth}
    \mbox{ }\\
    \vspace*{1.cm}\includegraphics[page=,width=0.92\textwidth]{PDF/DIS-IMF-ee.pdf}
  \end{minipage}
  \hfill%
  \begin{minipage}[t]{0.45\textwidth}
    \mbox{ }\\
      \includegraphics[page=,width=\textwidth]{PDF/HERAnew_10GEV_log.pdf}
        \end{minipage}
        \caption{\sl Left:  Electron-proton deep inelastic scattering in the target infinite momentum frame.
        Right: the gluon ($xg$), sea quark ($xS$), and
valence quark ($xu_{\rm v}$, $xd_{\rm v}$) distributions, 
as obtained from a DGLAP fit to the HERA data for $ep$ collisions,
are plotted as functions of $x$ for $Q^2=10$\,GeV$^2$. From~\cite{Aaron:2009aa}.
 \label{HERA-gluon}}
\end{figure}

Electron-proton ($ep$) or electron-nucleus ($eA$) collisions are particularly simple as they proceed via the exchange of a virtual photon ($\gamma^*$) which scatters of a quark from the hadron target and thus probes the quark distribution function  $F_q(x, Q^2)\equiv q(x, Q^2)$ on controlled resolution scales (see the left panel in Fig.~\ref{HERA-gluon}). The transverse resolution scale is set by the photon virtuality, $Q^2= -q^\mu q_\mu > 0$, while the longitudinal momentum fraction $x$ is fixed by the condition that the struck quark be (nearly) on-shell both before the collision, when it has a 4-momentum $p^\mu=xP^\mu_N$ (with $P^\mu_N$ the 4-momentum of the parent nucleon),  and after absorbing the photon, when it emerges in the final state with 4-momentum $p^\mu+q^\mu$. 
These conditions imply
\begin{equation} 
0 = (p+q)^2 = p^2 + q^2 + 2p\cdot q = -Q^2 +2x P_N\cdot q\quad\Longrightarrow\quad
x = \xbj \equiv \frac{Q^2}{2q\cdot P_N}\,.\end{equation}
We are interested in the high energy regime where the centre-of-mass (COM) energy squared $s\equiv (P_N+q)^2$ of the $\gamma^* N$ collision is much larger than the photon virtuality. Using
 $s= (P_N+q)^2 =  -Q^2 +2x P_N\cdot q$, one finds that this condition corresponds to $2x P_N\cdot q\gg Q^2$, or $x\ll 1$. So, DIS at high energy is probing ``wee'' quarks in the hadron wavefunction. In practice we will consider values $x \le 10^{-2}$, since $x_0=10^{-2}$ can be viewed as the upper boundary of the BFKL evolution: $\alpha_s\ln(1/x_0)\sim 1$.

 The total cross-section for the photon absorption by the target is suggestively written as  (we generally consider a nuclear target with mass number $A$; the proton case corresponds to $A=1$)
 \begin{equation}\label{sigmagamma}
 \sigma_{\gamma^* A}(x,Q^2)\,=\,
 \frac{4\pi^2 \alpha_{\rm em}}{Q^2}\,F_2(x,Q^2)\,,\end{equation}
 where the ``hard'' factor ${4\pi^2 \alpha_{\rm em}}/{Q^2}$ is the elementary cross-section for
photon absorption by a point-like fermion with unit electric charge, while 
the nuclear {\em structure function} $F_2(x,Q^2)$ counts the number of quarks and antiquarks of any flavour $f$ on the resolution scales alluded to above and weighted by the respective electric charges  squared  $e_f^2$:
 \begin{eqnarray}\label{F2}
 F_2^A(x,Q^2)\,=\,\sum_{f} e_f^2\,\big[xq^A_f(x,Q^2) + x\bar
 q_{f}^A(x,Q^2)\big]\,.
 \end{eqnarray}
 When $x\ll 1$ (the case of interest for us here), the (anti)quark PDFs truly count the number of {\it sea} quarks, i.e. the quarks and the antiquarks that are produced in pairs via radiative processes—mainly, via gluon decay: $g\to q\bar q$. Indeed, the distributions of the valence quarks are peaked around $x\sim 1/3$ and rapidly decrease with decreasing $x$. Vice-versa, the number of gluons is rapidly increasing with $1/x$, via the QCD evolution of BFKL type, as discussed in Sect.~\ref{sec:gluons}. 
This is indeed visible in the experimental results for the PDFs, as obtained from DGLAP fits to the HERA data for $ep$ collisions (see the right panel in  Fig.~\ref{HERA-gluon} and the recent review paper \cite{Cooper-Sarkar:2026yzt} for a thorough discussion of the DIS phenomenology at HERA).
 In a first approximation, all the sea quarks come from gluon decays, so one roughly has $xq_A(x,Q^2) + x\bar q_A(x,Q^2)\approx \alpha_s xG_A(x, Q^2)$.
 This argument shows that DIS at small $x$ rather directly probes the gluon distribution in the target. One can make this argument more explicit by working in a special Lorentz frame—the {\it dipole frame}.
  
\begin{figure}
\centerline{
\includegraphics[width=.9\textwidth]{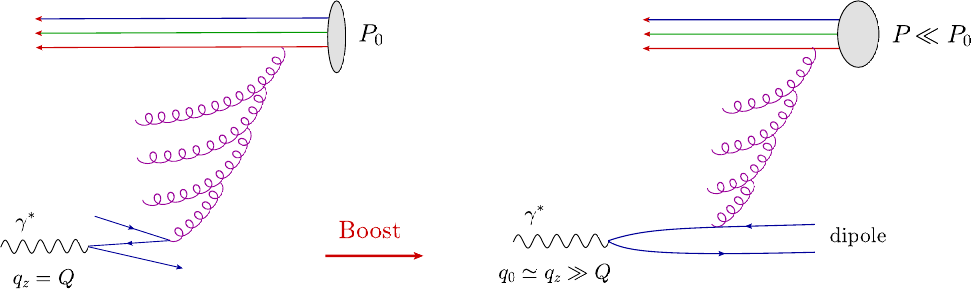}}
\caption{\sl The absorption of the photon by a sea quark seen in two different frames. Left: the Breit frame,
 where the photon is a  longitudinal standing wave: $q^\mu=(0, 0, 0, q_z)$ with $Q^2 = q_z^2$. Right: the dipole frame, as obtained from the Breit frame by a large boost in the positive $z$ direction.}
\label{fig:boost}
\end{figure}

 So far, our discussion was implicitly carried in a target  ``infinite momentum frame'' (like the Breit frame, cf. the caption of Fig.~\ref{fig:boost}), in which the nucleon has a large longitudinal momentum $P_N\gg M$, while the photon is relatively slow ($q_z\sim Q$). In this frame, the Feynman graph for a typical amplitude contributing to DIS at small $x$ has the structure shown in the left side of Fig.~\ref{fig:boost}: a BFKL-like gluon cascade initiated by one of the valence quarks ends with a gluon decay $g\to q\bar q$ and then the photon is absorbed  by either the quark, or the antiquark. As suggested by the drawing, we have chosen the hadron to be a left mover, $P_{0z}< 0$, while the photon has $q_z>0$.  Let us perform a large boost in  the positive $z$ direction: in the boosted frame, the nucleon is still a relativistic left mover, but with a much smaller longitudinal momentum ($M_N \ll |P_z| \ll |P_{0z}|$), while the photon is a relativistic right mover $q^\mu=(q_0, 0, 0, q_z)$ with $q_0\simeq  q_z\gg Q$ and $q_z^2-q_0^2 = Q^2$.
 
The picture of DIS is now different, as shown in the right hand side of Fig.~\ref{fig:boost}:
before the collision, the virtual photon decays into a virtual  $q\bar q$ pair which then scatters off a gluon from the target (itself produced via a BFKL cascade). 
Prior to the scattering, the $q\bar q $ pair is in a colour singlet state (since produced via photon decay), hence it forms a {\it colour dipole}. This dipole is truly {\it virtual}: in the absence of the collision, the $q\bar q$ pair would have  a finite lifetime $\Delta t_{\rm dip}$ before recombining back into the photon. This lifetime is however long, since enhanced by Lorentz time dilation. Importantly, it is much larger than the longitudinal extent $L \sim A^{1/3}/P_z$ of the nuclear target (which is roughly the same as the duration of the collision). One has indeed
\begin{equation}\label{cohtime}
\Delta t_{\rm dip}\,\simeq\,\frac{2q_z}{Q^2}\ \gg \ L 
 \simeq \frac{A^{1/3}}{P_z} \qquad\mbox{when}\qquad x\,\sim\, \frac{Q^2}{2q_zP_z}\,\ll\,A^{-1/3}\,.
 \end{equation}
Using $A^{1/3}\sim 6$ for $A\simeq 200$, one sees that the last condition is very well satisfied for any $x\le 0.01$.

This strong temporal separation justifies a  two-step factorisation of the cross-section for $\gamma^* A$ scattering: \texttt{(i)} the formation of the $q\bar q$ dipole via the photon decay $\gamma^*\to q\bar q$, and  \texttt{(ii)} the collision between the colour dipole and the small-$x$ gluons in the target. The first step is purely electromagnetic\footnote{This more precisely holds to leading order in $\alpha_s$. Beyond leading order, one needs to include additional QCD radiation in the photon wavefunction prior to the collision.}
 and can be easily computed in QED perturbation theory. The second step contains the information about QCD scattering. For sufficiently large $Q^2\gg \Lambda^2$, the dipole is a hard probe and its collision with the target can be computed in perturbation theory (up to non-perturbative aspects like the impact-parameter dependence; see below).

\begin{figure}[t]
 \centerline{   \includegraphics[width=0.42\textwidth]{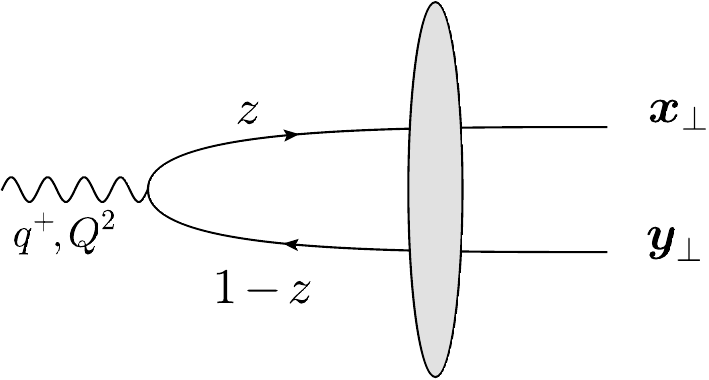}\qquad\qquad
   \includegraphics[width=0.48\textwidth]{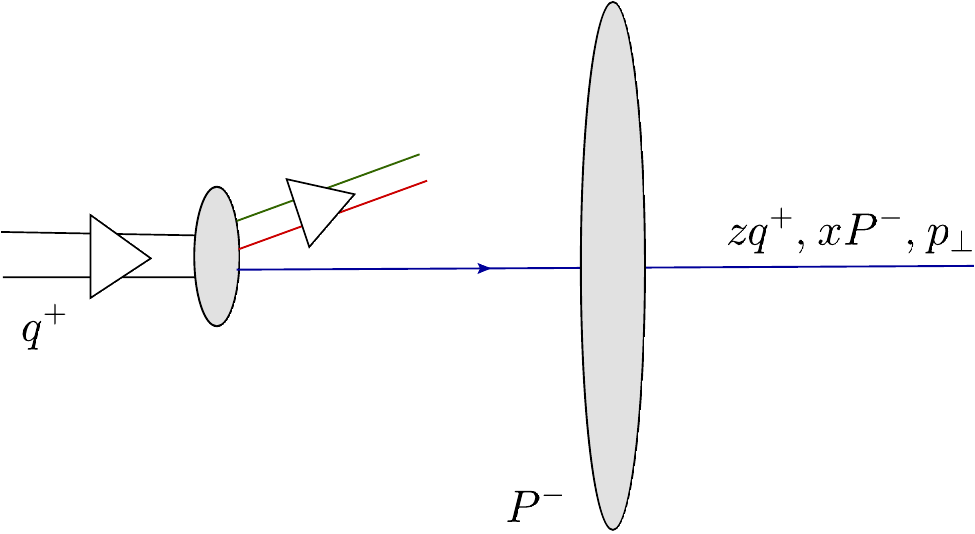}
}
 \caption{\sl Amplitudes for dilute-dense collisions  at small $x$. Left: Photon-nucleus collision in the dipole picture. Right: Quark production in a proton-nucleus collision. The nuclear target is represented as a shockwave. Each parton from the projectile which crosses this shockwave acquires a Wilson line like that in Eq.~\eqref{Wilson}. }
 \label{DDcoll}
\end{figure}

 For reasons to be shortly explained, the formula encoding this factorisation (known as the {\it dipole picture}) is most conveniently written in the transverse coordinate representation: it features the transverse coordinates $\bx_\perp$ and $\by_\perp$ of the quark and the antiquark, rather than the respective transverse momenta (see the amplitude in the left side of Fig.~\ref{DDcoll}). It reads~\cite{Mueller:1989st,Nikolaev:1990ja,Kovchegov:2012mbw} 
 \begin{equation}
\sigma_{T,L}^{\gamma^*A\to X} (x,Q^2)
=\int_0^1 \rmd z\int \rmd^2 {\bm r}\,
{
\left |\Psi_{T,L}(z, r; Q^2)\right |^2
}\;
\sigma_{\rm dipole}(x, {\bm r})
\; .
\label{DISdipole}
\end{equation}
The indices $T$ (transverse) and $L$ (longitudinal) refer to the polarisation of the virtual photon.
$\Psi_{T,L}(z, r;Q^2)$ are (light-cone) wavefunctions describing its $q\bar{q}$ fluctuation. Besides the photon polarisation ($T,L$) and virtuality ($Q^2$), they also depend upon the  dipole transverse size $r\equiv |\bx_\perp-\by_\perp|$ and upon the longitudinal momentum sharing by the two fermions: $z$ and $1-z$ are the fractions of the photon longitudinal momentum $q_z$ taken by the quark and, respectively, the antiquark. Explicit expressions for these wavefunctions will be presented in Sect.~\ref{sec:DIStot}. Their main physical effect is to limit the dipole size according to $z(1-z) Q^2r^2\lesssim 1$.

As already mentioned, the whole information about the hadronic collision is contained in the dipole cross-section $\sigma_{\rm dipole}(x, {\bm r}_\perp)$, that we shall now discuss in some detail. Via the optical theorem, the total cross-section can be related to the amplitude\footnote{Strictly speaking, the optical theorem features the {\it imaginary} part of the forward amplitude. But at high energy the amplitude is predominantly absorptive, hence imaginary, and we use the notation $T$ for its imaginary part.}
 $T_x(\bx_\perp, \by_\perp)$ for the elastic scattering:
 \begin{equation}\label{sigmaD} \sigma_{\rm dipole}(x,{\bm r})\,=\,2\int
\rmd^2{\bm b}\ T_x\Big({\bm b}+\frac{{\bm r}}{2},{\bm b}-\frac{{\bm r}}{2}\Big),\end{equation}
where ${\bm r}=\bx_\perp-\by_\perp$ (the dipole size)
and ${\bm b}=(\bx_\perp+\by_\perp)/2$ (the impact parameter w.r.t. the centre of the nucleus). 
To be elastic, a scattering requires a colourless exchange and in pQCD it starts with the exchange of two gluons. By taking the square of the inelastic amplitude in the r.h.s. of Fig.~\ref{fig:boost} and summing over all the open indices (spin, colour) for $q$ and $\bar q$ in the final state, one effectively builds an elastic amplitude involving a 2-gluon ($2g$) exchange. This  amplitude, which describes a single {\it elastic} scattering,  naturally 
probes the 2-point function of the gluon distribution, that is, the gluon PDF $xG_A(x, \mu^2)$ on the resolution scale $\mu^2\sim 1/r^2$. One has (see Sect.~\ref{sec:TMV}) for a derivation)
 \begin{align}\label{T2gxG}
T_{2g}(x, r)=\pi\alpha_s C_F\frac{r^2}{R_A^2}\,\frac{xG(x, 1/r^2)}{N_c^2-1}\ \sim \ \alpha_s C_F n(x, 1/r^2)\,.
\end{align}
where we assumed the nuclear target to be homogeneous in the transverse plane, within a disk of radius $R_A$. This amplitude vanishes like $r^2$ when $r\to 0$ (a property known as {\it colour transparency}) since a ``zero-size dipole'' is a pointlike object with zero colour charge, which has no hadronic interactions. 

The second estimate in Eq.~\eqref{T2gxG}, which relates the dipole amplitude to the gluon occupation number in the target, has been obtained by using Eq.~\eqref{nxQ}. This is quite useful since, for the scattering amplitude, there is an absolute upper limit $T(x,r)\le 1$ introduced {\it unitarity} (the expression of probability conservation for scattering). Via Eq.~\eqref{T2gxG} this implies an upper limit $n\lesssim 1/\alpha_s$ on the occupation number, which is suggestive of gluon saturation. Taken at face value,  the single-scattering (2-gluon exchange) approximation $T_{2g}(x, r)$ in Eq.~\eqref{T2gxG} will eventually violate unitarity for sufficiently large $r$ and/or small  $x$. But when the scattering becomes strong,  $T(x,r)\sim \order{1}$, the  single-scattering approximation is not correct any longer and one needs to include {\it multiple scattering}. Eq.~\eqref{T2gxG} therefore suggests a {\it duality} between multiple scattering effects on the scattering amplitudes and saturation effects (gluon recombination) in the wavefunction of the dense target. This is interesting since it looks difficult to expose both phenomena—multiple scattering and gluon saturation—within a same calculation (notably, because they become manifest in different gauges; see  Sect.~\ref{sec:CYM}).

This brings us to the main question: how to include the effects of multiple scattering in the calculation of scattering amplitudes? This is {\it a priori} a formidable problem, but the {\it eikonal approximation} to be discussed in the next section brings important simplifications.

\subsection{Eikonal approximation and the Wilson lines}

Before we proceed, it is preferable to introduce light-cone (LC) vector notations, which more compactly account for the kinematics of high-energy processes. For a generic  4-vector with  Minkowski  components $v^\mu=(v^0,v^1,v^2,v^3)$, the corresponding  LC  components are defined as $v^\mu=(v^+,v^-,{\bm v}_\perp)$, with
 \begin{equation} v^+\equiv \frac{1}{\sqrt{2}} (v^0+v^3),\qquad v^-\equiv {1 \over
 \sqrt{2}} (v^0-v^3),\qquad {\bm v}_\perp\equiv (v^1,v^2). \end{equation} 
In these new components, the scalar product reads $k\cdot x = k^+ x^- + k^- x^+ - \bk_\perp\cdot \bx_\perp$, so the on-shell condition $k^\mu k_\mu=0$ for a massless particle becomes $2k^+k^-=k_\perp^2$. A relativistic particle which is a right mover has a large ``plus'' momentum $k^+$ (which plays the role of the LC longitudinal momentum) and a small ``minus'' component, or LC energy, $k^-\simeq k_\perp^2/2k^+$. (Notice the hierarchy $k^+\gg k_\perp\gg k^-$.) The variable $x^-$ conjugated to $k^+$ plays the role of the LC longitudinal coordinate and is naturally small ($x^-\sim 1/k^+$),  as a consequence of Lorentz contraction. Vice-versa, $x^+$ is the LC time and the respective dependence is very slow, by Lorentz time dilation. For a left mover, the roles of the ``plus'' and ``minus'' components get interchanged.

Using these new notations, let us revisit the $\gamma^* A$ collision: the photon carries a 4-momentum  $q^\mu=(q^+, q^-, \bm{0}_\perp)$ with $q^+\gg Q$ and $q^-=-Q^2/2q^+$, while the struck nucleon has $P^\mu_N\simeq \delta^{\mu-}P_N^-$.  The longitudinal fractions of the quark ($z$) and the antiquark ($1-z$) are now defined w.r.t. $q^+$: $z=k_1^+/q^+$ and  $1-z=k_2^+/q^+$. Importantly, 
$k_1^+$ and $k_2^+$ are not modified by the scattering off the target, because the gluons from the nucleus carry no ``plus'' momenta.

We are now prepared to explain the eikonal approximation.  For the DIS problem at hand, this assumes that the transverse coordinates, $\bx_\perp$ and $\by_\perp$, of the quark and the antiquark are not modified by their scattering off the nuclear target. This relies on the same condition which justifies the dipole picture in the first place: the fact that the lifetime of the $q\bar q$ fluctuation is much larger than the longitudinal extent of the target. Consider the quark for definiteness. The maximal transverse deviation that the quark can suffer during the collision can be estimated as $\Delta x_\perp \sim \theta L$, where $ \theta \sim k_\perp/zq^+$ is the deflection angle and $L\sim A^{1/3}/P_N^-$ is the longitudinal width of the target. This deviation $\Delta x_\perp$ is irrelevant so long as it remains smaller than the quark transverse wavelength $\lambda_\perp\sim 2/k_\perp$ (the quantum uncertainty in its transverse position). The condition $\Delta x_\perp \ll \lambda_\perp$ can be rewritten as
\begin{equation}\label{eik}
\Delta x^+_{q}\,\simeq\,\frac{2z q^+}{k_\perp^2}\ \gg \ L 
 \simeq \frac{A^{1/3}}{P^-}\,.
\end{equation}
This is essentially the same condition as  Eq.~\eqref{cohtime} (rewritten in terms of LC components) since,  typically, $k_\perp^2\sim z(1-z) Q^2$, as we shall see in Sect.~\ref{sec:DIStot}.

The high energy collision cannot change the quark helicity either. Its only effect is a {\it colour precession}—a ``rotation'' of the quark colour state, which is implemented by a unitary matrix:
\begin{equation}\label{colrot}
\psi_\alpha(\by_\perp)\,\to\, V_{\alpha\beta}(\by_\perp)\psi_\beta(\by_\perp),\qquad
\bar\psi_\alpha(\by_\perp)\,\to\, V_{\alpha\beta}^*(\by_\perp)\bar\psi_\beta(\by_\perp)=
\bar\psi_\beta(\by_\perp)V_{\beta\alpha}^\dagger(\by_\perp)
\,.
\end{equation}
Here, $\psi_\alpha(\by_\perp)$ and $\bar\psi_\alpha(\by_\perp)$ are the quark and antiquark quantum fields, $\alpha,\beta =1,\dots,N_c$ are colour indices for the fundamental representation of SU$(N_c)$, and the matrix elements like $V_{\alpha\beta}(\bx)$ refer to the following {\em Wilson line} (a unitary matrix in the fundamental representation, with generators $t^a$):
  \begin{equation}\label{Wilson}
V(\bx_\perp)\,\equiv {\rm T}\exp\Big\{ig\int \rmd x^+
 A^-_a(x^+,x^-=0,\bx_\perp) t^a\Big\}.\end{equation}
 Here, $A^-_a(x)$ is the minus component of the quantum gauge field describing gluons in the target. 
The symbol T indicates the ordering of the operators  $A^-_a(x^+)t^a$, from right to left,
in increasing order of their $x^+$ arguments. The expression in the exponential is the eikonal version of the QCD action $\int\rmd^4 z j^\mu_a(z) A_\mu^a(z)$ describing the interaction between the quark colour current $j^\mu_a$ and the target gluon field $A^\mu_a$: for a right-moving ultrarelativistic quark, $ j^\mu_a(z) \simeq
gt^a\delta^{\mu+} \delta(z^-)\delta^{(2)}(\bz_\perp-\bx_\perp)$. The integral over the LC time
$x^+$ extends over the trajectory of the quark, but in practice it is restricted (by the support of $A^-_a$) to the longitudinal extent of the target (see the discussion in Sect.~\ref{sec:CYM}). 

Using Eq.~\eqref{colrot}, one can now construct the $S$-matrix for the  elastic scattering of a  colour dipole,
\begin{equation} 
S_x(\bx_\perp,\by_\perp)\,\equiv\,
 \frac{1}{N_c} \left\langle{\rm tr}\big[ V(\bx_\perp)
 V^\dagger(\by_\perp)\big]\right\rangle_x\,=\,1-T_x(\bx_\perp,\by_\perp)
 \label{Sdipole}
 \end{equation}
 where the colour trace occurs because the quark and the antiquark must be in the same colour state both before and after the collision (to make together a colour-singlet). Their final colour state is not measured and is therefore summed over; their initial colour state was fixed, but unknown, so it is averaged over (thus explaining the overall factor $1/N_c$). The external brackets $\langle\cdots\rangle_x$ denote the  average over the colour fields of the target;  in general, this is given by a quantum path integral, but its calculation can be simplified at high energy  (see Sect.~\ref{sec:CGCgen}). As shown by the notation, this average depends upon the longitudinal momentum fraction $x\equiv k^-/P_N^-$  of the gluons participating in the scattering. Within the approximations of interest, this dependence enters only via the high-energy (B-JIMWLK) evolution of the target, down to the value $x=\xbj$ probed by DIS. As also shown in Eq.~\eqref{Sdipole}, the dipole elastic amplitude is obtained from the respective $S$-matrix simply as $T=1-S$. 

The single-scattering (or 2-gluon exchange) approximation $T_{2g}$ can be obtained by expanding Eq.~\eqref{Sdipole} to quadratic order in the gauge fields $A^-$. Within the relevant approximations, this leads indeed to the estimate in Eq.~\eqref{T2gxG}, as will be shown in  Sect.~\ref{sec:TMV}. The terms involving higher powers of $A^-$ in the expansion of  Eq.~\eqref{Sdipole} describe multiple (elastic) scattering. Importantly, the multiple scattering series is resummed within {\it unitary} matrices, thereby ensuring that the unitarity limit $|T|\le 1$ is indeed satisfied. At high energy and for the processes to be discussed here,  $T$ is real (see however~\cite{Kovchegov:2003dm,Hatta:2005as,Jeon:2005cf}) and the unitarity (or ``black disk'') limit  $T=1$ can be saturated in practice.

%
%

 \subsection{Forward particle production in proton-nucleus collisions}
\label{sec:pA}

The second example of dilute-dense collisions to be discussed here is a proton-nucleus ($pA$) collision, and more precisely the scattering between a quark ($q$) from the proton and the dense gluon distribution in the nuclear target, as illustrated in the right side of Fig.~\ref{DDcoll}.  We shall assume that the quark is put on its mass-shell by the scattering, thus ignoring its final-state evolution via hadronisation. We work in a frame where both $p$ and $A$ are ultrarelativistic, like the laboratory frame at the LHC. (We shall later specialise to the COM frame.) The proton has  4-momentum $q^\mu\simeq \delta^{\mu+}q^+$, the nucleus has $P_N^\mu\simeq \delta^{\mu-}P_N^-$ per nucleon, and the final quark carries a 4-momentum $p^\mu=(p^+, p^-, \bp_\perp)$ with $p^+=zq^+$ (the ``plus'' momentum inherited from the parent proton) and $p^-=xP_N^-$ (the ``minus'' component transferred from the nucleus via the collision). The on-shell condition $2p^+p^-=p_\perp^2$ implies $zx=p_\perp^2/s$, where $s\equiv (q+P_N)^2= 2q^+P_N^-$. As announced, we shall specialise to the COM frame, where $q^+=P_N^-$. In this frame we introduce the {\it rapidity} of the final quark, via
\begin{equation}\label{xgpA}
y\equiv \frac{1}{2}\ln\frac{p^+}{p^-}=\ln\frac{\sqrt{2}p^+}{p_\perp}\quad\Longrightarrow\quad
p^\pm=\frac{p_\perp}{\sqrt{2}}\,\rme^{\pm y}\quad\Longrightarrow\quad z = \frac{p_\perp}{\sqrt{s}}\,\rme^{y}\ \
  \& \ \  x= \frac{p_\perp}{\sqrt{s}}\,\rme^{-y}
\,,\end{equation}
where to obtain the final estimates for $z$ and $x$ we also used $p^+=zq^+$ and $p^-=xP_N^-$, together with 
$q^+=P_N^- = \sqrt{s/2}$ in the COM frame. Notice that large and positive values for $y$ (``forward rapidities'')  correspond to the situation where $x \ll z < 1$, i.e. the gluon distribution in the nucleus is probed in the dense regime at small $x$, whereas  the proton is probed in the dilute regime at largish $z$ (so the participating quark is likely a valence quark). Since moreover $y=\ln(2/\theta)$, with $\theta=\sqrt{2}p_\perp/p^+$ the polar angle of the final quark, it is clear that particles produced at forward rapidities (in the fragmentation region of the dilute projectile) propagate at small angles $\theta\ll 1$ w.r.t. the collision axis.

The above discussion naturally suggests the following structure for the cross-section for the (semi-inclusive) production of a quark in $pA$ collisions at forward rapidity $y\gtrsim 1$:
\begin{equation}\label{quarkprod}
 \frac{\rmd \sigma^{pA\to qX}}{\rmd y\,\rmd^2\bp_\perp}\,=\,z q_p(z,\mu^2)\,
  \frac{\rmd \sigma^{qA\to qX}}{\rmd y\,\rmd^2\bp_\perp}\,,\end{equation}
 where $z q_p(z,\mu^2)$ is the quark distribution in the proton for longitudinal fraction $z$ and factorisation scale $\mu^2\sim p_\perp^2$ and the second factor is the cross-section for the partonic subprocess. The latter must account for the {\it multiple scattering} between the quark and the dense gluon system in the nucleus. As in the previous discussion of DIS, this scattering can be described in the eikonal approximation.   In this approximation, the amplitude for the quark to acquire a transverse momentum $\bp_\perp$
after crossing the target is simply the Fourier transform of the Wilson line in Eq.~\eqref{Wilson}:
 \begin{equation}
{\mathcal M}_{ij}(\bp_\perp)[A^-]\,\equiv\,\int\rmd^2\bx_\perp
 \,\rme^{-i\bx_\perp\cdot\bp_\perp}\,V_{ij}(\bx_\perp).
 \end{equation}
To compute the cross-section, one must take the modulus squared  $|{\mathcal M}_{ij}(\bp_\perp)|^2$, 
sum over the final colour states and average over the initial ones and, finally, average over all the configurations of the gluon fields $A^-_a(x)$ in the target. One thus finds~\cite{Dumitru:2002qt} (see also Refs.~\cite{Kovchegov:1998bi,Kovchegov:2001sc,Dumitru:2001jn,Dumitru:2001ux,Kharzeev:2003wz,Iancu:2004bx,Albacete:2003iq,Blaizot:2004wu} for semi-inclusive {\it gluon} production in $pA$ and the review papers~\cite{JalilianMarian:2005jf,Albacete:2014fwa} for more details and further references)
\begin{equation} \label{ptbroad}
 \frac{\rmd \sigma^{qA\to qX}}{\rmd y\,\rmd^2\bp_\perp}
 =\int
 \frac{\rmd^2\bx_\perp \rmd^2\by_\perp}{2\pi^2}
  \,\rme^{-i(\bx_\perp-\by_\perp)\cdot\bp_\perp}\,
   \frac{1}{N_c} \left\langle{\rm tr}\big[ V(\bx_\perp)
 V^\dagger(\by_\perp)\big]\right\rangle_x\,.
\end{equation}
Eq.~\eqref{ptbroad} features the same dipole $S$-matrix as the total cross-section for DIS, cf. Eq.~\eqref{Sdipole}. Yet, in the present case, there is no physical $q\bar q$ dipole partaking in the scattering: instead, a ``mathematical'' dipole, built with the physical quark in the direct amplitude and the same quark (effectively viewed as an antiquark) in the complex conjugate amplitude, has emerged when computing the cross-section.
The $p_\perp$--spectrum predicted by Eq.~\eqref{ptbroad} will be studied in Sect.~\ref{sec:pT}.

The expression for the  $pA\to qX$ cross-section given by Eqs.~\eqref{quarkprod} and \eqref{ptbroad} is generally referred to as {\em hybrid factorisation} \cite{Dumitru:2005gt}. It combines collinear factorisation on the proton side with a CGC treatment of the quark-nucleus collision, that  generalises $k_T$-factorisation by including the effects of multiple scattering. For hadron production, Eq.~\eqref{quarkprod} must be extended to include the quark-to-hadron fragmentation function. Besides the high-energy (BK) evolution of the dipole $S$-matrix in Eq.~\eqref{ptbroad}, the complete leading-order result for the cross-section for hadron production must also include the DGLAP evolutions for the quark PDF and for the fragmentation function. 

\section{The Colour Glass Condensate}
\label{sec:CGC}
\setcounter{equation}{0}

As previously established, small-$x$ gluons form a high-density medium characterised by a weak coupling $\alpha_s(Q_s)\ll 1$, large phase-space occupation numbers $n(x, k_\perp^2)\sim 1/\alpha_s\gg 1$, and prominent non-linear phenomena such as gluon saturation. Despite the smallness of the coupling, this dense system cannot be studied using standard QCD perturbation theory—defined as the expansion of the quantum path integral in powers of $\alpha_s$—because non-linear collective effects must be systematically resummed to all orders. 
Nevertheless, these specific physical conditions enable a profound simplification that opens the way for rigorous calculations using advanced perturbative techniques: highly occupied bosonic systems that are weakly coupled admit a semi-classical description in terms of \textit{strong classical fields} dressed by small quantum fluctuations. This framework naturally captures the \textit{coherence} among the numerous quantum modes occupying the same phase-space cell.

Just as a dense system of photons in a laser beam is more efficiently described as a classical electromagnetic wave rather than by tracking individual quanta, the high gluon occupancy at small $x$ allows us to model this gluonic medium via a classical colour field $A_{\mu}^{a}(x)$. This field is inherently \textit{strong}, scaling as $A_\mu^a\sim 1/g$, which reflects the fact that the gluon occupation number (fundamentally related to the two-point correlator of $A_\mu^a$) is of order $1/\alpha_s$. Consequently, the corresponding equations of motion—the classical Yang-Mills equations sourced by appropriate colour charges—must be solved \textit{exactly}. The non-linear dynamics encoded within these exact classical solutions offer a transparent realization of complex, non-perturbative phenomena like gluon saturation.

Unlike electromagnetic fields, which physically manifest as macroscopic waves or laser beams, classical colour fields $A_{\mu}^{a}(x)$ cannot propagate freely in nature due to colour confinement. Instead, the semi-classical framework for small-$x$ gluons takes the form of a \textit{stochastic} field theory, where the statistical average of the single field operator is guaranteed to vanish, $\langle A_{\mu}^{a}(x)\rangle = 0$. This effective field theory, widely known as the \textit{colour Glass Condensate} (CGC)~\cite{Iancu:2002xk,Mueller:2001fv,Iancu:2003xm,JalilianMarian:2005jf,Weigert:2005us,Gelis:2010nm,Kovchegov:2012mbw,Morreale:2021pnn}, provides a systematic tool to compute gauge-invariant $n$-point correlation functions ($n\ge 2$) that directly enter the calculation of high-energy cross-sections.

\subsection{The CGC effective theory}
\label{sec:CGCgen}

The CGC effective theory is rooted in a semi-classical separation of the high-energy degrees of freedom into two distinct classes: "slow" gluons at a given small fraction $x$, which are described by a classical colour field $A_{\mu}^{a}(x)$, and "fast" partons (valence quarks and harder gluons with $x'\gg x$), which act as static \textit{colour sources}. To maintain consistency across different scales, the theory is endowed with a renormalisation group equation (RGE) that shows how quantum fluctuations are integrated into classical sources as $x$ decreases. The core components of this framework are outlined below:

\begin{description}
\item[\texttt{(I)} The classical Yang-Mills (YM) equation:] The dynamics of the slow gluon fields is governed by the classical equation of motion:
\begin{equation}\label{YM}
 D_\nu^{ab} F_b^{\nu\mu}(x) = J^\mu_a(x)\,,
\end{equation}
where $F_a^{\nu\mu} = \partial^\nu A^\mu_a - \partial^\mu A^\nu_a - g f^{abc} A^\nu_b A^\mu_c$ is the non-Abelian field strength tensor encompassing the chromoelectric and chromomagnetic fields, and $D^\nu_{ab} = \partial^\nu \delta^{ab} - g f^{abc} A^\nu_c$ is the covariant derivative. The colour current $J^\mu_a(x)$ generated by the fast sources is largely constrained by high-energy kinematics. For a left-moving parent hadron ($P^\mu = \delta^{\mu-}P^-$), the fast partons are highly collinear, restricting the current to its minus component:
\begin{equation}
J^\mu_a(x) \simeq \delta^{\mu -} \rho_a(x^+,\bx_\perp)\,.
\end{equation}
Due to Lorentz time dilation, the colour charge density $\rho_a(x^+,\bx_\perp)$ is effectively independent of the longitudinal coordinate $x^-$ over the characteristic interaction timescale of the slow fields. Furthermore, Lorentz contraction localizes this source density into a thin shockwave near $x^+=0$. To the slow small-$x$ modes, the configuration of the fast sources appears static or "frozen" during the interaction.

\item[\texttt{(II)} The CGC weight function:] A specific solution $A_{\mu}^{a}(x)$ to Eq.~\eqref{YM} corresponds to a single event characterized by a localized configuration of fast colour charges. While frozen during a specific scattering event, this configuration varies stochastically from event to event, ensuring that the statistical average vanishes, $\langle\rho_a\rangle = 0$ (and consequently $\langle A_{\mu}^{a}\rangle=0$). To compute physical observables, one must specify the statistical distribution of the charge density $\rho_a$. This information is encoded in the \textit{CGC weight function} $W_Y[\rho]$—a gauge-invariant, positive-definite functional probability distribution normalized to unity ($\int {\mathcal D}\rho \, W_Y[\rho] = 1$). As indicated by the notation, $W_Y[\rho]$ depends explicitly on the rapidity scale $Y \equiv \ln(1/x)$ at which the effective theory is constructed. The expectation value of any gauge-invariant observable is evaluated by first solving the YM equation~\eqref{YM} for a frozen source configuration $\rho_a(x)$, computing the observable using this classical solution $A^\mu[\rho]$, and finally averaging over all source configurations with the weight function $W_Y[\rho]$. For instance, the dipole $S$-matrix is evaluated as:
\begin{equation}\label{Saverage}
S_Y(\bx_\perp,\by_\perp) = \int {\mathcal D}\rho \, W_Y[\rho]\, \frac{1}{N_c}\,{\rm tr}\big( V(\bx_\perp) V^\dagger(\by_\perp)\big)\,.
\end{equation}

\item[\texttt{(III)} The JIMWLK equation:] The evolution of the CGC weight function $W_Y[\rho]$ with decreasing $x$ (or increasing rapidity $Y$) is governed by a functional renormalization group equation known as the \textit{JIMWLK equation}~\cite{JalilianMarian:1997jx,JalilianMarian:1997gr,Kovner:2000pt,Iancu:2000hn,Iancu:2001ad,Ferreiro:2001qy}. Derived by integrating out quantum gluon layers in the leading logarithmic approximation in the presence of the classical background field ($x'\gg x$), JIMWLK constitutes the non-linear generalization of the BFKL evolution, to which it reduces in the weak-field limit ($gA^\mu_a\ll 1$). Its non-linearities encode mutual interactions between gluons from different generations, leading to saturation. The equation takes a Hamiltonian form:
 \begin{equation}\label{jimwlk}
 \frac{\partial W_Y[\rho]}{\partial Y} = H_{\rm JIMWLK}\left[\rho,\frac{\delta}{\delta \rho}\right] W_Y[\rho]\,,
 \end{equation}
 where the self-adjoint Hamiltonian $H_{\rm JIMWLK}$ is a second-order functional differential operator depending on $\rho$ through Wilson lines:
 \begin{equation}\label{Hjim}
 H_{\rm JIMWLK} = \frac{1}{2} \frac{\delta}{\delta \rho^a} \chi_{ab}[V, V^\dagger] \frac{\delta}{\delta \rho^b}\,.
 \end{equation}
 The explicit form of the kernel $\chi_{ab}$ is detailed in~\cite{Iancu:2001ad}. Physically, the two functional derivatives describe the emission and subsequent absorption of a small-$x$ gluon by preexisting colour sources, while the Wilson lines resum its multiple eikonal scatterings off the background field.
  
  \end{description}

 Mathematically, the JIMWLK framework maps onto a functional generalisation of the Fokker-Planck equation~\cite{Blaizot:2002xy}, which inherently guarantees the preservation of both the normalization and the positivity of $W_Y[\rho]$. Consequently, the evolution can be reformulated as a functional Langevin equation~\cite{Weigert:2000gi,Blaizot:2002xy} describing a random walk on the $SU(N_c)$ manifold of Wilson lines, which lends itself to numerical simulations on a two-dimensional transverse lattice. This numerical strategy (further discussed in Sect.~\ref{sec:Langevin}) has been successfully employed to construct exact solutions to the JIMWLK equation for $N_c=3$~\cite{Rummukainen:2003ns,Lappi:2011ju,Dumitru:2011vk,Lappi:2012vw,Schenke:2016ksl,Marquet:2016cgx,Marquet:2017xwy}.

By taking a derivative w.r.t. $Y$  in Eq.~\eqref{Saverage}, substituting Eq.~\eqref{jimwlk}, and performing a functional integration by parts, the evolution of a scattering operator can be written as (with  $H \equiv H_{\rm JIMWLK}$)
\begin{equation}\label{evolO} 
\frac{\partial}{\partial Y} \langle \hat{S} \rangle_Y = \int {\mathcal D}\rho \, \big(H W_Y[\rho]\big)\, \hat{S}[\rho] = \int {\mathcal D}\rho \, W_Y[\rho]\, \big(H \hat{S}[\rho]\big) = \big\langle H \hat{S} \big\rangle_Y\,,
\end{equation}
where $\hat{S} \equiv \frac{1}{N_c}{\rm tr}\big( V(\bx_\perp) V^\dagger(\by_\perp)\big)$ is the dipole $S$-matrix operator for a single event. 
Equation~\eqref{evolO}  shows that the high-energy evolution can equivalently be computed by acting with the JIMWLK Hamiltonian directly on the scattering observable. In this dual picture, the soft gluon fluctuation is viewed as being emitted and reabsorbed by the projectile partons (represented by the Wilson lines), effectively shifting the perspective from the evolution of the dense target to that of the dilute projectile. 

While this equivalence is a natural consequence of boost invariance at leading order—where the rapidity increment $\rmd Y$ can be arbitrarily assigned to either the projectile or the target—this simple mapping breaks down at next-to-leading order (NLO) and beyond. The evolution parameter $Y$ is not the true center-of-mass rapidity separation $y_T - y_P$; rather, it relates to the longitudinal momentum fractions of the evolving partons, which are inherently asymmetric between the left-moving target (where $Y=\ln(1/x)$ with $x=k^-/P^-$) and the right-moving projectile ($Y=\ln(1/z)$ with $z=k^+/q^+$). This kinematic mismatch becomes critical in the presence of highly disparate transverse scales (e.g., when the transverse size $r$ of the dipole projectile is much smaller than the inverse saturation momentum $1/Q_s$ of the target), in which case it introduces large radiative corrections, that need to be resummed to all orders  (see Sect.~\ref{sec:collBK}).

The action of $H_{\rm JIMWLK}$ on Wilson lines can be explicitly evaluated~\cite{Weigert:2000gi,Iancu:2001ad,Ferreiro:2001qy}, transforming Eq.~\eqref{evolO} from a functional equation into a set of ordinary differential equations known as the \textit{Balitsky hierarchy}~\cite{Balitsky:1995ub}. This system is not closed: a newly emitted gluon can scatter off the target background field and its scattering introduces an additional Wilson line in the adjoint representation. Consequently, scattering operators involving a progressively larger number of Wilson lines mix with each other under evolution, forming an infinite hierarchy.

Although formally equivalent at leading order, target and projectile evolutions offer distinct practical advantages. Probing a dilute projectile is conceptually simpler, enabling streamlined derivations of the Balitsky~\cite{Balitsky:1995ub} and BK~\cite{Kovchegov:1999yj,Kovchegov:1999ua} equations, as well as the JIMWLK Hamiltonian~\cite{Weigert:2000gi,Mueller:2001uk} (see Sect.~\ref{sec:BK}). Conversely, the target-centric JIMWLK equation provides deeper insights into the mechanisms of gluon saturation. It accommodates highly efficient mean-field approximations (such as Gaussian truncations)~\cite{Iancu:2001md,Iancu:2002xk,Iancu:2002aq,Kovchegov:2008mk,Marquet:2010cf,Dumitru:2011vk,Iancu:2011ns,Iancu:2011nj} (cf. Sect.~\ref{sec:Gauss}) and maps onto an equivalent Langevin formulation~\cite{Weigert:2000gi,Blaizot:2002xy} for exact numerical simulations at finite $N_c$  (see Sect.~\ref{sec:Langevin}). Both formulations have been extended to multi-particle production~\cite{Hentschinski:2005er,Kovner:2006ge,Kovner:2006wr,Iancu:2011ns}, including long-range rapidity correlations~\cite{Iancu:2013uva}. Next-to-leading order (NLO) corrections are now fully established for both the BK~\cite{Balitsky:2008zza,Balitsky:2013fea} and JIMWLK frameworks~\cite{Kovner:2013ona,Kovner:2014lca,Lublinsky:2016meo}. As detailed in Sects.~\ref{sec:rcBK} and \ref{sec:collBK}, these NLO corrections are numerically large, making their inclusion essential for quantitative comparisons with the phenomenology.

To conclude, it is  instructive to recall the physical motivation behind the terminology \textit{Colour Glass Condensate} (CGC) used to define this dense small-$x$ gluonic medium~\cite{Iancu:2000hn}:
\begin{description}
\item[\textbf{Colour:}] Reflects the fact that gluons carry non-Abelian $SU(N_c)$ colour charge. Crucially, non-linear colour coherence dominates the approach to saturation, meaning that simpler, colourless degrees of freedom—such as the inclusive gluon number, or colour dipole distributions~\cite{Mueller:1993rr}—are insufficient to fully capture the high-density dynamics.
\item[\textbf{Glass:}] Highlights the wide separation of timescales between the "slow" small-$x$ modes and their "fast" valence sources. Due to Lorentz time dilation, the large-$x$ valence partons appear static or "frozen" over the characteristic lifetime of the soft fields, mimicking a disordered system that behaves as a solid on short timescales but flows as a fluid over larger ones—the defining characteristic of a \textit{glass}.
\item[\textbf{Condensate:}]  Reflects the macroscopically large phase-space occupation numbers $n(x,k_\perp) \sim 1/\alpha_s$, which allow the saturated gluons to  behave coherently and thus be described as a strong, classical background field.
\end{description}


\subsection{The classical colour field}
\label{sec:CYM}

We shall now construct explicit solutions to the classical YM equation \eqref{YM}. In a wide class of gauges, one can systematically find a solution satisfying the following structural properties:
\begin{equation}\label{YMprop}
F^{ij}_a = 0, \qquad A^+_a = 0, \qquad A^-_a, \, A^i_a : \text{ static}\,,
\end{equation}
where "static" implies independence from the light-cone time $x^-$ of the left-moving nucleus. 
(This reflects the specific structure of the colour current  which has just a ``minus'' component and is static.)
These conditions restrict the independent field degrees of freedom to a single component once a gauge-fixing condition is enforced. Two specific gauge choices are of primary interest for us here:

\begin{description}
\item[\textit{(i)} The covariant gauge ($\partial_\mu A^\mu_a = 0$):] This choice is highly convenient for computing scattering amplitudes. Combined with Eq.~\eqref{YMprop}, it enforces $\partial_i A^i_a = 0$, which yields $A^i_a = 0$. The classical field is thus reduced to a single non-zero component, $A^\mu_a(x) = \delta^{\mu -} A^-_a(x^+,\bx_{\perp})$, where $A^-_a$ satisfies the two-dimensional Poisson equation:
\begin{equation}\label{EQTA} 
- \nabla^2_\perp A^-_a(x) = \rho_a(x^+,\bx_\perp)\,. 
\end{equation} 
This is readily integrated to give:
\begin{equation}\label{alpha} 
A^-_a(x^+,\bx_\perp) = \int \rmd^2\by_\perp\,G (\bx_\perp-\by_\perp)\,\rho_a(x^+,\by_\perp)\,,
\end{equation} 
where $G(\bx_\perp-\by_\perp)$ represents the two-dimensional Coulomb propagator:
\begin{equation}\label{Coulomb} 
G (\br_\perp) \equiv \int\frac{\rmd^2\bk_\perp}{(2\pi)^2} \,\frac{\rme^{i\bk_\perp\cdot\br_\perp}}{k_\perp^2} = \frac{1}{4\pi}\,\ln\frac{1}{r_\perp^2\Lambda_{\rm QCD}^2}\,.
\end{equation}
The infrared cutoff $\Lambda_{\rm QCD}$ reflects the non-perturbative boundary where long-range gauge fields are screened by colour confinement. Because Eq.~\eqref{alpha} is strictly linear in $\rho_a$ and local in $x^+$, fields generated by sources at different longitudinal positions do not overlap or mutually interact. In this gauge, gluon saturation is not explicitly visible within the target wavefunction itself; instead, the non-linear dynamics are entirely transferred into the eikonal Wilson lines that describe multiple scattering with an external projectile, as visible in Eq.~\eqref{Wilson}.

\item[\textit{(ii)} The target light-cone gauge ($\mathcal{A}^-_a = 0$):] In this gauge, the transverse fields $\mathcal{A}^i_a$ are non-vanishing and can be derived via a local gauge rotation of the covariant-gauge solution \eqref{alpha}:
\begin{equation}\label{COVLC}
\mathcal{A}^\mu = U \left(A^\mu + \frac{i}{g}\partial^\mu \right) U^\dagger.
\end{equation}
We use matrix notations appropriate to the fundamental representation, e.g. $A^\mu \equiv A^\mu_at^a$, and the gauge rotation $U\equiv U(x^+,\bx_{\perp})$ is a unitary matrix, determined by the condition $\mathcal{A}^- = 0$:
\begin{equation}\label{UTA}
U(x^+,\bx_{\perp}) = \text{T} \exp \left \{ig \int^{\infty}_{x^+} \rmd z^+\,A^-_a (z^+,\bx_{\perp}) t^a \right \}\,,
\end{equation}
where the boundary condition $U(x^+\to \infty,\bx_{\perp})=1$ eliminates any residual gauge freedom. Eq.~\eqref{UTA} represents a direct generalisation of the eikonal Wilson line operator \eqref{Wilson}. Modelling the nucleus as a localised shockwave near $x^+=0$, the transverse fields become:
\begin{equation}\label{tpg}
\mathcal{A}^i(x^+,\bx_{\perp}) = \frac{i}{g}\, U(x^+,\bx_{\perp})\,\partial^i U^{\dagger}(x^+,\bx_{\perp}) \simeq \theta(-x^+)\frac{i}{g}\,V(\bx_\perp)\partial^i V^\dagger (\bx_\perp)\,.
\end{equation}
Eq.~\eqref{tpg} shows that the potentials ${A}^i(x^+,\bx_{\perp})$ are delocalised in $x^+$. Consequently, fields originating from distinct  sources overlap extensively and the non-linear saturation effects become manifest as direct modifications to the gauge fields, encoded within the non-linear functional dependence of the Wilson line $V(\bx_\perp)$ on $\rho_a$.

\end{description}

In summary, multiple scattering and gluon saturation are gauge-dual manifestations of the same underlying high-density QCD dynamics. In the covariant gauge, the classical field equations are linear and local, while non-linearities enter through the scattering matrix of the projectile. In the light-cone gauge, the fields are delocalized and explicitly non-linear, providing a direct partonic picture of saturation within the target wavefunction. In both representations, the physical cross-sections are ultimately governed by the properties of Wilson lines.

Finally, we clarify the structure of the associated field strength tensor $F^{\mu\nu}_a$. According to Eq.~\eqref{YMprop}, the only non-vanishing components are $F^{i-}_a$, computed as $F^{i-}_a = \partial^i A^-_a$ in the covariant gauge and as $\mathcal{F}^{i-}_a = -\partial^- \mathcal{A}^i_a$ in the light-cone gauge. Regardless of the gauge choice, these fields are strictly confined to the two-dimensional shockwave plane, scaling as $F^{i-}_a \sim \delta(x^+)$. Expressing the field strength tensor in terms of standard 3-dimensional components reveals that the chromoelectric ($\bm{E}_a$) and chromomagnetic ($\bm{B}_a$) fields are strictly transverse to each other and to the direction of propagation ($\bm{E}_a \perp \bm{B}_a \perp \hat{\mathbf{z}}$), with identical magnitudes $|\bm{E}_a|=|\bm{B}_a|$. These configurations form the non-Abelian generalization of the Weizsäcker-Williams virtual photon fields in classical electrodynamics.

\subsection{The McLerran-Venugopalan (MV) model}
\label{sec:MV}

The McLerran-Venugopalan (MV) model~\cite{McLerran:1993ni,McLerran:1993ka} is the original and conceptually simplest formulation of a CGC-like effective field theory. It describes the gluon distribution of a large nucleus with mass number $A\gg 1$ probed at momentum fractions that are small, $x\ll 1$, but not sufficiently low to trigger significant high-energy radiative evolution ($\abar\ln(1/x)\ll 1$). In this regime, typically assumed to hold around $x \sim 0.01$, the primary colour sources within the nucleus are restricted to the $A N_c$ valence quarks. Beyond its tremendous  pedagogical value, the classical MV model provides the standard, non-perturbative initial condition for the evolution equations with decreasing $x$.

We examine the nucleus in an ultrarelativistic frame where it moves as an energetic left-mover with a nucleon light-cone momentum $P^\mu \simeq \delta^{\mu -} P^- \gg M$. By the uncertainty principle, a soft small-$x$ gluon from the nuclear wavefunction carries a longitudinal momentum $k^- = xP^-$, meaning it is strongly delocalised over a light-cone distance $\Delta x^+ \sim 1/(xP^-)$. When the kinematics satisfy $1/x \gg A^{1/3}$, this longitudinal wavelength vastly exceeds the width of the Lorentz-contracted nuclear disk, $\Delta z^+ \sim A^{1/3}/P^-$. Consequently, such a soft gluon samples and overlaps with all nucleons located within a longitudinal tube slicing through the entire core of the nucleus. 

The transverse area of this coherent interaction tube scales as $\Delta x_\perp^2 \sim 1/k_\perp^2$, where $k_\perp$ represents the gluon transverse momentum. For semi-hard probes ($k_\perp \gg \Lambda_{\rm QCD}$), this tube is relatively  thin, ensuring that it primarily captures valence quarks originating from distinct nucleons. Because these valence quarks are physically isolated from each other by colour confinement within their parent nucleons, they act as mutually \textit{independent} colour sources. For the purpose of computing soft gluon fields, these discrete partons can be modeled macroscopically as a random colour charge density $\rho_a(x^+,\bx_\perp)$. Global gauge symmetry dictates that the statistical 1-point function vanishes, $\langle\rho_a(x^+,\bx_\perp)\rangle = 0$. Invoking the central limit theorem, the 2-point correlation function maps onto a local \textit{Gaussian distribution} weighted by the local charge density squared $\lambda(x^+,\bx_\perp)$:
\begin{equation}\label{MV_corr}
\langle\rho_a(x^+,\bx_\perp) \rho_b(y^+,\by_\perp)\rangle = \delta_{ab} \, \delta(x^+-y^+) \, \delta^{(2)}(\bx_\perp-\by_\perp) \, \lambda(x^+,\bx_\perp)\,.
\end{equation}

Equivalently, the initial CGC weight function $W_{\rm MV}[\rho]$ takes a local Gaussian form. The longitudinal coordinate $x^+$ is bounded by the contracted nuclear width, while the transverse coordinate $\bx_\perp$ is confined to the nuclear disk of radius $R_A$. Assuming a homogeneous profile in the transverse plane ($\lambda$ independent of $\bx_\perp$ for $x_\perp \le R_A$), the evaluation of high-energy observables depends exclusively on the longitudinally integrated transverse charge density squared:
\begin{equation}\label{muMV}
\mu^2_A \equiv \int\rmd x^+\lambda(x^+) = \frac{g^2 C_F A N_c}{(N_c^2-1)\pi R_A^2} = \frac{2\alpha_s A}{R_A^2} \sim A^{1/3}\,.
\end{equation}
In Eq.~\eqref{muMV}, $g^2 C_F$ represents the  coloor charge squared of a single quark, $A N_c$ is the total number of valence quarks, and the denominator normalises the density over the number of colour degrees of freedom $N_g = N_c^2-1$ and the nuclear transverse area $\pi R_A^2$. The geometric scaling law $\mu^2_A \propto A^{1/3}$ (since $R_A \propto A^{1/3}$) emerges directly from integrating over the longitudinal depth of the heavy-ion target, providing the famous nuclear "oomph" factor that amplifies non-linear saturation effects.

\subsection{Dipole scattering in the MV model}
\label{sec:TMV}

Using the Gaussian weight function of the MV model alongside the classical Yang-Mills solutions in the covariant gauge, Eq.~\eqref{alpha}, it is straightforward to compute the $S$-matrix for the elastic scattering of a quark-antiquark dipole within the eikonal approximation~\eqref{Sdipole}. Because successive collisions off uncorrelated colour sources are statistically independent, the multi-scattering series exponentiates into a simple form, $S = \rme^{-T_{2g}}$, where $T_{2g}$ represents the amplitude for a single elastic scattering via a two-gluon exchange. This amplitude is obtained by expanding the eikonal Wilson lines in Eq.~\eqref{Sdipole} to second order in the field $A^-_a$, using ${\rm tr} (t^at^b)=\delta^{ab}/2$, and performing the Gaussian average as in~\eqref{MV_corr}:
\begin{align}\label{T2g}
 T_{2g}(\bx_\perp,\by_\perp) &= \frac{g^2}{4N_c} \left\langle\big(A^-_a(\bx_\perp) - A^-_a(\by_\perp)\big)^2\right\rangle_0 \nonumber \\[0.2cm]
 &= g^2 C_F \int\frac{\rmd^2\bk_\perp}{(2\pi)^2} \frac{\mu^2_A}{k_\perp^4} \Big[1 - \rme^{i\bk_\perp\cdot\br}\Big] \simeq \frac{\alpha_s C_F}{4}\,r^2\mu^2_A\ln\frac{1}{r^2\Lambda^2}\,,
\end{align}
where $\br \equiv \bx_\perp - \by_\perp$ is the dipole transverse separation and the confinement scale 
$\Lambda\equiv \Lambda_{\rm QCD}$ has been introduced to regulate a logarithmic infrared divergence in the original integral. The final, logarithmic approximation follows from expanding the phase factor $\rme^{i\bk_\perp\cdot\br}$ to quadratic order. This expansion is justified since the intermediate momentum range ($\Lambda^2 \lesssim k_\perp^2 \lesssim 1/r^2$) generates a large collinear logarithm, $\ln(1/r^2\Lambda^2)$, which represents the dominant contribution for sufficiently small dipoles ($r^2\Lambda^2 \ll 1$).  

The amplitude Eq.~\eqref{T2g}
depends only upon the dipole size $r$ because we assumed the target to
be homogeneous and isotropic in the transverse plane. This is a reasonable approximation
so long as the dipole hits the nucleus at relatively central impact parameters.
Using Eq.~\eqref{T2g} alongside our previous estimate \eqref{xGxR} for the nuclear gluon distribution $xG(x,Q^2)$ generated by valence quarks, one can verify that $T_{2g}(r) \propto \alpha_s r^2 xG(x, 1/r^2)$, as anticipated in Eq.~\eqref{T2gxG}.

Exponentiating this two-gluon exchange result yields the full dipole $S$-matrix for multiple elastic scatterings off the nuclear valence quarks:
\begin{equation}\label{SMV}
S(r) = \rme^{-T_{2g}(r)} = \exp\left\{-\frac{r^2 Q^2_A}{4}\ln\frac{1}{r^2\Lambda^2}\right\},
\end{equation}
where we have defined the baseline scale $Q_A^2 \equiv \alpha_s C_F \mu^2_A$. Equation~\eqref{SMV} 
 illustrates some of the general properties anticipated in Sect.~\ref{sec:DIS}.
The exponent (i.e. the amplitude $T_{2g}(r)$ for 2-gluon exchange) becomes
of $\order{1}$ when $2/r= Q_s(A)$, with  the nuclear saturation momentum $Q_s(A)$ implicitly defined by the following equation:
\begin{equation}\label{QSMV} 
Q_s^2(A) = Q_A^2 \ln\left(\frac{Q_s^2(A)}{\Lambda^2}\right) \propto A^{1/3}\ln A^{1/3}\,.
\end{equation}

For a very small dipole satisfying $r Q_s(A) \ll 1$, the scattering is weak and the forward scattering amplitude $T(r) \equiv 1 - S(r) \simeq T_{2g}(r)$ vanishes quadratically as $r \to 0$, a property known as \textit{colour transparency}. Conversely, a large dipole with $r Q_s(A) \gtrsim 1$ undergoes multiple scattering, causing the total amplitude to approach the \textit{unitarity} (or ``black-disk'') limit  $T \to 1$. Thus, $Q_s(A)$ emerges naturally as the characteristic scale governing the unitarization of the dipole scattering amplitude. According to the discussion in Sect.~\ref{sec:DIS}, this momentum must correspond to the scale where the gluon occupation number in the target saturates at a value of order $1/\alpha_s$. This will be verified in the next section via a direct computation of the classical gluon distribution.

\subsection{The gluon occupation number: unveiling gluon saturation}
\label{sec:sat}

As explained in Sect.~\ref{sec:gluons}, the gluon  occupation number provides a useful criterion for the on-set of saturation. It is therefore  interesting to compute this quantity within the MV model. For small-$x$ gluons inside a large nucleus, this is defined as
\begin{align}\label{nxk}
 n(x, \bk_\perp, \bb_\perp) 
 \equiv \frac{(2\pi)^3}{2(N_c^2-1)}\ x\frac{\rmd N_g}{\rmd x \,\rmd^2\bk_\perp\rmd^2\bb_\perp}\,\simeq \,  \frac{(2\pi)^3}{2(N_c^2-1)}\, \frac{xf_g(x, k_\perp)}{\pi R_A^2}\,, 
\end{align} 
where $\bb_\perp$ is the gluon position in the transverse plane, as measured from the centre of the nucleus.
This definition assumes that the gluon transverse momentum $\bk_\perp$ and coordinate $\bb_\perp$ can be measured simultaneously. Such a semi-classical description is only justified for sufficiently hard modes ($k_\perp \gg \Lambda_{\rm QCD}$), which are localized over spatial sizes $\Delta x_\perp \sim 1/k_\perp \ll R_A$. However, a similar treatment would be inappropriate for the longitudinal phase space, as the quantum uncertainty $\Delta z \sim 1/k_z = 1/(xP_z)$ vastly exceeds the Lorentz-contracted nuclear width. Thus, the integrated quantity $x (\rmd N_g / \rmd x)$ acts as a proxy for the would-be classical longitudinal distribution
${\rmd N_g}/({\rmd z\rmd k_z})$.

The second equality in Eq.~\eqref{nxk} holds for a nucleus which is homogeneous in $\bb_\perp$ and involves the UGD introduced in Eq.~\eqref{unint}. At small $x$, this is related to the following 2-point function of the transverse gauge fields in the  LC gauge $A^-=0$~\cite{Jalilian-Marian:1996mkd,McLerran:1998nk,Kovchegov:1998bi,Iancu:2002xk,Iancu:2002aq,Dominguez:2011wm}:
\begin{align}\label{xfg}
xf_g(x, k_\perp^2) &= 8\pi \int\frac{\rmd^2\bx_\perp\, \rmd^2 \by_\perp}{(2\pi)^4} \, \rme^{-i\bk_\perp \cdot (\bx_\perp - \by_\perp)} \int \rmd x^+ \rmd y^+ \left\langle {\rm tr} \left[ \mathcal{F}^{i-}(x^+,\bx_\perp)\mathcal{F}^{i-}(y^+,\by_\perp) \right]\right\rangle_{x} \nonumber \\[0.2cm]
&= \int\frac{\rmd^2\bx_\perp\, \rmd^2 \by_\perp}{(2\pi)^4} \, \rme^{-i\bk_\perp \cdot (\bx_\perp - \by_\perp)} \, \frac{-2}{\alpha_s} \left\langle {\rm tr} \left[ V(\bx_\perp) \big(\partial^i V^\dagger(\bx_\perp)\big) V(\by_\perp) \big(\partial^i V^\dagger(\by_\perp)\big) \right]\right\rangle_{x}\,,
\end{align}
where we used  matrix notations ($\mathcal{F}^{i-}\equiv \mathcal{F}^{i-}_at^a$, etc.) together with $\mathcal{F}^{i-} = -\partial^- \mathcal{A}^i$ and Eq.~\eqref{tpg}.

We will now compute $xf_g(x, k_\perp) $ within the MV model.  To that aim, it is preferable to consider the 2-point function in the first line of Eq.~\eqref{xfg}. We first relate the LC-gauge field  ${\cal F}^{i-}$ to the corresponding field $F^{i-}=\del^i A^-$ in the covariant gauge via a gauge rotation, cf. Eq.~\eqref{COVLC}:
\begin{equation}\label{LCF}
{\cal F}^{i-}\,=\,U F^{i-} U^\dagger\quad\Longrightarrow\quad {\cal F}^{i-}_a = \tilde U^\dagger_{ab}\del^i A^-_b,\end{equation}
with $A^-_a(x^+,\bx)$ given by Eq.~\eqref{alpha} and $ \tilde U^\dagger$ the gauge rotation matrix in the adjoint representation, as obtained by replacing $t^a\to T^a$ with $T^a_{bc}=if^{abc}$ in Eq.~\eqref{UTA}.

 Due to rotational symmetry, to the time-ordering of the fields in the Wilson lines, and to the locality of the MV correlator \eqref{MV_corr} in  $x^+$, it is easy to check that the only allowed contraction is~\cite{Jalilian-Marian:1996mkd,Iancu:2002xk,Iancu:2002aq} (we consider a homogeneous target)
\begin{align}
 \left\langle {\cal F}^{i-}_a(x^+,\bx){\cal F}^{i-}_a(y^+,\by)\right\rangle
&\, = \left\langle \big(\tilde U^\dagger_{ab}\del^i A^-_b\big)(x^+,\bx)\, \big(\tilde U^\dagger_{ac}\del^i A^-_c\big)(y^+,\by) \right\rangle
\\*[0.2cm]
	&\,=\,\delta(x^+-y^+)\left\langle {\rm Tr} \big[\tilde U(x^+,\bx)\tilde U^\dagger(x^+,\by)\big] \right\rangle
	\lambda(x^+)
	\Big(-\nabla^2 \gamma_A(\bx-\by)\Big),   \nonumber
\end{align}
where the trace Tr refers to the adjoint representation and (cf. Eqs.~\eqref{alpha}, \eqref{Coulomb}  and \eqref{MV_corr})
\begin{equation}
\gamma_A(\bx-\by)\equiv \int\rmd^2\bz \,G(\bx-\bz) G(\bz-\by) \quad\Longrightarrow\quad
 -\nabla^2 \gamma_A(\br)= \frac{1}{4\pi}\,
 \ln\frac{1}{r^2\Lambda^2}\,.\end{equation}
The trace of the product of two adjoint Wilson lines looks like the dipole $S$-matrix in Eq.~\eqref{Sdipole}, but applied to a gluon-gluon dipole ($gg$). Its estimate in the MV model is similar to Eq.~\eqref{SMV}:
\begin{equation}\label{SMVadj}
\frac{1}{N_c^2-1}\left\langle {\rm Tr} \big[\tilde U(x^+,\bx) \tilde U^\dagger(x^+,\by)\big] \right\rangle
=\exp\left\{-\frac{r^2 N_c}{4}\ln\frac{1}{r^2\Lambda^2}\int_{x^+}^\infty \rmd z^+ \lambda(x^+)\right\},
\end{equation}
where $r_\perp \equiv |\bx_\perp-\by_\perp|$. Performing the double longitudinal integrations over $x^+$ and $y^+$ in Eq.~\eqref{xfg} and using the  definition  \eqref{muMV} for $\mu_A^2$ together with the relation \eqref{nxk} between the gluon occupation number and the UGD, one finds~\cite{Jalilian-Marian:1996mkd,Kovchegov:1998bi}:
\begin{align}\label{xfgMV}
 n(x, k_\perp) =
	\frac{1}{ \pi \alpha_s N_c}\int\rmd^2\br 
	 \,\rme^{-i\bk\cdot\br}\,
	 \frac{1}{r^2}\,\left[1-\tilde S(r)\right],
\end{align}
where $\tilde S(r)$ is the elastic $S$-matrix for a $gg$ dipole, given by an expression similar to Eq.~\eqref{SMV}, but with the scale $Q_A^2$ replaced by $\tilde Q^2_A\equiv\alpha_s N_c \mu_A^2=(N_c/C_F)Q^2_A$.The saturation momentum $\tilde Q_s(A)$ for an adjoint dipole is defined analogously to Eq.~\eqref{QSMV}. 

\begin{figure}
	\begin{center}
		\includegraphics[width=0.47\textwidth]{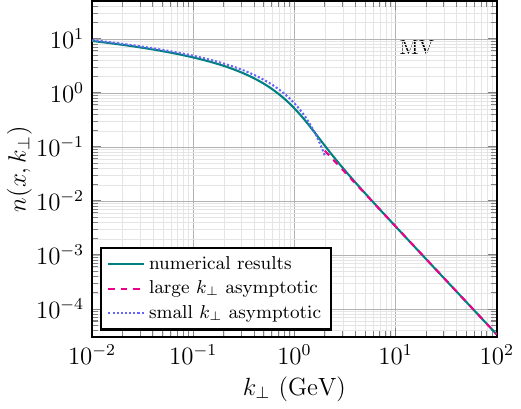}\qquad
		\includegraphics[width=0.47\textwidth]{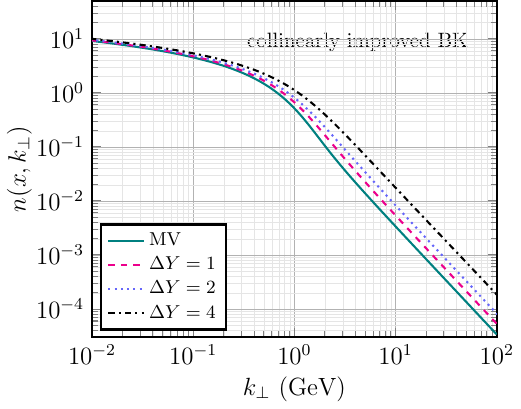}
			\end{center}
	\caption{\small  The gluon occupation number in the MV model (left plot) and its high energy evolution in the Gaussian approximation to the JIMWLK equation, as discussed in Sect.~\ref{sec:Gauss} (right plot). An overall factor $(1/\alpha_s N_c)$ was left out when making the plots. The MV model result is obtained by numerical integration in Eq.~\eqref{xfgMV} with  $\tilde Q^2_A=0.68$~GeV$^2$ (yielding $ \tilde Q_s^2(A)\simeq 2.3$~GeV$^2$). For the right plot, we used numerical solutions to the BK equation with running coupling and collinear improvement~\cite{Ducloue:2019ezk}.}
\label{fig:WW}
\end{figure}

Eq.~\eqref{xfgMV} makes it natural to distinguish between two physical regimes:

\begin{description}
\item[\textit{(i)} The dilute regime ($k_\perp\gg \tilde Q_s(A)$):] The Fourier integral in Eq.~\eqref{xfgMV} is controlled by small dipole sizes $r\ll 1/ \tilde Q_s(A)$, for which we can linearise the $gg$ dipole $S$-matrix (corresponding to a single scattering). This yields 
\begin{align}\label{noccup-high}
n(x, k_\perp)
\,\simeq\,
\frac{\mu_A^2}{k_\perp^2}\
\quad\mbox{when}\quad k_\perp\gg \tilde Q_s(A)\,,
\end{align}
which is 
the power-law bremsstrahlung spectrum produced by independent valence quarks.

\item[\textit{(ii)} The saturation regime ($k_\perp \ll \tilde Q_s(A)$):] The integral samples large dipoles ($r_\perp \gtrsim 1/\tilde Q_s(A)$), where multiple scatterings dominate. Approximating the exponent in $\tilde S(r_\perp)$ via its local saturation scale, $\tilde S(r_\perp) \simeq \exp\big[-r_\perp^2 \tilde Q_s^2(A)/4\big]$, Eq.~\eqref{xfgMV} integrates exactly to:
\begin{equation}\label{noccup-sat}
n(x, k_\perp^2) \simeq \frac{1}{\alpha_s N_c}\,\Gamma\left(0, \frac{k_\perp^2}{\tilde Q_s^2(A)}\right) \quad \text{for } k_\perp \lesssim \tilde Q_s(A)\,,
\end{equation}
where $\Gamma(0,z) \equiv \int_z^\infty \rmd t \, (\rme^{-t}/t)$ is the incomplete Gamma function. At very low momenta $k_\perp \ll \tilde Q_s(A)$, one can use $\Gamma(0,z) \simeq \ln(1/z)$ for $z\ll 1$ to find
 \begin{equation}\label{noccup-low}
n(x, k_\perp^2) \simeq \frac{1}{\alpha_s N_c}\ln\left(\frac{\tilde Q_s^2(A)}{k_\perp^2}\right) \quad \text{for } k_\perp \ll \tilde Q_s(A)\,.
\end{equation}
\end{description}

Eq.~\eqref{noccup-low} is the expression of \textit{gluon saturation}. As $k_\perp \to 0$, the power-law divergence $1/k_\perp^2$ of the bremsstrahlung spectrum~\eqref{noccup-high} is replaced by a mild, logarithmic growth. The phase-space occupation number caps at a universal value of order $1/\alpha_s$ since further emissions of soft, long-wavelength gluons are dynamically blocked by non-linear effects. Another interesting feature of the saturation spectrum \eqref{noccup-sat} is the fact that it is {\it compact}: it vanishes exponentially when extrapolated to large values $k_\perp\gg \tilde Q_s(A)$. (Indeed,  $\Gamma(0,z)\simeq e^{-z}/z$ for $z\gg 1$.) This feature is a consequence of the transverse momentum redistribution via multiple scattering among the saturated gluons.



The above estimates are confirmed by a numerical calculation of the integral in  Eq.~\eqref{xfgMV}, with the results displayed in the left plot of Fig.~\ref{fig:WW}. For comparison, we also display the two limiting behaviours, Eq.~\eqref{noccup-high} at large $k_\perp$ and Eq.~\eqref{noccup-sat} at low $k_\perp$. As for the right plot in  Fig.~\ref{fig:WW}, this shows the evolution of the gluon occupation number with increasing $\Delta Y\equiv \ln(x/x_0)$, as numerically computed within a Gaussian approximation to the JIMWLK equation to be discussed in Sect.~\ref{sec:Gauss}. These plots  confirm that the saturation momentum $ \tilde Q_s(A,Y)$ increases with $Y$ (see Sect.~\ref{sec:QsY}).

\section{High-energy evolution of the CGC}
\label{sec:evol}

As discussed in Sect.~\ref{sec:CGCgen}, the high-energy evolution becomes important when the gluon distribution is probed at sufficiently small values of $x$, such that  $\abar\ln(1/x)\gtrsim 1$ with $\abar\equiv \alpha_s N_c/\pi$. If the gluon density is so large that saturation effects become important, the high-energy evolution is governed by the non-linear JIMWLK equation for the CGC weight function~\cite{JalilianMarian:1997jx,JalilianMarian:1997gr,Kovner:2000pt,Iancu:2000hn,Iancu:2001ad,Ferreiro:2001qy}. When applied to dilute-dense collisions, this functional equation can be equivalently rewritten as an infinite hierarchy of ordinary integro-differential equations—the Balitsky equations~\cite{Balitsky:1995ub,Weigert:2000gi,Mueller:2001uk} —
which describe the high-energy evolution of the dilute projectiles.
 In practice, it is much simpler to follow the evolution of a dilute hadronic system, since this avoids an explicit treatment of gluon saturation. Non-linear effects are still present—they refer to the multiple scattering between partons from the projectile and the dense gluonic system in the target ---, but they can be easily accommodated (at least in the eikonal approximation) via background field techniques. 
 
 In this section, we shall study the evolution of the simplest colourless projectile—a quark-antiquark dipole—and thus derive the first equation in the Balitsky hierarchy. At large $N_c$, this becomes a {\it closed}, non-linear, equation—the BK equation~\cite{Kovchegov:1999yj,Kovchegov:1999ua} ---, which is useful in practice.  Based on it, we shall discuss the  evolution of the saturation momentum,  including the important effects of a running coupling. We will then discuss a particular class of higher-order corrections (the collinear double logarithms), which are large and potentially harmful. We will explain their origin and describe methods for keeping their effects under control. The last two subsections will be devoted to the (functional)  JIMWLK evolution. After briefly discussing a mean field approximation (an extension of the MV model to larger rapidities), we will explain the Langevin formulation of the JIMWLK evolution. Besides its utility in practice (for constructing numerical solutions), this formulation also offers a transparent picture of the high-energy evolution in the presence of gluon saturation.

\subsection{The Balitsky-Kovchegov equation}
\label{sec:BK}

The BK equation describes the high-energy evolution of the elastic $S$-matrix for a quark-antiquark ($q\bar q$) colour dipole, \eqref{Sdipole}, in the multicolour limit $N_c\gg 1$. This limit possesses two primary virtues. First, it permits the formulation of a \textit{closed}  equation, which is significantly easier to handle in practice than an infinite hierarchy. Second, it leads to a transparent physical interpretation of the high-energy evolution in terms of successive dipole splittings, widely known as Mueller's dipole picture~\cite{Mueller:1993rr}.

To uncover this picture and establish the BK equation, let us examine the change in the $S$-matrix $S(\bx,\by)$ associated with the emission of a soft gluon ($g$) carrying a small longitudinal momentum fraction $z=k^+/q^+\ll 1$ by either the quark ($\bx$) or the antiquark ($\by$) leg of the parent dipole. We compute this process in the light-cone gauge $A^+_a=0$, where the high-energy evolution of the right-moving dipole admits a partonic interpretation: the soft gluon represents a nearly on-shell excitation—a \textit{Weizsäcker-Williams gluon}—characterised by a transverse polarisation and a relatively long lifetime $\Delta x^+\simeq 2k^+/k_\perp^2\gg 1/P^-$. For the scattering process to remain elastic, this fluctuation must be both emitted and reabsorbed within the dipole system. Since the gluon is soft, its emission does not alter the transverse coordinate of the emitting parton (and similarly for its subsequent absorption).

\begin{figure}[t] 
\centering
\includegraphics[width=0.9\textwidth]{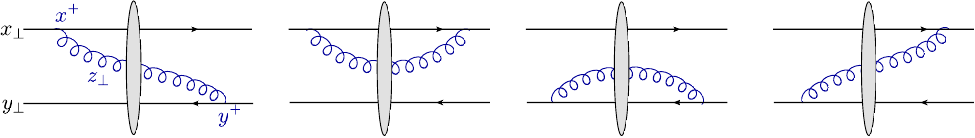}
\caption{\sl Diagrams illustrating one step in the high-energy evolution of a $q\bar q$ dipole via the emission of a soft gluon that interacts with the target (represented as a shockwave). The emission and absorption times $x^+$ and $y^+$, as well as the gluon transverse coordinate $\bz$ at the time of scattering, are arbitrary and must be integrated over.}
\label{1gluon_real}
\end{figure}

Consider first the case where the gluon is emitted before the shockwave and reabsorbed after it, meaning that the gluon itself actively participates in the collision (see Fig.~\ref{1gluon_real}). Its scattering can be computed within the eikonal approximation, which mathematically amounts to inserting a Wilson line in the adjoint representation $\tilde V({\bz})$ at the gluon transverse position $\bz$. Hence, the change in the elastic dipole $S$-matrix due to this specific $q\bar q g$ configuration reads (where the overall factor $1/N_c$ arises from averaging over the initial fundamental colour states):
\begin{align}\label{Sqqg}
\frac{1}{N_c}\left\langle\tilde V^{ab}(\bz) \, \rmtr \Big(V(\bx) t^b\, V^{\dagger}(\by) t^a \Big)\right\rangle
& =\frac{1}{N_c}\left\langle\rmtr \Big(V(\bx) V^{\dagger}(\bz) t^{a}V(\bz)V^{\dagger}(\by) t^a \Big)\right\rangle \nonumber\\[0.2cm]
& =\frac{N_c}{2}\left\langle\hat{S}_{\bx\bz} \hat{S}_{\bz\by} -\frac{1}{N_c^2}\, \hat{S}_{\bx\by}\right\rangle 
 \simeq \frac{N_c}{2}\,\langle\hat{S}_{\bx\bz}\rangle \langle\hat{S}_{\bz\by}\rangle\,,
\end{align}
where $\hat{S}_{\bx\by}\equiv ({1}/{N_c}) \, \rmtr\big( V(\bx){V}^{\dagger}(\by) \big)$ represents the dipole $S$-matrix \textit{operator}. In deriving Eq.~\eqref{Sqqg}, we have successively the following Fierz identities to relate adjoint and fundamental representations:
\begin{equation}\label{Fierz} 
\tilde V^{ac}(\bz)t^c= V^\dagger(\bz) t^a V(\bz),\qquad t^a_{\alpha\beta}\,t^a_{\gamma\delta}\,=\,{1 \over 2}\,\delta_{\alpha\delta}\delta_{\beta\gamma}\,-\, {1 \over 2N_c}\, \delta_{\alpha\beta}\delta_{\gamma\delta}.
\end{equation}
The final factorisation step in \eqref{Sqqg} holds strictly in the large-$N_c$ limit. In this regime, the expectation value of a product of gauge-invariant operators factorises into the product of their individual expectation values, as colour fluctuations between the two colour singlets are suppressed by powers of $1/N_c^2$. Consequently, the $q\bar q g$ system scatters precisely like two independent, colour-singlet $q\bar q$ dipoles with transverse coordinates $(\bx,\bz)$ and $(\bz,\by)$.

To complete the evaluation of the diagrams in Fig.~\ref{1gluon_real}, we need the differential probability for emitting a gluon with longitudinal momentum $k^+$ at  transverse coordinate $\bz$. This is obtained by squaring the emission vertex and summing over the gluon polarisation states, yielding~\cite{Mueller:1993rr,Mueller:2001fv}:
\begin{equation}\label{Pdip}
\rmd P = \frac{\alpha_s}{\pi^2}\,\mcal{M}_{\bx\by\bz}\,\rmd^2\bz\,\rmd Y\quad\text{with}\quad \mcal{M}_{\bx\by\bz}\,\equiv\, \frac{(\bx-\by)^2}{(\bx-\bz)^2(\bz-\by)^2}\,,
\end{equation}
where the differential rapidity is defined via $\rmd k^+/k^+=\rmd Y$. The functional form of the \textit{dipole kernel} $\mcal{M}_{\bx\by\bz}$ is generated by the overlapping emission amplitudes:
\begin{equation}\label{Mdef}
\mcal{M}_{\bx\by\bz} = \left(\frac{x^i-z^i}{(\bx-\bz)^2}-\frac{y^i-z^i}{(\by-\bz)^2}\right) \left(\frac{x^i-z^i}{(\bx-\bz)^2}-\frac{y^i-z^i}{(\by-\bz)^2}\right).
\end{equation}
Crucially, this kernel vanishes when the parent dipole size approaches zero ($|\bx-\by|\to 0$). This behavior manifests quantum colour transparency, arising from the exact destructive interference between emissions sourced by the quark and the antiquark.

\begin{figure}[t] 
\centering
\includegraphics[width=0.9\textwidth]{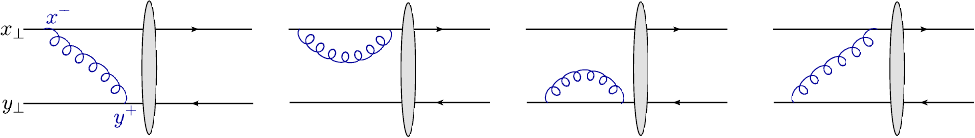}
\caption{\sl Dipole evolution via a virtual soft gluon emission that does not cross the shockwave: the gluon is both emitted and reabsorbed prior to the collision. An identical set of diagrams accounts for fluctuations occurring entirely after the collision.}
\label{1gluon_virt}
\end{figure}

Next, we consider the graphs where the soft gluon does not cross the shockwave, meaning it is both emitted and absorbed either entirely before the collision (see Fig.~\ref{1gluon_virt}) or entirely after it. Then the colour structure simplifies considerably because only the parent dipole participates in the scattering:
\begin{equation}\label{Sqq}
\frac{1}{N_c}\left\langle \, \rmtr \Big(t^a V(\bx) V^{\dagger}(\by) t^a \Big)\right\rangle = \frac{C_F}{N_c}\left\langle\rmtr \big(V(\bx) V^{\dagger}(\by)\big)\right\rangle \simeq \frac{N_c}{2}\,\langle\hat{S}_{\bx\by}\rangle\,,
\end{equation}
where we used the large-$N_c$ limit of the quadratic Casimir operator, $C_F=(N_c^2-1)/2N_c\simeq N_c/2$.

Remarkably, before taking the $N_c\to\infty$ limit explicitly, the sub-leading terms in $1/N_c^2$ cancel exactly between the real contributions \eqref{Sqqg} and the virtual contributions \eqref{Sqq}. Hence, for a \textit{generic} number of colours $N_c$, we deduce the following evolution equation (defining $\abar\equiv \alpha_s N_c/\pi$):
\begin{equation}\label{BKgen}
\frac{\del \lan \hat{S}_{\bx\by} \ran_Y}{\del Y}= \frac{\abar}{2\pi}\, \int \rmd^2{\bz}\, \mcal{M}_{\bx\by\bz} \big\lan \hat{S}_{\bx\bz} \hat{S}_{\bz\by} -\hat{S}_{\bx\by} \big\ran_Y\,.
\end{equation}
Equation \eqref{BKgen} is not yet a \textit{closed} equation: its right-hand side couples the single-dipole $S$-matrix to the $S$-matrix of a two-dipole system, $\big\lan \hat{S}_{\bx\bz} \hat{S}_{\bz\by}\big\ran_Y$. For a finite value of $N_c$, this expression does not factorise due to  colour correlations among the target fields (see e.g., \cite{Iancu:2011nj} for an explicit calculation within the MV model, and \cite{Kovchegov:2008mk} for numerical studies of factorisation violations). This two-dipole operator obeys its own evolution equation
(see e.g. Appendix F of Ref.~\cite{Triantafyllopoulos:2005cn}), which in turn couples to quadrupole operators (see Eq.~\eqref{quadrupole}), generating the infinite B-JIMWLK hierarchy.

The situation becomes simpler in the large-$N_c$ limit, where the two-dipole correlator factorizes, allowing Eq.~\eqref{BKgen} to reduce to a closed, non-linear,  equation known as the \textit{BK equation}~\cite{Kovchegov:1999yj}:
\begin{align}\label{BK}
 \frac{\del {S}_Y(\bx,\by)}{\del Y}\, =\, \frac{\abar}{2\pi}
 \int\rmd^2{\bz}\,
 \mcal{M}_{\bx\by\bz}
 \Big\{ {S}_Y(\bx,\bz){S}_Y(\bz,\by)
 - {S}_Y(\bx,\by)\Big\},
 \end{align}
where we denote $S_Y(\bx,\by) \equiv \langle\hat{S}_{\bx\by}\rangle_Y$. Notice that the colour transparency of small dipoles is preserved by the evolution: ${S}_Y(\bx,\by)\to 1$ when $|\bx-\by|\to 0$, since the dipole kernel vanishes in that limit. This property, in turn, ensures that the short-distance (ultraviolet) poles in the dipole kernel are harmless in practice. For instance, when $(\bx-\bz)^2\to 0$, we have ${S}_Y(\bx,\bz)\to 1$ and $S_Y(\bz,\by)\to S_Y(\bx,\by)$, which causes the combination of $S$-matrices inside the brackets to vanish identically.

For a physical discussion, it is instructive to rewrite the BK equation in terms of the forward dipole scattering amplitude, defined as ${T}_Y(\bx,\by)\equiv 1 - {S}_Y(\bx,\by)$:
\begin{align}\label{BKT}
 \frac{\del {T}_Y(\bx,\by)}{\del Y} = \frac{\abar}{2\pi}
 \int\rmd^2{\bz}\,
 \mcal{M}_{\bx\by\bz}
 \Big\{ T_Y(\bx,\bz) + T_Y(\bz,\by) - T_Y(\bx,\by) - T_Y(\bx,\bz) T_Y(\bz,\by)\Big\}.
 \end{align}
This rewriting makes the dipole picture manifest: the quantity $(\abar/2\pi) \mcal{M}_{\bx\by\bz}$ represents the differential probability for a dipole splitting event; the first two terms inside the brackets, $T_Y(\bx,\bz)$ and $T_Y(\bz,\by)$, describe the independent elastic scattering of the newly formed daughter dipoles off the target; the last, quadratic, term $- T_Y(\bx,\bz) T_Y(\bz,\by)$ accounts for their simultaneous scattering (absorptive corrections); finally, the linear negative term $- T_Y(\bx,\by)$ expresses the reduction in the probability for the parent dipole to remain intact at the time of scattering.

Importantly, the combination of dipole amplitudes on the right-hand side of Eq.~\eqref{BKT} vanishes when $T_Y\to 1$. Consequently, the quantum evolution automatically slows down and stops when the scattering amplitude reaches the unitarity (or ``black disk'') limit, $T_Y=1$. In the opposite regime where the scattering is weak and the target is dilute ($T_Y\ll 1$), one can neglect the quadratic term. In this linear approximation, Eq.~\eqref{BKT} reduces to the dipole-picture version of the celebrated BFKL equation~\cite{Lipatov:1976zz,Kuraev:1977fs,Balitsky:1978ic,Mueller:1993rr}:
\begin{align}\label{BFKL}
 \frac{\del {T}_Y(\bx,\by)}{\del Y} = \frac{\abar}{2\pi}
 \int\rmd^2{\bz}\,
 \mcal{M}_{\bx\by\bz}
 \Big\{ T_Y(\bx,\bz) + T_Y(\bz,\by) - T_Y(\bx,\by)\Big\}.
 \end{align}
Taken at face value, this linear equation predicts an exponential growth of the amplitude with rapidity, which eventually leads to violations of the unitarity bound $T_Y\le 1$. Yet, because its solutions can be constructed in analytic form, the BFKL equation is a very useful tool for studying the approach towards unitarity with increasing $Y$. This will be discussed in the next section.

\comment{The emission vertices. For a single emission, by either the quark or the antiquark,  this is computed as\footnote{In the gauge $A^+_a=0$, the polarisation 4-vector for a transverse gluon reads $\epsilon^\mu_\lambda= (0, \epsilon^-_\lambda,\bme_\lambda)$ with $\epsilon^-_\lambda=(\bme_\lambda \cdot\bk)/k^+$ to satisfy the condition $k_\mu \epsilon^\mu_\lambda =0$.} (see e.g. \cite{Mueller:2001fv})
\begin{equation}\label{Vdip} {\cal
 V}^a_\lambda(k^+,\bx, \by, \bz)
 =\,\frac{2i gt^a} {\sqrt{(2\pi)^3 2k^+}}\left[\frac{\bme_\lambda \cdot
 (\bx-\bz)}{(\bx-\bz)^2}-\frac{\bme_\lambda \cdot
 (\by-\bz)}{(\by-\bz)^2}\right]
 \end{equation}
where $a$ and $\lambda$ represent colour and (transverse) polarisation indices for
the emitted gluon at $\bz$. The respective absorption vertex is  the complex conjugate of \eqref{Vdip}.
The differential probability for the process in Fig.~\ref{1gluon_real} is obtained as
by multiplying the emission and absorption vertices, summing over the polarisation indices (with the help of $\sum_\lambda \epsilon_\lambda^i
(\epsilon_\lambda^{j})^*=\delta^{ij}$) and discarding the colour matrices (since they were already included in  \eqref{Sqqg}). One thus finds}

\subsection{Saturation line and geometric scaling}
\label{sec:QsY}

In the gauge duality between gluon saturation in the  target  and multiple scattering for the projectile dipole, the saturation momentum $Q_s(Y)$ is defined as the inverse dipole size at which the scattering becomes strong:\footnote{We consider a homogeneous and isotropic hadronic target, for which ${T}_Y(\bx_\perp,\by_\perp) = T(r_\perp,Y)$ with $r_\perp \equiv |\bx_\perp-\by_\perp|$.}
\begin{equation}\label{sat_cond}
T(r_\perp,Y) = \kappa \quad \text{when } r_\perp = \frac{2}{Q_s(Y)}\,,
\end{equation}
where $\kappa < 1$ is an arbitrary parameter of order unity (typically $\kappa = 1/2$). While the precise value of $Q_s(Y)$ depends on $\kappa$, its high-energy behavior at large rapidity $Y$ is \textit{universal}—independent of the initial condition, the nature of the projectile, or the chosen value for  $\kappa$. In particular, one can use a small number $\kappa\ll 1$ and still correctly deduce the asymptotic behaviour at large $Y$.  This implies that the \textit{saturation line}—defined by the trajectory $\rho_s(Y) \equiv \ln[Q_s^2(Y)/Q_0^2]$, where $Q_0$ is a convenient non-perturbative reference scale (e.g., $Q_0 = Q_A$ for a large nucleus)—can be determined by evaluating the BK solution in the weak scattering regime ($T \sim \kappa \ll 1$), where the linear BFKL evolution applies.

To resolve the non-locality of the BFKL equation \eqref{BFKL} in transverse space, it is advantageous to perform a Mellin transform from the dipole coordinate $r_\perp$ to the complex variable $\gamma$~\cite{Iancu:2002tr,Mueller:2002zm}. In Mellin space, the BFKL equation admits exact analytic solutions. At large rapidities, the inverse Mellin transform can be evaluated within the saddle-point approximation. Combining the saddle-point condition with the strong scattering condition $T(\rho_s(Y), Y) = \kappa$ forms a system of two coupled equations that determine both the saturation line $\rho_s(Y)$ and the saddle-point $\gamma_s(Y)$. Performing a Gaussian integration around this  saddle point finally yields the structure of the dipole amplitude in the dilute-to-dense transition~\cite{Iancu:2002tr,Mueller:2002zm,Munier:2003vc,Munier:2003sj}:
\begin{equation}\label{Trho} 
T(\rho,Y) \simeq \big(\rho-\rho_s(Y)+\kappa\big)\, \rme^{-\gamma_s (\rho-\rho_s(Y))}\, \exp\left\{-\frac{(\rho-\rho_s(Y))^2}{2\beta_s \bar{\alpha}_s Y} \right\}\,,
\end{equation}
where $\rho \equiv \ln(4/r_\perp^2 Q_0^2)$ and the  saturation line
\begin{equation}\label{rhos} 
\rho_s(Y) \equiv \ln\left(\frac{Q_s^2(Y)}{Q_0^2}\right) = c_s \bar{\alpha}_s (Y-Y_0) - \frac{3}{2\gamma_s} \ln\left(\frac{Y}{Y_0}\right)\,. 
\end{equation}
Here, the reference rapidity scale $Y_0$ satisfies $Q_s^2(Y_0)=Q_0^2$. The coefficients $c_s$, $\gamma_s$, and $\beta_s$ are determined by the BFKL characteristic function $\chi(\gamma)$ expressing the BFKL kernel in Mellin space:
\begin{equation}\label{chigamma}
\chi(\gamma) \equiv 2\psi(1) - \psi(\gamma) - \psi(1-\gamma)\,, \qquad \psi(\gamma) \equiv \frac{\rmd \ln \Gamma(\gamma)}{\rmd\gamma}\,.
\end{equation}
The saturation exponent $\gamma_s$ is implicitly defined as the unique root of the transcendental relation:
\begin{equation}\label{gammasBFKL}
\gamma_s = \frac{\chi(\gamma_s)}{\chi'(\gamma_s)} = 0.627\dots\,, \qquad c_s \equiv \frac{\chi(\gamma_s)}{\gamma_s} = 4.883\dots\,, \qquad \beta_s \equiv \chi''(\gamma_s) = 48.518\dots\,.
\end{equation}
Eq.~\eqref{Trho} is strictly valid in the asymptotic high-energy limit ($Y\gg Y_0$) and within a limited region within the dilute regime, at $0 < \rho - \rho_s(Y) \ll Y$. In this domain, the logarithmic trajectory \eqref{rhos} is universal. For larger dipoles satisfying $\rho \lesssim \rho_s(Y)$ ($r_\perp > 2/Q_s(Y)$), the amplitude smoothly approaches towards the black-disk limit $T = 1$, a non-linear effect that transcends the linear structure of Eq.~\eqref{Trho}.

The Gaussian factor in Eq.~\eqref{Trho} describes diffusion in the logarithmic variable $\rho$ around the saturation line $\rho_s(Y)$: with increasing rapidity, the typical deviation grows like $|\rho-\rho_s(Y)|\propto \sqrt{Y}$. For values of $\rho$ sufficiently close to the saturation boundary such that this diffusive expansion remains negligible ($0 \le \rho - \rho_s(Y) \ll \sqrt{2\beta_s \bar{\alpha}_s Y}$), Eq.~\eqref{Trho} reduces to the following expression 
\begin{equation}\label{TGS} T(\rho,Y) \, \simeq \,\big(\rho-\rho_s(Y)+\kappa\big)\,
 {\rm e}^{-\gamma_s (\rho-\rho_s(Y))}\,\simeq\,\left(\ln\frac{4}{r^2Q_s^2(Y)}+\kappa\right)\left[\frac{r^2Q_s^2(Y)}{4}\right]^{\gamma_s}
,\end{equation}
which exhibits {\it geometric scaling}: $T(\rho,Y) \simeq \mcal{T}(\rho-\rho_s(Y))$ or, equivalently, $T(r,Y)\simeq
\mcal{T}(r^2Q^2_s(Y))$. That is, the scattering amplitude depends upon the dipole size and upon the rapidity only via the dimensionless variable  $r^2Q^2_s(Y)$.  While such a scaling is {\it a priori} natural in the saturation region at $r\gtrsim 1/Q_s(Y)$, 
where the target is nearly ``black'' ($T\simeq 1$) and the saturation momentum is the only scale in the problem, 
its persistence in the weak scattering regime at $r\ll 1/Q_s(Y)$ is very non-trivial. This can be related to the 
``anomalous dimension'' $1-\gamma_s \simeq 0.37$ visible  in Eq.~\eqref{TGS}, that is generated by the BFKL evolution. The fact that $\gamma_s < 1$  slows down the growth of the scattering amplitude with $r$ in the dilute regime (one has $T\sim r^{2\gamma_s}$ instead of $T\sim r^2$) and thus removes the need for an infrared cutoff  $\sim\Lambda$ in the approach towards saturation.  (This should be contrasted with the MV model, where such a cutoff is indeed needed, see Eq.~\eqref{SMV}.) This improved infrared behaviour reflects the partial screening of the colour charges of the valence partons by the soft emitted gluons~\cite{Mueller:2002pi,Iancu:2002aq}.

\begin{figure}[t]
 \centerline{   \includegraphics[width=0.48\textwidth]{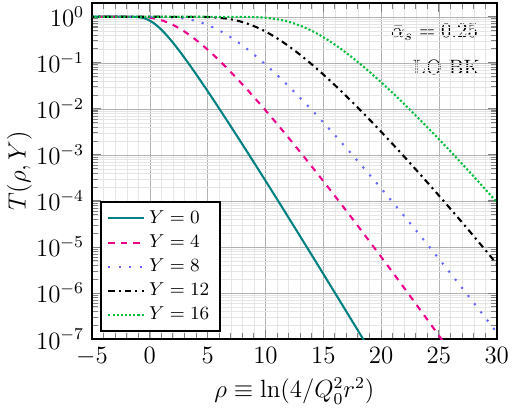}\qquad
   \includegraphics[width=0.48\textwidth]{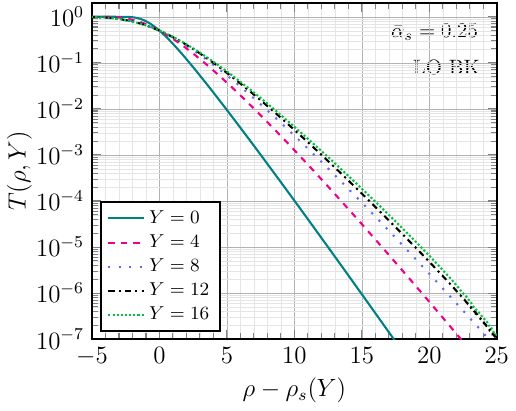}
}
 \caption{\sl Left: Numerical solutions to the BK equation \eqref{BKT} with fixed coupling $\abar=0.25$ and
 initial condition from the MV model at $Y=0$. (In the MV model,  $Q_s(A)=1$~GeV and $Q_A=0.34$~GeV.)
 Right: The same as in the left figure, but plotted as a function of  $\rho-\rho_s(Y)$,  to illustrate geometric scaling. }
 \label{BKfront}
\end{figure}

 Eq.~\eqref{Trho}  also shows that {\it deviations} from geometric scaling are to be expected when the difference $\rho-\rho_s(Y)$ is sufficiently large.   For $\rho-\rho_s(Y)\sim \sqrt{2\beta_s \bar\alpha Y}$, they are a consequence of BFKL diffusion, as 
 described by the Gaussian factor in \eqref{Trho}. For much larger  $\rho-\rho_s$, one can rely on the double-logarithmic approximation (DLA) discussed in Sect.~\ref{sec:gluons} (recall Eq.~\eqref{fgDLA}). These arguments imply the following, piecewise, approximation for the dipole amplitude $T(\rho, Y)$:
\begin{equation}\label{TYrho}
  T(\rho, Y)\simeq   \begin{cases}
        \displaystyle{\rme^{-\rho}\,\rme^{2\sqrt{\abar Y\rho}}} 
        &
        \text{ for\,  $\rho-\rho_s(Y)\gg \sqrt{2\beta_s\abar Y}$\qquad\quad (DLA),}
        \\*[0.3cm]
        \displaystyle{\rme^{-\gamma_s(\rho-\rho_s(Y))} }&
        \text{ for\,  $1\lesssim \rho-\rho_s(Y) \lesssim
\sqrt{2\beta_s\abar Y}$\qquad (geometric scaling),}
        \\*[0.3cm]
         \displaystyle{1} &
        \text{ for\,  $\rho\lesssim \rho_s(Y)$\qquad\qquad\qquad (black disk limit).}
         \\*[0.1cm] 
    \end{cases}
\end{equation}

These analytic considerations are illustrated by the numerical solutions to the BK equation \eqref{BK} shown in Fig.~\ref{BKfront}. In logarithmic units, the dipole amplitude $T(\rho, Y)$ looks like a front  (or ``traveling wave''), which with increasing $Y$ propagates towards larger values of $\rho$. This propagation is driven by the rapid BFKL growth in the dilute tail at $\rho >\rho_s(Y)$. The saturation scale $\rho_s(Y)$ is numerically extracted from the condition that $T(\rho, Y)=1/2$ when $\rho =\rho_s(Y)$. In the right plot, the amplitude is displayed in terms of the difference $\rho-\rho_s(Y)$, to illustrate geometric scaling: at sufficiently large $Y$, the successive saturation fronts fall nearly on top of each other.

The speed of the front $\lambda_s(Y)\equiv\rmd \rho_s/\rmd Y$ (the ``saturation exponent'') is numerically extracted as well and shown in Fig.~\ref{BKfront-run} (right plot). 
This numerical result should be compared with the analytic estimate in Eq.~\eqref{rhos}, valid at large $Y$:
\begin{equation}\label{lambdas} \lambda_s(Y)\equiv\frac{\rmd  \rho_s(Y)}{\rmd Y}\,=\,
 c_s\bar\alpha\,-\, \frac{3}{2\gamma_s Y}+ \order{\frac{1}{Y^{3/2}}}\,. \end{equation}
 As shown by the numerics,  the  asymptotic value $\lambda_s\simeq 4.88\abar$ is  reached only for  $\abar Y \gtrsim 5$.
 
 The universal features of BK solutions at large \(Y\)—most notably the emergence of traveling waves—gained immense clarity through the uncovering of a profound correspondence~\cite{Munier:2003vc,Munier:2003sj,Iancu:2004es,Munier:2009pc} linking high-energy QCD evolution to reaction-diffusion problems in statistical physics~\cite{vanSaarloos2003}. Within this framework, the BK evolution near the saturation line maps directly onto the F-KPP (Fisher, Kolmogorov, Petrovsky, and Piskunov) equation~\cite{Munier:2003vc,Munier:2003sj}. Furthermore, the stochastic nature of the reaction-diffusion problem provides a classical analog for gluon number fluctuations in high-energy QCD—giving rise to the celebrated "Pomeron loops," for which a complete, first-principles description remains a major theoretical challenge~\cite{Iancu:2003zr,Mueller:2004sea,Iancu:2004es,Iancu:2004iy,Mueller:2005ut,Kovner:2005nq,Dumitru:2007ew,Hentschinski:2026nfe,Triantafyllopoulos:2005cn}.

At this level, one may observe a tension between the BK prediction for the saturation exponent, $\lambda_s 
\simeq 4.88\abar \sim 1$ for $\abar= 0.25$,  and the corresponding value extracted from dipole-model fits to the HERA data~\cite{GolecBiernat:1998js,GolecBiernat:1999qd,Stasto:2000er,Bartels:2002cj} (and also from the small-$x$ behaviour of the gluon PDF), which is {\it much} slower, in the ballpark of 0.3.
As we shall see, this discrepancy can be solved after including higher order corrections to the BK equation, which dramatically slow down the evolution.


\subsection{Running coupling effects in  the BK evolution}
\label{sec:rcBK}

The BK and BFKL equations are non-local in transverse coordinates (or momenta) and the effect of the non-locality gets amplified with increasing $Y$, via the BFKL diffusion (cf. Eq.~\eqref{Trho}). For the BFKL equation, this diffusion proceeds towards both lower and larger values of $\rho\sim \ln k_\perp^2 \sim \ln(1/r^2)$ and the infrared diffusion is favoured by the running of the coupling (which becomes larger with decreasing $\rho$). Because of that, the BFKL evolution with running coupling eventually runs out of control—it probes very soft momenta, or large dipole sizes, for which unitarity is violated and perturbation theory fails to apply. This problem is cured by gluon saturation in the case of the non-linear BK equation, where the diffusion is restricted to the right of the saturation boundary, that is, to $\rho >\rho_s(Y)$. Since $\rho_s(Y)$ increases with $Y$, whereas the running coupling $\alpha_s(\rho)$ decreases with $\rho$, this constrained diffusion introduces a positive feedback which slows down the evolution with increasing $Y$: the relevant values of $\alpha_s(\rho)$ become smaller and smaller, leading to a decrease in the saturation exponent $\lambda_s$.

To study this, we shall use the one-loop version of the running coupling, that is,
\begin{align}
	\label{runcoup}
	\alpha_s(\mu^2) = \frac{1}{\beta_0 \ln (\mu^2/\Lambda^2)}
	\qquad \textrm{with} \qquad
	\beta_0=\frac{11 N_c - 2 n_f}{12\pi}\,,
\end{align}
with $n_f$ the number of active flavours. Also, it is convenient to redefine the logarithmic variable $\rho$ by using $\Lambda$ as the reference scale: $\rho\equiv \ln(4/r^2\Lambda^2)$ and similarly $\rho_s(Y)\equiv \ln(Q_s^2(Y)/\Lambda^2)$. The running coupling corresponding to a dipole of size $r$ is therefore $\abar(\rho)=b_0/\rho$ with $b_0\equiv N_c/\pi\beta_0$.

For values of $\rho$ within the scaling window,  one can approximate
$\abar(\rho)\approx \abar(\rho_s(Y))$.  Indeed, the width of that window is relatively small already for a fixed coupling and it becomes even smaller after using a running coupling (see Eq.~\eqref{gswindowrun} below). With this approximation, the saturation exponent takes the same form as at fixed coupling, cf. Eq.~\eqref{lambdas}, except for the replacement $\abar\to  \abar(\rho_s(Y))$.  Then the saturation line with running coupling can be computed as~\cite{Iancu:2002tr,Mueller:2002zm,Triantafyllopoulos:2002nz,Munier:2003sj}
   \begin{equation}\label{eq-dxsat}
    \frac{\dif \rho_s}{\dif Y}\, \simeq \,c_s\abar(\rho_s) \,=\,
    \frac{c_s b_0}{\rho_s}\quad \Longrightarrow\quad \rho_s(Y) =
    \sqrt{{2c_s}b_0(Y-Y_0)+\rho_0^2}\,.
   \end{equation}
with the initial condition $\rho_s(Y_0)=\rho_0\equiv\ln(Q_0^2/\Lambda^2)$. More refined calculations permit to also estimate the width of the geometric scaling window~\cite{Triantafyllopoulos:2002nz,Iancu:2004bx}: one finds
\begin{equation}\label{gswindowrun}
 0\,<\, \rho-\rho_s(Y)\,\lesssim\,  \rho_s^{1/3}(Y)\,\qquad\mbox{(geometric scaling window for a running coupling)}.
    \end{equation}
Eq.~\eqref{eq-dxsat} can be more explicitly written as
\begin{equation}\label{QsRC}
Q_s^2(Y)=\Lambda^2\,\rme^{ \sqrt{2c_sb_0(Y-Y_0)+\rho_0^2}}
\simeq   \begin{cases}
        \displaystyle{Q_0^2\,\rme^{c_s\bar\alpha(Q_0^2) (Y-Y_0)}} 
        &
        \text{when\,  $\sqrt{2c_sb_0(Y-Y_0)}\ll \rho_0$,}
        \\*[0.3cm]
        \displaystyle{\Lambda^2\,\rme^{ \sqrt{2c_sb_0(Y-Y_0)}}}
 &
        \text{ when\,  $\sqrt{2c_sb_0(Y-Y_0)}\gg \rho_0$.}
                 \\*[0.2cm] 
    \end{cases}
\end{equation}
Interestingly, the asymptotic estimate in the second line is independent of $\rho_0$, hence of the initial condition. This has the important consequence that, for sufficiently large $Y$, also the saturation {\it momentum} (and not only the saturation {\it exponent}) becomes a {\it universal} quantity: it approaches the same asymptotic value for any hadronic target (a proton or a large nucleus)~\cite{Mueller:2003bz}. Eq.~\eqref{QsRC} also shows that the running of the coupling significantly slows down the evolution: the respective saturation exponent $\lambda_s(Y) \simeq (c_s b_0/2Y)^{1/2}$ decreases with $Y$.

\begin{figure}[t]
 \centerline{   \includegraphics[width=0.48\textwidth]{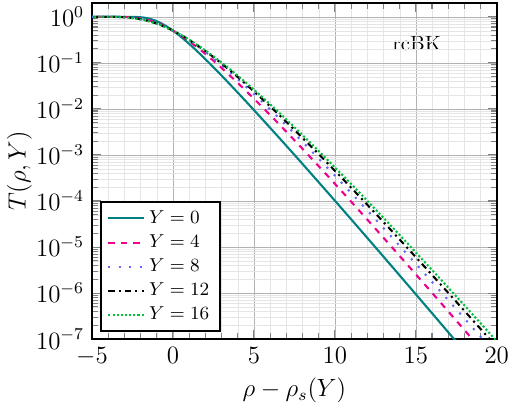}\qquad
   \includegraphics[width=0.48\textwidth]{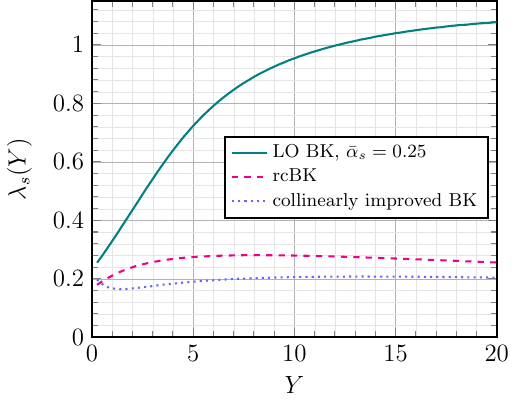}
}
 \caption{\sl Left: Numerical solutions to the BK equation  with running coupling (RC) for $n_f=3$, $\Lambda=0.2$~GeV and with the smallest dipole size prescription. The amplitude is plotted as a function of the difference $\rho-\rho_s(Y)$, with  the saturation scale obeying $T(\rho_s(Y), Y)=1/2$. 
  Right: The saturation exponent $\lambda_s(Y)\equiv\rmd \rho_s/\rmd Y$ numerically extracted from solutions to three versions of the BK equation: (i) Eq.~\eqref{BKT} with fixed coupling $\abar=0.25$, (ii)  Eq.~\eqref{BKT} with RC, and (iii) the ``collinearly-improved'' equation~\cite{Ducloue:2019ezk} with RC (see Sect.~\ref{sec:collBK}).}
 \label{BKfront-run}
\end{figure}

These simple estimates are consistent with numerical solutions to the BK equation using a running coupling $\alpha_s(\mu^2)$ endowed with a physical prescription for the scale $\mu^2$. This prescription can be inferred from a study of the NLO corrections to the BK equation~\cite{Balitsky:2006wa,Balitsky:2008zza,Balitsky:2013fea}. The NLO corrections proportional to $\beta_0$ can be minimised by choosing  $\mu^2 = 4/r_{\rm min}^{2}$, where  $r_{\rm min}={\rm min}\{|\bx-\by|,|\bx-\bz|,|\bz-\by|\}$  is the  smallest dipole size. (Alternative prescriptions, which involve smooth functions of the dipole sizes, have been proposed in Refs.~\cite{Kovchegov:2006wf,Kovchegov:2006vj,Balitsky:2006wa,Iancu:2015joa}.) In Fig.~\ref{BKfront-run} we show the numerical results obtained with this prescription for  the dipole amplitude and the saturation exponent. With increasing $Y$, the amplitude exhibits the same ``travelling wave'' pattern as in Fig.~\ref{BKfront}, thus confirming the persistance of geometric scaling in the presence of a running coupling. The saturation exponent is dramatically reduced by the running of the coupling and so is its variation with $Y$.

The running coupling version of the BK equation, a.k.a. ``rcBK'', proved out to be tremendously successful in applications to the phenomenology—for  both electron-proton DIS at HERA~\cite{Albacete:2009fh,Albacete:2010sy} and proton-nucleus collisions at RHIC and the LHC~\cite{Albacete:2010pg,Albacete:2012xq,JalilianMarian:2011dt,Rezaeian:2012ye,Lappi:2012nh,Lappi:2013zma,Albacete:2018ruq}.

\subsection{Time-ordering and the collinearly-improved BK equation}
\label{sec:collBK}

The running coupling corrections discussed in the previous section are not the only higher-order contributions to be essential for a robust description of the small-$x$ evolution. The full next-to-leading order (NLO) corrections to the BK equation were computed by Balitsky and Chirilli~\cite{Balitsky:2008zza,Balitsky:2013fea}. While representing a seminal theoretical milestone, the initial implementation of this full NLO kernel revealed severe instabilities: the emergence of large, negative NLO corrections rendered the numerical evolution unstable and unphysical in practice~\cite{Lappi:2015fma}. 

This complication was not entirely unexpected, as a closely related instability had previously been identified in the linear BFKL equation at NLO. In that linear framework, systematic analyses successfully traced the mathematical origin of the problem and delivered a self-consistent prescription based on collinear resummations~\cite{Kwiecinski:1997ee,Salam:1998tj,Ciafaloni:1998iv,Ciafaloni:1999yw,Ciafaloni:2003rd,Vera:2005jt}. In both linear and non-linear regimes, the large negative NLO corrections stem from regions of phase space where the scattering is weak and the system remains dilute. However, the momentum-space techniques developed to stabilise the linear BFKL equation cannot be directly transposed to the non-linear BK equation.\footnote{A notable exception is the application of stabilized NLO BFKL to compute the saturation exponent prior to full unitarization; see Ref.~\cite{Triantafyllopoulos:2002nz}.} Resolving this instability for NLO BK required innovative methods tailored to coordinate-space non-linear evolution~\cite{Beuf:2014uia,Iancu:2015vea,Ducloue:2019ezk,Boussarie:2025bpq}. A crucial step in this program was the proper identification of the "evolution time" governing the high-energy QCD evolution.

To contextualise this issue, recall that in asymmetric dilute-dense scatterings, target expectation values—such as the dipole $S$-matrix \eqref{Saverage}—are evaluated at a target rapidity scale $Y = \ln(1/x)$, where $x \equiv k^-_g/P^-$ is the minus longitudinal momentum fraction of the target gluons. Accordingly, the functional JIMWLK evolution integrates out quantum fluctuations in successive layers of $x$. However, in deriving the leading-order BK equation (Sect.~\ref{sec:BK}), we focused on soft gluon emissions within the wave function of the dilute projectile dipole, that are ordered according to the plus fraction $z \equiv k_g^+/q^+$. Iterating the LO BK equation thus spans an internal projectile rapidity space $\eta \equiv \ln(1/z)$, which differs structurally from the target rapidity $Y$. 

To better appreciate this rapidity mismatch, consider Deep Inelastic Scattering (DIS) at high virtuality $Q^2 \gg Q_0^2$, where $Q_0 \sim \Lambda_{\rm QCD}$ is a baseline, non-perturbative hadronic scale. The $q\bar q$ fluctuation of the virtual photon probes the target gluon distribution at a Bjorken scale $x \simeq \xbj = Q^2/(2q^+P^-)$, setting the target evolution time to $Y = \ln(1/\xbj)$. Evolving the projectile dipole to NLO accuracy opens a larger kinematic phase space:
\begin{equation}\label{etadef}
\eta \equiv \ln\left(\frac{q^+}{q_0^+}\right) = \ln\left(\frac{2q^+P^-}{Q_0^2}\right) = \ln\left(\frac{1}{\xbj}\right) + \ln\left(\frac{Q^2}{Q_0^2}\right) = Y + \rho\,,
\end{equation}
where $q_0^+ \equiv Q_0^2/2P^-$ acts as the physical lower bound on the plus momentum of the emissions by the projectile. (Gluons with  $k_g^+\ll q_0^+$ would be absorbed by the target and cannot contribute to the evolution.) When $Q^2 \gg Q_0^2$, there is a substantial kinematic mismatch $\eta - Y = \rho \equiv \ln(Q^2/Q_0^2)$, raising the fundamental question of whether $Y$ or $\eta$ constitutes the correct evolution time.

The discussion of the CGC effective theory in Sect.~\ref{sec:CGCgen} strongly indicates that the target rapidity $Y$ is the physically meaningful variable.This is corroborated by the following argument on {\it light-cone time ordering}: for a soft gluon fluctuation in the projectile wave function to participate in the scattering, its lifetime $\Delta x^+ \sim 2k_g^+/k_{g\perp}^2$ must be shorter than the lifetime of the parent $q\bar q$ pair ($\Delta x^+_{q\bar q} \sim 2q^+/Q^2$) but longer than the longitudinal shockwave thickness ($\sim 1/P^- = 2q_0^+/Q_0^2$). These boundary conditions dictate the logarithmic phase space:
\begin{equation}\label{gluonlife}
\frac{2q^+}{Q^2} \gtrsim \frac{2k_g^+}{k_{g\perp}^2} \gtrsim \frac{2q_0^+}{Q_0^2} \implies \int_{q_0^+ (k_{g\perp}^2/Q_0^2)}^{q^+ (k_{g\perp}^2/Q^2)} \frac{\rmd k_g^+}{k_g^+} = \ln\left(\frac{q^+}{q_0^+}\right) - \ln\left(\frac{Q^2}{Q_0^2}\right) = \eta - \rho = Y\,.
\end{equation}
Thus, the limits on the logarithmic phase-space for $k_g^+$ are dynamically constrained by the transverse momentum scales of the process, ensuring that the true rapidity phase space accessible to the projectile emission matches exactly the target rapidity $Y$. This causal constraint is validated by explicit higher-order calculations of dipole impact factors~\cite{Taels:2022tza,Caucal:2022ulg,Caucal:2024bae}, which confirm that the soft gluon phase space is limited to $k_g^+ \ll q^+(k_{g\perp}^2/Q^2)$.

Importantly, an identical time-ordering condition must be satisfied by each successive emission within the BFKL cascades generated by the projectile: $\Delta x^+_{q\bar q} \gg \Delta x^+_1 \gg \Delta x^+_2 \gg \dots \gg 1/P^-$, with $\Delta x^+_i = 2k_i^+/k_{i\perp}^2$. In the standard leading-logarithmic approximation (LLA) underlying the BK equation, this strict causal constraint is replaced by a simplified ordering in longitudinal momentum: $q^+ \gg k_1^+ \gg k_2^+ \gg \dots \gg q_0^+$. This LLA substitution implicitly assumes that all transverse momentum scales are comparable ($k_{i\perp} \sim k_{i-1,\perp}$), which roughly aligns time-ordering with $k^+$-ordering. 

However, this assumption breaks down in highly asymmetric configurations where $Q^2 \gg Q_0^2$. Then typical evolution paths involve transverse momenta that drop sharply from one gluon generation to the next one ($k_{i\perp}^2 \ll k_{i-1,\perp}^2$). Under these kinematics, the strict time-ordering constraint is far more restrictive than simple $k^+$-ordering, meaning the LLA framework severely overestimates the available phase space at leading order. This spurious phase space is precisely the source of the NLO instability: the unphysical contributions artificially included at LO are compensated at NLO (and beyond) by large, negative corrections.

To illustrate this mechanism and connect with the full NLO kernel derived in~\cite{Balitsky:2008zza,Balitsky:2013fea}, we consider two successive gluon emissions strongly ordered in both longitudinal momenta ($q^+\gg k_1^+ \gg k_2^+\gg q_0^+$) and transverse momenta ($Q^2\gg k_{1\perp}^2\gg k_{2\perp}^2\gg Q_0^2$). In the coordinate dipole picture, this configuration corresponds to two successive dipole splittings where the transverse sizes of the daughter dipoles increase sharply at each step. For convenience, we formulate this dynamics in terms of transverse momenta, leveraging the reciprocal relation $k_\perp^2\sim 1/r^2$.

If light-cone time-ordering is neglected, these two emissions map onto two standard iterations of the LO BK equation evaluated in the double-logarithmic approximation, cf. Eq.~\eqref{fgDLA}. In contrast, the exact NLO framework automatically enforces the physical time-ordering condition, $\Delta x^+_1\gg \Delta x^+_2$. To isolate the net effect introduced by this causal constraint—which is absent at LO and must be interpreted as a genuine NLO correction—we integrate out the kinematics of the intermediate gluon 1 for a fixed configuration of the measured gluon 2, subject to a step function enforcing time-ordering:
\begin{equation}\label{gluon1}
\abar\int_{k_{2\perp}^2}^{Q^2}\frac{\rmd k_{1\perp}^2}{k_{1\perp}^2} \int_{k_2^+}^{q^+} \frac{\rmd k_1^+}{k_1^+}\,\Theta\left(\frac{k_1^+}{k_{1\perp}^2}-\frac{k_2^+}{k_{2\perp}^2}\right) = \abar\ln\left(\frac{q^+}{k_2^+}\right)\ln\left(\frac{Q^2}{k_{2\perp}^2}\right) -\frac{\abar}{2} \ln^2\left(\frac{Q^2}{k_{2\perp}^2}\right)\,.
\end{equation}
The first term on the r.h.s., proportional to the rapidity logarithm $\ln(q^+/k_2^+)$, recovers the standard unconstrained LLA result. The second, negative term represents the correction introduced by time-ordering. Within standard $k^+$-ordered perturbation theory, the first term is absorbed into one step of the LO evolution, while the second term behaves as an $\mathcal{O}(\alpha_s)$ correction to the dipole kernel for the subsequent emission. This radiative correction is large since amplified by a \textit{double collinear logarithm}.

This simple kinematic argument correctly reproduces the dominant negative NLO contribution identified in~\cite{Balitsky:2008zza,Balitsky:2013fea}. Restricting the full expression to double-logarithmic kinematics yields:
\begin{equation}\label{colleq}
\frac{\partial T(r, \eta)}{\partial \eta} = \abar \int_{r^2}^{1/Q_0^2} \rmd z^2\, \frac{r^2}{z^4} \left(1 -\frac{\abar}{2} \ln^2\left(\frac{z^2}{r^2}\right) \right) T(z,\eta).
\end{equation}
This is valid when the measured daughter dipole is relatively large ($z^2\gg r^2$) but still within the weak scattering domain ($z^2 Q_0^2\ll 1$, where $T(z,\eta)\ll 1$). The double logarithm is generated by integrating out the phase space of an unobserved intermediate dipole. Via the uncertainty-principle identifications $r^2\sim 1/Q^2$ and $z^2\sim 1/k_{2\perp}^2$, this term matches the large negative contribution isolated in Eq.~\eqref{gluonlife}. In turn, this correction is the origin of the numerical instability that plagues the strict NLO BK equation~\cite{Lappi:2015fma,Iancu:2015vea,Lappi:2016fmu}.

Because the LO BK equation generates an arbitrary number of successive gluon emissions that violate time-ordering, higher-order corrections will develop an alternating series in powers of $(\abar \rho^2)^n$. This whole series can be resummed into a modified version of the evolution equation that explicitly incorporates time-ordering from the outset. Several such "collinearly-improved" versions of the BK equation have been established in the literature, based on distinct methods to enforce time-ordering~\cite{Beuf:2014uia,Iancu:2015vea,Ducloue:2019ezk}. While technically equivalent regarding the resummation of the leading double collinear logarithms, they may differ in their treatment of the subleading terms. Crucially, all these frameworks resolve the instability problem, as demonstrated by explicit numerical solutions~\cite{Iancu:2015joa,Albacete:2015xza,Lappi:2016fmu,Ducloue:2019ezk,Ducloue:2019jmy}.


The most intuitive formulation among these resummation strategies consists in employing the target rapidity $Y$ as the evolution time~\cite{Ducloue:2019ezk}. With this choice, the functional form of the LO BK equation remains unchanged: when evolving in $Y$ rather than $\eta$, successive emissions are implicitly ordered in $k^-\sim 1/\Delta x^+$, which automatically aligns them with physical time-ordering. This variable choice represents the standard baseline for high-energy phenomenology; for instance, in DIS, the BK solution is evaluated at $Y=\ln(1/\xbj)$. Implementing this change of variable---from $\eta$ to $Y \equiv \eta - \rho$---within 
the NLO BK equation constructed in the $\eta$-representation~\cite{Balitsky:2008zza,Balitsky:2013fea} amounts to a profound re-organisation of the perturbation theory: this transformation mixes rapidity and transverse coordinates, causing the problematic double collinear logarithms to cancel to all orders. 

Consequently, the remaining NLO corrections become genuinely small, $\mathcal{O}(\alpha_s)$ effects. More precisely, they encompass two sub-leading residual structures: single, DGLAP-like collinear logs, and double \textit{anti-collinear} logarithms.\footnote{Anti-collinear double logarithms are generated by gluon emissions with increasing transverse momenta (or decreasing dipole sizes) in the projectile wave function. These configurations are highly atypical but can accumulate at very large values of $Y$, though they remain sub-dominant in practical kinematic ranges.} These components are far less harmful and can be systematically resummed into a stabilised, collinearly-improved version of the BK equation in the $Y$-representation~\cite{Ducloue:2019ezk,Ducloue:2019jmy}. Numerical solutions of this equation incorporating a running coupling yield a saturation exponent $\lambda_s(Y)$ (shown in the right panel of Fig.~\ref{BKfront-run}) that is quasi-independent of $Y$ and stabilises at a value of $\lambda_s \simeq 0.2$, in good agreement with experimental data.

All collinearly-improved versions of the BK equation have been tested against HERA data for small-$x$ electron-proton DIS. They provide very good descriptions of the total DIS cross-section for $x\le 0.01$ over a wide range of virtualities $Q^2$~\cite{Ducloue:2019jmy,Beuf:2020dxl,Casuga:2025etc,Casuga:2026xxt,Korcyl:2026nrz}. State-of-the-art fits~\cite{Beuf:2020dxl,Casuga:2025etc,Casuga:2026xxt} combine these improved evolution equations with the exact NLO photon impact factor, which accounts for the radiative emission of a gluon on top of the $q\bar q$ pair~\cite{Balitsky:2010ze,Balitsky:2012bs,Beuf:2011xd,Beuf:2016wdz,Beuf:2017bpd}. Furthermore, when integrated with recent NLO results incorporating massive quarks~\cite{Beuf:2021srj,Beuf:2021qqa,Beuf:2022ndu}, the collinearly-improved fits deliver an excellent description of the HERA inclusive cross-sections for heavy flavours, including charm ($c$) and bottom ($b$) quark production~\cite{Hanninen:2022gje}.

\subsection{JIMWLK evolution in the Gaussian approximation}
\label{sec:Gauss}


The methods discussed in the previous sections allow for an efficient calculation of the high-energy evolution of the dipole scattering amplitude in the multi-colour limit $N_c\gg 1$. But what about the physical case $N_c=3$?
And what about more complicated $n$-point functions like the quadrupole \eqref{quadrupole}? At large $N_c$, the quadrupole too obeys a closed, non-linear, equation, whose r.h.s. involves just the dipole and the quadrupole itself~\cite{Dominguez:2011gc,Iancu:2011ns}. Yet, given that the quadrupole depends upon 4 transverse coordinates, this equation is highly non-local and direct numerical solutions seem hopeless. 
In what follows,  we shall describe two powerful methods that can deal with such complicated problems. 

The first one, which is known as {\it the Gaussian approximation} (GA)~\cite{Iancu:2002aq,Blaizot:2004wv,Kovchegov:2008mk,Marquet:2010cf,Iancu:2011ns,Iancu:2011nj,Dominguez:2011wm} and will be discussed in this section, allows one to relate generic $n$-point functions of the Wilson lines to the dipole $S$-matrix, itself extracted from solutions to the BK equation. 

Specifically, the GA assumes that, even after including the effects of the JIMWLK evolution, the CGC weight function can still be approximated by a Gaussian in $\rho_a$, like in the MV model, but with a more general expression for the charge-charge correlator, which is non-local in the transverse plane and depends upon $Y$. That is, Eq.~\eqref{MV_corr} is generalised by replacing $\delta^{(2)}(\bx-\by)\lambda(x^+)\to \lambda_Y(x^+, \bx, \by)$ in the r.h.s. The function $\lambda_Y(x^+, \bx, \by)$ is not {\it a priori} known, but it can be related to the dipole $S$-matrix $S_Y(\bx,\by)$ via a slight generalisation of Eq.~\eqref{SMV}, which reads
\begin{align}\label{SdipGauss}
S_Y(\bx,\by)=\rme^{-C_F\Gamma_Y(\bx,\by)}\,,
  \end{align}
with the function $\Gamma_Y(\bx,\by)$ defined by the two following equations (recall Eqs.~\eqref{alpha}--\eqref{Coulomb}):
\begin{align}
 \Gamma_Y(\bx,\by)\,\equiv&\ \frac{g^2}{2}\int\rmd x^+\left[\gamma_Y(x^+, \bx,\bx)+\gamma_Y(x^+, \by,\by)-2\gamma_Y(x^+, \bx,\by)\right],\nn*[0.2cm]
 \gamma_Y(x^+, \bx,\by)&\,\equiv \int \rmd^2\bz  \rmd^2\bu\,G (\bx-\bz)\,\lambda_Y(x^+, \bz, \bu)
 \,G (\bu-\by)\,.
 \end{align}
 Importantly, in this GA, all the expectation values  of gauge-invariant products of Wilson lines depend upon the charge-charge correlator exclusively via the function $ \Gamma_Y(\bx,\by)$. (See Refs.~\cite{Blaizot:2004wv,Kovchegov:2008mk,Marquet:2010cf,Iancu:2011nj,Dominguez:2011wm}
for explicit calculations). 
This suggests the following strategy: \texttt{(i)} compute the dipole $S$-matrix $S_Y(\bx, \by)$ by numerically solving the BK equation (possibly, with running coupling and collinear improvement), then  \texttt{(ii)} use Eq.~\eqref{SdipGauss} to deduce a corresponding result for $ \Gamma_Y(\bx,\by)$ and, finally,  \texttt{(iii)} use this result to compute the $n$-point function of interest at rapidity $Y$. For instance, the Gaussian approximation to the quadrupole $S$-matrix \eqref{quadrupole} can be found in~\cite{Iancu:2011nj,Dominguez:2011wm}. Based on such results, Refs.~\cite{Fujii:2006ab,Dusling:2009ni,Lappi:2012nh,Albacete:2018ruq,Caucal:2025zkl} have computed di-hadron  production  in proton-nucleus collisions at forward rapidities in the proton fragmentation region.

As a simple application, consider the gluon occupation number $n(Y,k_\perp)$ defined in Eq.~\eqref{nxk}. In general, this quantity is related to a complicated 2-point function built with four Wilson lines and their derivatives, as shown in Eq.~\eqref{xfg}. The corresponding evolution equation has been constructed in~\cite{Dominguez:2011gc}, but it is too complicated to be solved in practice. Yet, in the GA, $n(Y, k_\perp)$ can be related to the (adjoint) dipole scattering amplitude, as shown in  Eq.~\eqref{xfgMV} for the example of the MV model. To study its high-energy evolution in this approximation, one can still use Eq.~\eqref{xfgMV}, but replace the MV model estimate  $\tilde S(r)$ for the dipole $S$-matrix by the respective solution to the BK equation. Strictly speaking, the BK equation refers to a colour dipole in the fundamental representation,  but the $S$-matrices in the two representations are simply related to each other: repeatedly using the Fierz identities \eqref{Fierz}, one can write $\tilde V_{ab}=2{\rm tr}\big(V^\dagger t^a Vt^b\big)$ and hence
\begin{align}\label{SMVadj2}
\tilde S_Y(\bx,\by)\equiv 
\frac{1}{N_c^2-1}\left\langle {\rm Tr} \big[\tilde V(\bx) \tilde V^\dagger(\by)\big] \right\rangle_Y
=\frac{1}{N_c^2-1}\left[N_c^2   S_Y(\bx,\by)  S_Y(\by,\bx)-1\right].
\end{align}
 Using Eq.~\eqref{xfgMV} with $\tilde S(r, Y)\simeq S^2(r, Y)$ (at large $N_c$)  alongside the BK solution for $S(r,Y)$ (with running coupling and collinear improvement), one finds the saturation fronts visible in the right panel of Fig.~\ref{fig:WW}. These are similar to the corresponding fronts for the dipole amplitude (see Fig.~\ref{BKfront-run}), except for the fact that the unitarity limit $T(\rho, Y)=1$ for $\rho<\rho_s(Y)$ is now replaced by the saturation condition \eqref{noccup-low}  at $k_\perp < \tilde Q_s(Y)$.
 
While the GA  to the high energy evolution is very useful in practice, one may wonder about its justification and limitations. Direct calculations of $n$-point functions with $n=4$ and 6 from numerical solutions to the JIMWLK equation (see the next section) have been possible for special configurations, which depend upon a single transverse size~\cite{Dumitru:2011vk}. They show a good agreement with the respective predictions of the Gaussian approximations, notably for the quadrupole ($n=4$). Analytic arguments based on the JIMWLK equation have been presented in~\cite{Iancu:2011nj}. They support a GA with the following structure for the charge-charge correlator:
\begin{equation}
\lambda_Y(x^+, \bx, \by)=\Theta(x_{\rm M}^+(Y)-|x^+|)\lambda(x^+, \bx, \by)\quad\mbox{with}\quad
x_{\rm M}^+(Y) = x_0^+\rme^{Y-Y_0}.\end{equation}
This shows that the high-energy evolution proceeds via the symmetric expansion of the colour charge  distribution towards both larger and smaller values of $x^+$ (the original values $x_0^+$ and $Y_0$ refer to the MV model). This is in agreement with the fact that gluons with smaller values of $x$ are more delocalised along the $x^+$ direction (the longitudinal direction for the left-moving target). This picture is also useful for understanding the Langevin representation of the JIMWLK evolution to be discussed in the next section.

Despite its success in the aforementioned examples, the GA has been shown not to be accurate for correlators which start with an exchange of four gluons \cite{Lappi:2026kxi} (as was already conjectured in \cite{Iancu:2011nj}) and for certain transverse-momentum dependent parton distributions (see Sect.~\ref{sec:dijet})~\cite{Cougoulic:2026sww}. Interestingly, in all these cases the JIMWLK result is systematically larger than the GA one.

\subsection{The Langevin formulation of the JIMWLK evolution} 
\label{sec:Langevin}

To elucidate the connection between the JIMWLK Hamiltonian and the functional Fokker-Planck equation, we first recall the emergence of the latter within a more familiar classical context: the Brownian motion of a microscopic particle suspended in a viscous fluid. Under continuous random collisions with the surrounding molecules, the  trajectory $\bx(t) = (x^i(t))$ (with $i=1,2,3$) of the  particle satisfies the stochastic Langevin equation:
\begin{equation}\label{eq:Langevin-formal} 
\frac{\rmd x^i}{\rmd t} = \nu^i(t)\,, \qquad \langle \nu^i(t)\rangle = 0\,, \qquad \langle \nu^i(t)\nu^j(t') \rangle = D\delta^{ij}\delta(t-t')\,.
\end{equation}
The "noise" $\nu^i(t)$ represents the random velocity fluctuations induced by the fluid molecules. It is modeled as a Gaussian white noise with a vanishing expectation value and a local two-point correlation function proportional to the macroscopically defined diffusion coefficient $D$, assuming a homogeneous and isotropic medium.

The statistical expectation value $\langle \mcal{O}[\bx]\rangle (t)$ of any observable $\mcal{O}[\bx]$ at time $t$ can be evaluated by integrating Eq.~\eqref{eq:Langevin-formal} to obtain $\bx(t) \equiv \bx(t)[\nu]$ as a explicit functional of the historical noise profile, followed by a Gaussian average over $\nu(t')$. For example, the mean squared displacement evolves linearly:
\begin{align}
x^i(t) = x^i_0 + \int_0^t\rmd t' \nu^i(t') \implies \langle \bx^2\rangle(t) = \bx^2_0 + D\int_0^t \rmd t_1 \int_0^t \rmd t_2 \langle\nu^i(t_1)\nu^i(t_2)\rangle = \bx^2_0 + 3Dt\,.
\end{align}

Alternatively, this stochastic dynamics can be modeled via the probability density $P(\bx,t)$ of finding the particle at coordinate $\bx$ at time $t$, given the boundary condition $P(\bx,0) = \delta^{(3)}(\bx-\bx_0)$. This distribution satisfies the deterministic Fokker-Planck equation:
\begin{equation}\label{eq:FP} 
\frac{\partial P (\bx, t)}{\partial t} = \frac{D}{2}\,\frac{\partial^2 P(\bx, t)}{\partial x^i\partial x^i} \implies P(\bx,t) = \frac{1}{(2\pi Dt)^{3/2}}\,\exp\left\{-\frac{(\bx-\bx_0)^2}{2Dt}\right\}\,.
\end{equation}
Integrating the observable directly over this probability density naturally recovers the Langevin result, $\langle \bx^2\rangle(t) \equiv \int\rmd^3 x \,\bx^2 P(\bx,t) = \bx^2_0 + 3Dt$.

This mathematical equivalence between the Langevin and Fokker-Planck descriptions holds for complex systems involving inhomogeneous or anisotropic media ($D\delta^{ij}\to D^{ij}(\bx)$), where numerical implementations of the Langevin equation usually provide the most practical solution method. 
This is also the case for the JIMWLK equation in Eqs.~\eqref{jimwlk}--\eqref{Hjim}, to which we now return. Comparing  Eqs.~\eqref{Hjim} and \eqref{eq:FP}, it becomes clear that the JIMWLK Hamiltonian is a functional generalisation of the Fokker-Plank operator which acts  in the space of Wilson lines: the rapidity $Y$ plays the role of time, the particle position $\bx$ is replaced by the  group-valued Wilson line field $V(\bx)$ (where the spatial coordinate $\bx\equiv \bx_\perp$ refers to the two dimensional transverse plane),
 and the functional derivatives $\delta/\delta \rho_a(\bx_\perp)$ serve as the functional analogues of the ordinary partial derivatives $\partial/\partial x^i$.\footnote{More precisely, these functional derivatives act as Lie derivatives generating colour rotations on the Wilson lines~\cite{Weigert:2000gi,Blaizot:2002xy}.}

\begin{figure}
	\begin{center}
		\includegraphics[width=0.45\textwidth]{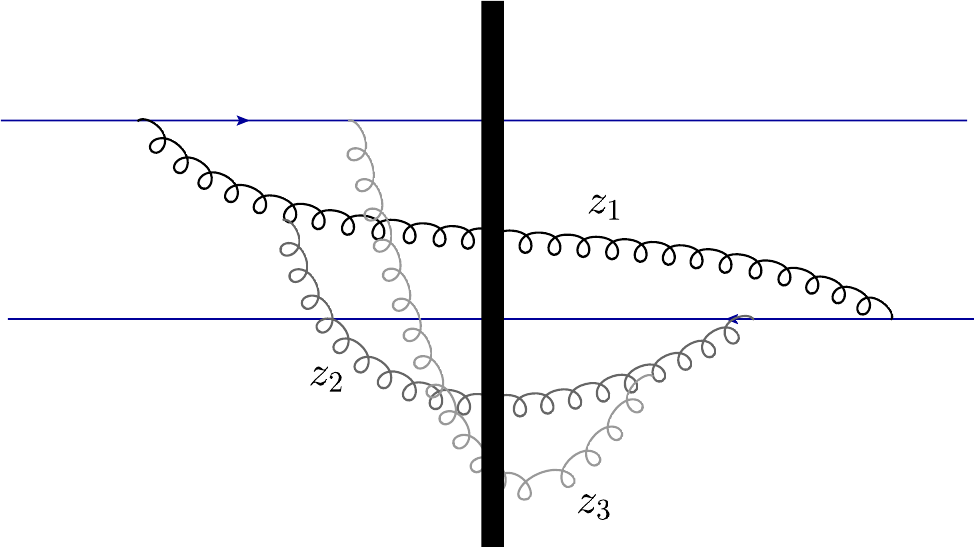}\qquad
		\includegraphics[width=0.47\textwidth]{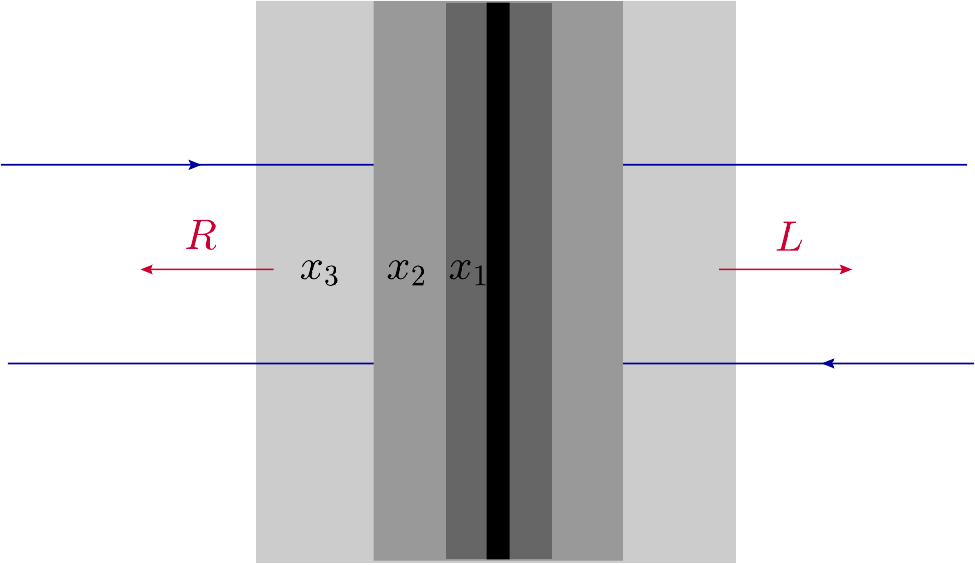}
			\end{center}
	\caption{\small Two alternative pictures for the high-energy evolution of the elastic amplitude for dipole-hadron scattering. Left: projectile evolution. The dipole sources successive gluon emissions with lower and lower ``plus'' longitudinal momentum fractions ($1\gg z_1\gg z_2 \gg z_3\dots$) and shorter and shorter
	lifetimes ($\Delta x^+_n\propto z_n$). Right: target evolution. The large-$x$ colour sources emit gluons with smaller and smaller ``minus'' fractions  ($1\gg x_1\gg x_2 \gg x_3\dots$), thus generating additional layers in the target field $A^-$, which are located at larger and larger values of $x^+$ ($\Delta x^+_n \sim 1/x^-_n$).
}
\label{fig:JIMWLK}
\end{figure}

This structural mapping enables the reformulation of the JIMWLK framework as a functional Langevin equation describing a random walk on the $SU(N_c)$ manifold of Wilson lines~\cite{Weigert:2000gi,Blaizot:2002xy}. The formalism dictates the stochastic update $V_{Y}(\bx) \to V_{Y+\rmd Y}(\bx)$ generated by integrating out a thin quantum layer of small-$x$ gluons within the rapidity step $\alpha_s \rmd Y \ll 1$. Because this interval is infinitesimal, the amplitude of the quantum fields satisfies $g a^\mu_a \ll 1$, allowing us to neglect gluon self-interactions and focus exclusively on their eikonal propagation through the strong target background field $A^\mu_a \sim 1/g$. Consequently, the quantum integration depends entirely on the background-field gluon propagator, $G^{\mu\nu}_{ab}(x,y)[A] \equiv \langle {\rm T}[a^\mu_a(x) a^\nu_b(y)]\rangle_A$. The functional Langevin equation mimics these quantum background fluctuations by introducing a stochastic noise term tailored to reproduce this exact propagator structure. This is achieved by the following stochastic equation, which uses a discretised version of the rapidity variable ($Y=n\epsilon$) \cite{Iancu:2011nj,Lappi:2012vw}
 \begin{align}
 \label{Langevin}
 V_{n}(\bx) = \exp[\rmi \epsilon g \alpha^{L}_{n,\bx}]\,
 V_{n-1}(\bx) \exp[-\rmi \epsilon g \alpha^{R}_{n,\bx}],
 \end{align}
where the left ($L$) and right ($R$), infinitesimal, colour fields read
 \begin{align}
 \label{alphan}
 \alpha^{L}_{n,\bx} = 
 \frac{1}{\sqrt{4\pi^3}}\int \rmd{\bz}\, \mcal{K}^i_{\bx\bz} \,\nu^{ia}_{n,\bz}\, t^a,
 \qquad \alpha^{R}_{n,\bx} = 
 \frac{1}{\sqrt{4\pi^3}} \int \rmd{\bz} \,\mcal{K}^i_{\bx\bz} \,\nu^{ia}_{n,\bz} \tilde V^{ab}_{n-1,\bz}\, t^b.
 \end{align}
The kernel $\mcal{K}^i_{\bx\bz} $ and the  Gaussian white noise
$\nu_n$ appearing in these equations are defined as
 \begin{align}
 \label{noise}
 {\mathcal K}_{\bx \bz}^i\equiv\frac{(\bx-\bz)^i}{(\bx-\bz)^2}\,,\qquad
 \big\langle \nu^{ia}_{m,\bx}\, \nu^{jb}_{n,\by} \big\rangle
 = \frac{1}{\epsilon}\,\delta^{ij} \delta^{ab} \delta_{mn} \delta^{(2)}(\bx-\by).
 \end{align} 
These equations can be understood as follows: the quantum gluons that are integrated out during the $n$-th step carry low longitudinal momenta $k^- = xP^-$, so they are delocalised in $x^+$ over distances much larger than the support of the shockwave field $A^-$ generated at the previous steps (see Fig.~\ref{fig:JIMWLK}). Accordingly, the quantum field additions $a^-$ are pushed to the outer boundaries of the shockwave, either before (the $R$ field in \eqref{Langevin}), or after (the $L$ field in \eqref{Langevin}) the pre-existing field $A^-$. Being emitted prior to the background field, the ``R'' gluon can interact with it, which dictates the insertion of the adjoint Wilson line $ \tilde V^{ab}_{n-1,\bz}$ in $\alpha^{R}_{n,\bx} $ in Eq.~\eqref{alphan}. 
 The normalization of the white noise in Eq.~\eqref{noise} is chosen such that the noise-averaged left-right correlator $\langle \alpha^{L}_{n,\bx} \alpha^{R}_{n,\by} \rangle$ exactly reproduces the $G^{--}_{ab}(x,y)[A]$ component of the background-field gluon propagator  (integrated over negative $y^+$ for the ``right'' gluon emissions and over positive $x^+$ for the ``left'' emissions, respectively).

It is a straightforward exercise to verify that, by using Eq.~\eqref{Langevin}
together with the corresponding equation for $V_{n}^\dagger(\by)$ and keeping only the linear terms in the expansion of their product ${\rm tr}\big(V_{n}(\bx) V_{n}^\dagger(\by)\big)$ in powers of $\epsilon$, one recovers the BK equation Eq.~\eqref{BKgen}, as expected. More generally, all equations in the Balitsky hierarchy can
be generated in this way.

However, the main virtue of the Langevin formulation in Eq.~\eqref{Langevin} is to allow for explicit numerical calculations for  $N_c=3$, via classical lattice simulations in 2+1 dimensions (the transverse space and the rapidity)~\cite{Rummukainen:2003ns,Lappi:2011ju}. Starting at $Y=0$ with a collection of Wilson lines $\{V_0(\bx)|\bx\in  \mathbb{R}^2\}$  which are distributed, say,
according to the MV model and iterating the Langevin equation 
\eqref{Langevin}--\eqref{noise}, one obtains a
statistical ensemble of Wilson lines $\{V_Y(\bx)|\bx\in \mathbb{R}^2\}$ at the rapidity $Y$ of interest.
A physical observable is then evaluated as the statistical average over this ensemble of the respective 
(gauge-invariant) product of Wilson lines. This method has been successfully implemented in practice~\cite{Rummukainen:2003ns,Lappi:2011ju,Dumitru:2011vk,Lappi:2012vw,Schenke:2016ksl,
Marquet:2016cgx,Marquet:2017xwy}, but so far only for 2-point functions and special limits of some higher-point functions. It is {\it a priori} formulated for the leading-order evolution, but extensions to running coupling and to collinear improvement have been proposed as well~\cite{Lappi:2012vw,Hatta:2016ujq}. However, the NLO version of the  JIMWLK Hamiltonian~\cite{Kovner:2013ona,Kovner:2014lca,Lublinsky:2016meo} is considerably more complicated and so far it seems to be incompatible with a Langevin formulation.

\section{Selected applications to dilute-dense collisions}

In this section we shall use the CGC effective theory to study three dilute-dense collisions: the two processes that have already been introduced in Sect.~\ref{sec:DD} (single quark production in $pA$ collisions and, respectively, the total cross-section for DIS) and the more exclusive version of DIS in which the quark-antiquark produced by the photon fluctuation is measured in the final state. 

\subsection{Transverse momentum broadening in $pA$ collisions: the Cronin peak}
\label{sec:pT}

As a first application of Eq.~\eqref{SMV}, let us compute the transverse momentum spectrum of a quark produced in proton-nucleus ($pA$) collision at forward rapidity. The CGC formulation of this process has  already been presented in Sect.~\ref{sec:pA}, from which we quote the relevant formula, cf. Eq.~\eqref{ptbroad}:
\begin{equation} \label{ptbroadMV}
 \frac{\rmd N}{\rmd^2\bp} \equiv  \frac{1}{\sigma_{qA}}
 \frac{\rmd \sigma^{qA\to qX}}{\rmd y\,\rmd^2\bp_\perp}
 =\int
 \frac{\rmd^2\br}{(2\pi)^2}
  \,\rme^{-i \br\cdot\bp}\,S(x, r)\,,
\end{equation}
where $\sigma_{qA}=2\pi R_A^2$ is the total $pA$ cross-section and we assumed a homogeneous nucleus, for simplicity.

We focus on the MV model, where the dipole $S$-matrix $S(x,r)$ is independent of $x$ and given by Eq.~\eqref{SMV}.  For low momenta $p_\perp\lesssim Q_s(A)$, the integral in Eq.~\eqref{ptbroadMV} is controlled by $r\sim 1/Q_s(A)$ and then we can  replace  $r\to 1/Q_s(A)$ within the slowly varying logarithm. After also using Eq.~\eqref{QSMV}, one concludes that the dipole $S$-matrix can be approximated by the Gaussian\footnote{This approximation, that is often referred to as the {\it GBW model}~\cite{GolecBiernat:1998js,GolecBiernat:1999qd}, should only be used for relatively large dipole sizes $ r\gtrsim 1/Q_s(A)$, as shown by its above derivation from the MV model.} $S(r)\simeq\,\rme^{-{r^2Q_s^2(A)}/{4}}$. Then the integral can be easily evaluated and the resulting spectrum is a Gaussian too:
  \begin{equation} 
 \label{ptGauss}
 \frac{\rmd N}{\rmd^2\bp_\perp} \,\simeq\,\frac{1}{\pi Q_s^2(A)}\,\rme^{-{p_\perp^2}/{Q_s^2(A)}}\quad
 \mbox{for}\quad p_\perp\lesssim Q_s(A).
 \end{equation}
This Gaussian spectrum describes transverse momentum broadening via multiple scattering 
off the valence quarks. The average transverse momentum squared   $\langle p_\perp^2\rangle\simeq Q_s^2(A)$  grows linearly with the longitudinal extent $\sim A^{1/3}$ of the nuclear target, as for a diffusive process.

 For larger  transverse momenta $p_\perp\gg Q_s(A)$, the integral in Eq.~\eqref{ptbroadMV} is controlled by small dipole sizes $r\ll 1/Q_s$, for which the scattering is weak and one can use the 2-gluon exchange approximation  $S\simeq 1-T_{2g}(r)$. Then the Fourier transform in  \eqref{ptbroadMV} is best computed by using the integral representation of $T_{2g}(r)$, as shown in the second line of Eq.~\eqref{T2g}. This yields
 \begin{equation} 
 \label{highptMV}
 \frac{\rmd N}{\rmd^2\bp_\perp} \,\simeq\,\frac{\alpha_s C_F}{\pi}\frac{\mu^2_A}{p_\perp^4}\quad
 \mbox{for}\quad p_\perp\gg Q_s(A).\,.
 \end{equation}
 The power tail $1/{p_\perp^4}$ at large $p_\perp$  can be recognised as the product of the Coulomb propagators for two exchanged gluons (one in the direct amplitude, the other one in the complex conjugate amplitude). Physically, this is the effect of a single hard scattering off any of  the valence quarks.
  
\begin{figure}
	\begin{center}
		\includegraphics[width=0.45\textwidth]{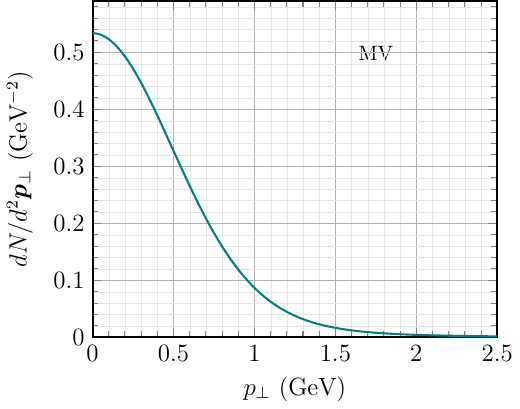}\qquad
		\includegraphics[width=0.45\textwidth]{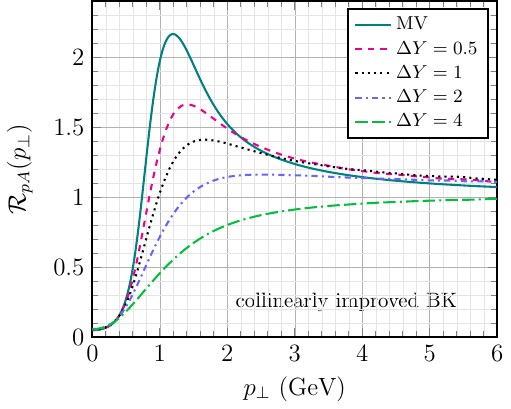}
			\end{center}
	\caption{\small  Left: the quark spectrum ${\rmd N}/{\rmd^2\bp_\perp}$ produced via quark-nucleus scattering as predicted by the MV model, Eq.~\eqref{ptbroadMV},  for a nuclear target with saturation momentum $Q_s(A)=1$~GeV. 	Right: the nuclear modification factor $\mcal{R}_{pA}(p_\perp)$ for quark production, Eq.~\eqref{RpA}. The curve labelled ``MV'' has been obtained by using Eq.~\eqref{ptbroadMV} for both the nucleus and the proton, with  $Q_s(A)=1$~GeV and $Q_A^2=4 Q_p^2$. The other curves have been obtained from the more general expression \eqref{ptbroad} together with numerical solutions of the BK equation (with a running coupling and collinear improvement~\cite{Ducloue:2019ezk}), with the initial conditions at $\Delta Y=0$ taken from the MV model (for both the proton and the nucleus).
}
\label{fig:pT}
\end{figure}

A numerical calculation of the Fourier transform in Eq.~\eqref{ptbroadMV} (with the choice $Q_s(A)=1$~GeV for the target saturation momentum) leads to the result shown in Fig.~\ref{fig:pT} (left panel). As anticipated by our analytic arguments, this exhibits Gaussian broadening at low momenta $p_\perp\lesssim Q_s(A)$ and a rapidly decaying power tail at large momenta $p_\perp\gg Q_s(A)$.

To better visualise the effects of gluon saturation in a large nucleus, it is instructive to consider the ratio between the spectrum \eqref{ptbroadMV} for quark production in $pA$ collisions and the corresponding result for proton-proton ($pp$) collisions. It is moreover useful to normalise this ratio in such a way that it would be equal to one in the absence of nuclear effects, i.e. in the situation where the nucleus was an incoherent superposition of $A$ nucleons. Within the present approximations, the high-$p_\perp$ tail \eqref{highptMV} is indeed generated via incoherent scattering and the $A$-dependence enters only via the cumulative factor $\mu^2_A\sim A^{1/3}$.  For a proton target, Eq.~\eqref{highptMV} would change only via the replacement $\mu^2_A\to \mu_p^2$ with $\mu_p^2=\mu^2_A/A^{1/3}$. (We implicitly assume here that a proton target can be treated as a small nucleus with $A=1$.)
This motivates the following definition for the {\it nuclear modification factor}:
\begin{align}\label{RpA}
\mcal{R}_{pA}(p_\perp)\equiv\frac{1}{A^{1/3}}\,\frac{\frac{\rmd N}{\rmd^2\bp_\perp}\big|_{pA\to qX}}
{\frac{\rmd N}{\rmd^2\bp_\perp}\big|_{pp\to qX}}\,,
\end{align}
which within the present approximations approaches to unity at large $p_\perp\gg Q_s(A)$. Using Eqs.~\eqref{ptGauss} and \eqref{highptMV}, one easily finds the following,  piecewise, approximation:
\begin{align}
	\label{RpAMV}
	\mcal{R}_{pA}(p_\perp)\ \simeq
	\begin{cases}    
    \displaystyle{\ \frac{1}{A^{1/3}}\,\frac{Q_{s}^2(p)}{Q_{s}^2(A)}\sim 
      \frac{1}{A^{2/3}}}
        &\quad \text{for} \quad
    p_\perp\!\lesssim Q_{s}(p),
    \\*[0.35cm]
    \displaystyle{\ \frac{1}{A^{1/3}}\,\frac{p_{\perp}^4}{Q_{s}^2(A)Q_{p}^2}}
    &\quad \text{for} \quad
    Q_{s}(p) \ll p_\perp\!\lesssim  Q_{s}(A),
    \\*[0.35cm]
    \displaystyle{\ 1} 
        &\quad \text{for} \quad
    p_\perp\!\gg  Q_{s}(A).
    \end{cases}
   \end{align}
  
 These analytic estimates can be compared with the corresponding numerical results, as shown in  the right panel of Fig.~\ref{fig:pT} (the curve denoted ``MV''). The numerical evaluation confirms the general pattern suggested by Eq.~\eqref{RpAMV}:  $\mcal{R}_{pA}(p_\perp)$ starts be being small at low $p_\perp\lesssim Q_s(p)$, then it increases very fast with $p_\perp$ in the intermediate range between the two saturation momenta, and eventually approaches to 1 at large $p_\perp\gg Q_{s}(A)$. Interestingly, it also shows that the ratio exhibits a maximum with a value $\mcal{R}_{\rm max} > 1$ at $p_\perp\sim Q_s(A)$, so it approaches unity {\it from the above}.  This maximum, known as the {\it Cronin peak}, could have been anticipated too from Eq.~\eqref{RpA}:
the second line there implies
\begin{align}
	\label{Rpmax}
	\mcal{R}_{pA}(p_\perp=Q_s(A))\, \simeq\,
	\frac{1}{A^{1/3}}\,\frac{Q_{s}^2(A)}{Q_{p}^2}\, \simeq\,\ln{Q_s^2(A)\over
 \Lambda^2}\,>\,1\,,
\end{align}
where we used $Q_{A}^2= A^{1/3} Q_{p}^2$ together with \eqref{QSMV}. This simple estimate confirms that  $\mcal{R}_{pA}(p_\perp)$ is parametrically larger than 1 at $p_\perp=Q_s(A)$, so it must have a maximum in that region.

As explained in~\cite{Kharzeev:2003wz,Iancu:2004bx,Blaizot:2004wu}, the emergence of the Cronin peak is a consequence of ``classical saturation'' (in the sense of $p_\perp$-broadening), as described by the MV model. One expects this peak to be washed out by the evolution of the gluon distribution in the target with decreasing $x$~\cite{Kharzeev:2003wz,Iancu:2004bx, Albacete:2003iq}. This is illustrated too in the right panel of Fig.~\ref{fig:pT},  where the curves corresponding to non-zero values of  $\Delta Y\equiv \ln(x_0/x)$ have been obtained by evaluating the dipole $S$-matrix in Eq.~\eqref{ptbroad} from numerical solutions of the BK equation with initial conditions at $x=x_0$ taken from the MV model. These curves demonstrate the disappearance of the Cronin peak with increasing $\Delta Y$ and, more generally, the global suppression of the  $\mcal{R}_{pA}$ ratio: for sufficiently large $\Delta Y$, the ratio approaches unity from the below when increasing $p_\perp$. This suppression is understood as follows: with increasing $Y$, the gluon density can rise only in the dilute region at sufficiently large transverse momenta, $k_\perp > Q_s$. For momenta around the {\it nuclear} saturation scale, $k_\perp \sim Q_s(A,Y)$, the {\it proton} target is dilute, so its gluon distribution can rise very fast with $Y$, according to the BFKL equation. The {\it nuclear} distribution on the other hand rises only weakly, so the ratio $\mcal{R}_{pA}$ decreases with $Y$, as visible in Fig.~\ref{fig:pT}.

\begin{figure*}[t]
\begin{center}
\centerline{
\includegraphics[width=0.9\textwidth]{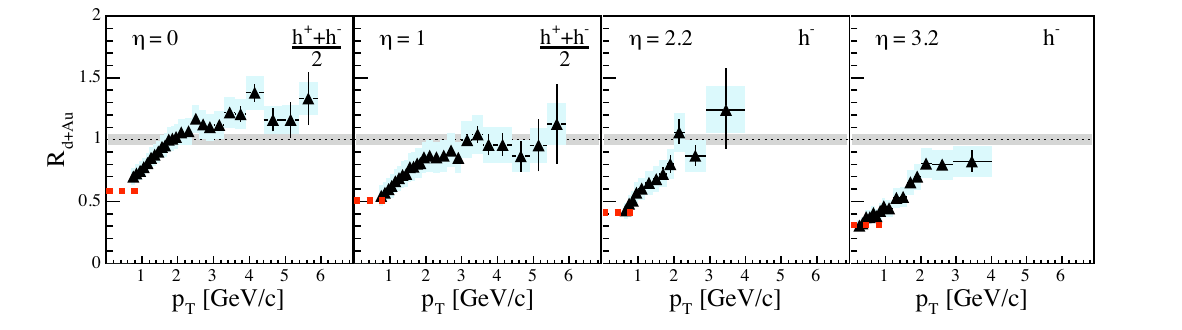}
} \caption{\sl 
The nuclear modification factor for charged hadron production in deuteron-gold (d+Au) collisions (compared to $pp$ collisions) at nucleon-nucleon COM energy $\sqrt{s_{NN}}=200$~GeV, as measured by the BRAHMS experiment at RHIC. The data are shown as a function of the hadron transverse momentum $p_T$ for various values of the pseudo-rapidity $\eta$, which increases from $\eta=0$ (central rapidities) to $\eta=3.2$ (forward rapidities). From~\cite{Arsene:2004ux}.
\label{fig:dAu}}
\end{center}
\end{figure*}

As discussed in Sect.~\ref{sec:pA}, this evolution can be probed experimentally by measuring particle production  at increasingly forward rapidities: one can decrease  $x$ by increasing  $y$, cf. Eq.~\eqref{xgpA}. Remarkably, the above scenario for the $\mcal{R}_{pA}$ ratio—the existence of a peak at low rapidity and the global suppression of the ratio with increasing $y$—has  indeed been observed in the deuteron-gold (d+Au) collisions at RHIC, for a COM energy $\sqrt{s_{NN}}=200$~GeV~\cite{Arsene:2004ux,Arsene:2004fa,Adcox:2004mh,Back:2004je,Adams:2005dq} (see Fig.~\ref{fig:dAu}). This is one of the most compelling experimental indications in favour of gluon saturation to date.

We conclude this section with a brief discussion of next-to-leading order (NLO) corrections and more complicated final states. Forward hadron production in $pA$ collisions is the first ``dilute-dense'' process for which the NLO corrections to the impact factor have been computed in the CGC effective theory~\cite{Chirilli:2011km,Chirilli:2012jd}. By ``impact factor'', one means the partonic cross-section with all the quantum evolutions excluded. In this case, the LO impact factor refers to the quark-nucleus scattering and is given by  Eq.~\eqref{ptbroad}. At NLO, the quark emits a (real or virtual) gluon, which can scatter too with the nucleus. This gluon is not measured in the final state, yet its emission modifies the cross-section for quark production.  By appropriately separating the (longitudinal and transverse) phase-space for the gluon emission, one can recognise the first step in the BK/JIMWLK evolution of the dipole $S$-matrix and also in the DGLAP evolution for both the quark PDF and the quark fragmentation function. These special contributions must be subtracted to deduce the NLO correction to the impact factor. This subtraction turns out to be tricky and leads to a paradox: the NLO cross-section becomes negative with increasing $p_\perp$ above $Q_s$~\cite{Stasto:2013cha}. Two conceptually different solutions to this paradox have been proposed in Refs.~\cite{Iancu:2016vyg} (see also \cite{Ducloue:2017mpb,Ducloue:2017dit}) and, respectively, \cite{Shi:2021hwx}. 

More complicated final states, describing the simultaneous production of several jets (or hadrons)
 in $pA$ collisions at forward rapidities, have been considered as well: two jets~\cite{JalilianMarian:2004da,Baier:2005dv,Marquet:2007vb,Blaizot:2004wv,Iancu:2013dta}, three jets~\cite{Iancu:2018hwa,Caucal:2026ymj}, a 
(real) photon plus a jet~\cite{Gelis:2002ki}, and a virtual photon decaying into a final lepton pair (the Drell-Yan process)~\cite{Gelis:2002fw,Stasto:2012ru}. Starting with a trijet final state and integrating out the kinematics of one of the jets, one has obtained the real NLO corrections to the cross-section for forward dijets~\cite{Iancu:2020mos,Caucal:2026ymj}. We finally mention  the impressive calculation of the NLO corrections to the cross-section for virtual photon plus jet production 
\cite{Taels:2023czt}.

\subsection{Deep inelastic scattering: total and diffractive cross-sections}
\label{sec:DIStot}

In this section, we employ the colour dipole picture alongside the CGC effective theory to provide semi-quantitative estimates for the total and diffractive cross-sections in photon-proton ($\gamma^*p$) and photon-nucleus ($\gamma^*A$) deep inelastic scattering (DIS) at small $x$. The primary ingredients for the discussion that follows are the generic expression \eqref{DISdipole} for the DIS cross-section, combined with the CGC estimate for the dipole scattering amplitude $T(x, r)$. We shall approximate the latter via the following piecewise parametrization (cf. the second and third lines in \eqref{TYrho}):
\begin{equation}\label{Txr}
  T(x, r)\simeq \begin{cases} 
            \displaystyle{\big[r^2Q_s^2(x)\big]^{\gamma_s}} & \text{for } r\ll 1/Q_s(x), \\[0.2cm]
            \displaystyle{1} & \text{for } r\gtrsim 1/Q_s(x),
    \end{cases}
\end{equation}
where $\gamma_s\simeq 0.63$ is the characteristic anomalous dimension governing the dilute-to-dense transition. This piecewise approximation remains robust provided that the photon virtuality $Q^2$ is not excessively large: it can comfortably exceed the target saturation momentum $Q_s^2(x,A)$, but it must remain within the \textit{geometric scaling window} specified for a running coupling by \eqref{gswindowrun}.

We start by exhibiting the light-cone wavefunctions describing the quark-antiquark ($q\bar q$) fluctuation of the virtual photon. For simplicity, we consider massless quarks and keep the sum over quark flavors implicit (see e.g., \cite{Kovchegov:2012mbw} for technical details):
\begin{align}
\left |\Psi_T(z,\br; Q^2)\right |^2 & = \frac{ N_c \alpha_{\rm em} e_f^2}{2\pi^2} [z^2+(1-z)^2]\,\bar Q^2 K_1^2(\bar Q r)\,, \label{PsiT} \\[0.2cm]
\left |\Psi_L(z,\br; Q^2)\right |^2 & = \frac{N_c \alpha_{\rm em} e_f^2}{2\pi^2}\,4 z^2(1-z)^2\, Q^2 K_0^2(\bar Q r)\,, \label{PsiL}
\end{align}
for a photon with transverse ($T$) and longitudinal ($L$) polarization, respectively. The modified Bessel functions $K_1(\bar Q r)$ and $K_0(\bar Q r)$, with the effective scale $\bar Q^2\equiv z(1-z)Q^2$, express the limitation on the dipole size $r$ introduced by the photon virtuality. They decay exponentially for large values of their argument ($\bar Q r \gg 1$), demonstrating that $r$ is effectively restricted to small spatial sizes $r\lesssim 1/\bar Q$.

On the other hand, non-linear effects from gluon saturation are expected to become prominent only for relatively large dipoles satisfying $r\gtrsim 1/Q_s$. Hence, to capture clear saturation signatures in the total DIS cross-section, one requires the kinematic condition $z(1-z) Q^2 \lesssim Q^2_s$. For parametric estimates of the double integral in \eqref{DISdipole}, one can approximate the asymptotic behavior of the Bessel functions via sharp $\Theta$-functions:
\begin{equation}\label{Bessel}
K_1(\bar Q r)\,\sim\,\frac{\Theta(1-\bar Q r)}{ \bar Q r}\,,\qquad K_0(\bar Q r)\,\sim\,\Theta(1-\bar Q r)\,.
\end{equation}
Given the fundamental kinematic constraint $z(1-z)\le 1/4$, it is convenient to distinguish between two distinct types of $q\bar q$ spatial configurations: \texttt{(i)} \textit{symmetric configurations}, for which $r^2 Q^2 < 4$, and \texttt{(ii)} \textit{aligned-jet configurations}, for which $r^2 Q^2 \gg 4$. In the first scenario, $\Theta(1-\bar Q r)=1$, meaning the integral over $z$ is unconstrained and dominated by symmetric valence fractions $z\sim 1-z\sim 1/2$; the photon longitudinal momentum $q^+$ is democratically shared between the quark and antiquark. In the second scenario, the phase space is restricted to $z(1-z) < 1/(r^2Q^2) \ll 1$, meaning that one of the two fermions carries the quasi-totality of $q^+$ and emerges along the original photon direction in the final state—this is the  ``aligned jet''.

To parametric accuracy, one can use the $\Theta$-functions in Eq.~\eqref{Bessel} to integrate over $z$ and thus explicitly separate the symmetric from the aligned-jet contributions (below, $ \sigma_0\equiv  N_c \alpha_{\rm em} e_f^2 \pi R_A^2$)
\begin{align}
\sigma_T(x,Q^2) &\sim \sigma_0 \left[ \int^{4/Q^2}_0 \frac{\rmd r^2}{r^2}\,T(x,r) + \frac{1}{Q^2}\int_{4/Q^2}^\infty \frac{\rmd r^2}{r^4}\,T(x,r) \right], \label{intzT} \\[0.2cm]
\sigma_L(x,Q^2) &\sim \sigma_0 \left[ Q^2 \int^{4/Q^2}_0 {\rmd r^2}\,T(x,r) + \frac{1}{Q^4}\int_{4/Q^2}^\infty \frac{\rmd r^2}{r^6}\,T(x,r) \right]. \label{intzL}
\end{align}

Consider first the case of a low-virtuality photon satisfying $Q^2\lesssim Q_s^2$. Under these conditions, the requirement $z(1-z) Q^2 \lesssim Q^2_s$ is satisfied for all values of $z$, allowing the double integral in \eqref{DISdipole} to naturally sample large $q\bar q$ configurations that are highly sensitive to saturation physics. Using Eqs.~\eqref{intzT}--\eqref{intzL}, one can easily verify that the dominant contributions to both $\sigma_T$ and $\sigma_L$ stem from symmetric dipoles ($z\sim 1/2$) with large sizes $r\gtrsim 1/Q_s(x)$, which undergo strong scattering: $T(x, r)\simeq 1$. For a transverse photon, the integral over $r$ receives a logarithmic enhancement from the range $1/Q_s\lesssim r < 1/Q$. For a longitudinal photon, the integration over $r$ is rather pinned at $r\sim 1/Q$.  One thus finds the following parametric estimates:
\begin{align}
\sigma_T (x,Q^2)\sim \sigma_0 \ln\frac{Q_s^2(x)}{Q^2}\,, \qquad \sigma_L(x,Q^2) \sim \sigma_0 \qquad \text{for } Q^2\lesssim Q_s^2. \label{DISlowQ}
\end{align}

For a  photon with  relatively large virtuality $Q^2\gg Q_s^2(x)$, the dominant contributions to both $\sigma_T$ and $\sigma_L$ arise from weak scattering, i.e. from small dipoles with $r\ll 1/Q_s(x)$ and $T(x,r)\ll 1$. So long as the anomalous dimension $\gamma_s$ is significantly smaller than unity—as is naturally the case at small $x$, where non-linear BK evolution stabilizes the exponent at $\gamma_s\simeq 0.63$ (cf. \eqref{gammasBFKL})—these dominant contributions stem from mostly symmetric dipoles and share the same parametric order: $\sigma_T\sim \sigma_L$. 
Conversely, when $\gamma_s\to 1$ (as in the MV model), the contribution to $\sigma_T$ originating from aligned-jet configurations is enhanced by a large logarithm $\ln(Q^2/Q_s^2)$ and dominates over the symmetric configurations. To see this explicitly, we evaluate the aligned-jet contribution to $\sigma_T$ via the second integral in \eqref{intzT}, restricted to the domain $r < 1/Q_s(x)$. For $\gamma_s < 1$, we find:
\begin{align}
\sigma_T(x,Q^2) &\sim \sigma_0 \int_{1/Q^2}^{1/Q_s^2(x)} \frac{\rmd r^2}{r^2}\,\frac{1}{r^2 Q^2}\,[r^2Q_s^2(x)]^{\gamma_s} \sim \frac{ \sigma_0}{1-\gamma_s}\left[\frac{Q_s^2(x)}{Q^2}\right]^{\gamma_s}\,, \label{DISscaling}
\end{align}
where the integral is dominated by its lower bound $r^2=1/Q^2$. A similar scaling behaviour is produced by the first integral in \eqref{intzT}, with the overall normalisation factor $1/(1-\gamma_s)$ replaced by $1/\gamma_s$. Furthermore, the two integrals in \eqref{intzL} for $\sigma_L$ yield analogous structures, weighted by factors of $1/(1+\gamma_s)$ and $1/(2-\gamma_s)$, respectively. As long as $1-\gamma_s$ is not parametrically small, all these components contribute at the same order. 
However, in the limit $\gamma_s\to 1$, the expression in \eqref{DISscaling} becomes divergent and the calculation must be reformulated to isolate the logarithmic enhancement:
\begin{align}
\sigma_T(x,Q^2) &\sim \sigma_0 \int_{1/Q^2}^{1/Q_s^2(x)} \frac{\rmd r^2}{r^2}\,\frac{1}{r^2 Q^2}\,r^2Q_s^2(x) \sim { \sigma_0} \frac{Q_s^2(x)}{Q^2}\ln\frac{Q^2}{Q_s^2(x)}\,. \label{DIShighQ}
\end{align}
In this same limit, the contributions from symmetric $q\bar q$ pairs to both $\sigma_T$ and $\sigma_L$ are parametrically of order $\sigma_0({Q_s^2(x)}/{Q^2})$, thus lacking the large logarithmic enhancement visible in Eq.~\eqref{DIShighQ}.

To summarise, for moderately small values of $x$ (say, $x\sim 10^{-2}$), where the classical MV model remains applicable, the total DIS cross-section at high virtuality $Q^2\gg Q_s^2(x)$ is controlled by aligned-jet configurations in the wavefunction of a transversely polarised photon. However, at smaller values of $x$, where the high-energy evolution becomes important  ($\alpha_s\ln(x_0/x)\gg 1$ with $x_0\sim 10^{-2}$), the total DIS cross-section $\sigma_{\rm tot}=\sigma_{T}+\sigma_{L}$ is rather controlled by predominantly symmetric $q\bar q$ fluctuations characterised by a typical transverse size $r\sim 1/Q$. Under these conditions, the longitudinal and transverse contributions become comparable in magnitude.


\begin{figure*}[t]
\begin{center}
\centerline{\includegraphics[width=0.4\textwidth]{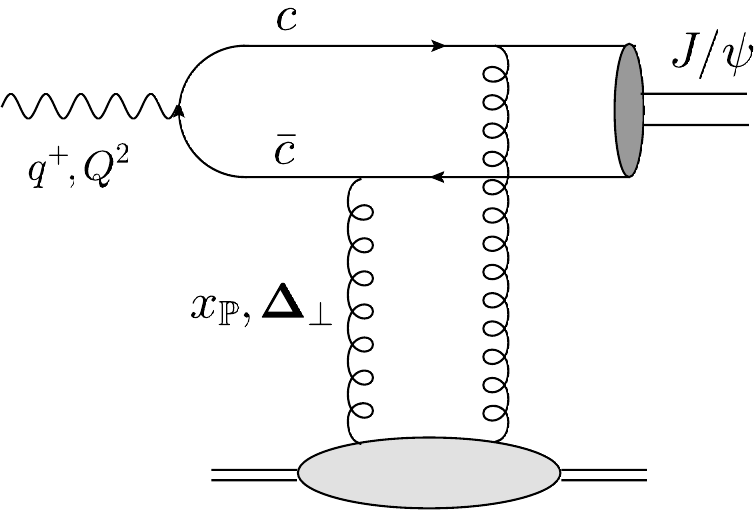}
\qquad
\includegraphics[width=0.5\textwidth]{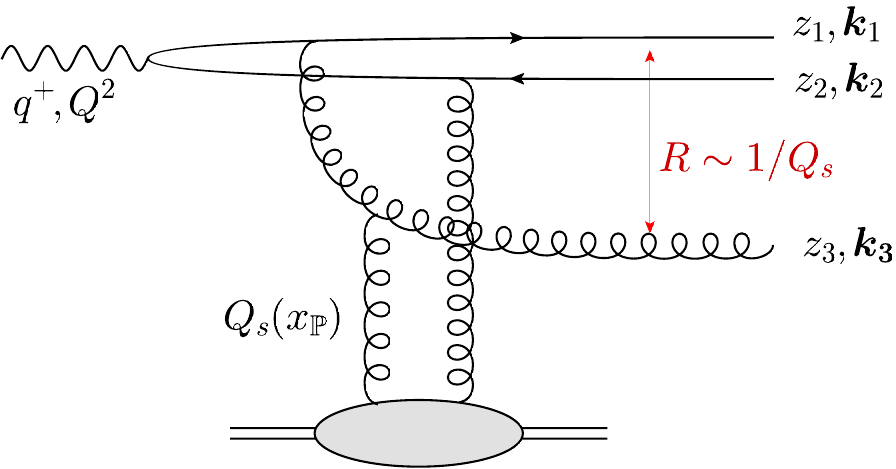}
} \caption{\sl Diffractive production in photon-nucleus collisions.
A minus longitudinal momentum $x_\mathbb{P}P_N^-$ and a transverse momentum $\bm{\Delta}_\perp$ are
transferred from the nucleus to the projectile.
 Left: exclusive production of a $J/\psi$ meson. The process is coherent if the nucleus does not break in the final state (meaning that the $c\bar c$ pair has elastically scattered off the nucleus as a whole), and incoherent in the opposite situation (the scattering involves subnuclear structures from the target, like individual nucleons or gluon hot spots). Right: coherent production of 2+1 jets: a hard $q\bar q$ dijet with $k_{1\perp}, k_{2\perp}\gg Q_s$ and a semi-hard gluon with $k_{3\perp}\sim Q_s$.
\label{fig:DIFFR}}
\end{center}
\end{figure*}
It is furthermore instructive to consider the \textit{diffractive} process, where the dipole-hadron scattering is purely elastic, leaving the outgoing $q\bar q$ pair as a colour singlet and the target intact. The corresponding diffractive cross-section is obtained from \eqref{DISdipole} via the substitution:
\begin{equation}\label{sigmaDel} 
\sigma_{\rm dipole}(x, r) \to \sigma_{\rm dipole}^{\rm el}(x,r) = \int \rmd^2{\mathbf{b}} \left |T\left({\mathbf{b}}+\frac{\br}{2},{\mathbf{b}}-\frac{\br}{2}\right)\right|^2 \simeq \pi R_A^2 \,T^2(x,r)\,.
\end{equation}
The appearance of the factor $T^2$ instead of $T$ implies that contributions from small $q\bar q$ dipoles ($r\ll 1/Q_s$), which scatter only weakly ($T\ll 1$), are strongly power-suppressed. This induces a major  bias in favour of large dipoles satisfying $r\gtrsim 1/Q_s$. Consequently, for a transversely polarised photon, the diffractive cross-section is controlled by strong scattering ($T\sim 1$) even at high virtuality $Q^2\gg Q_s^2$. To see this explicitly, we evaluate the aligned-jet contribution to the transverse diffractive cross-section by inserting $T \to T^2$ into the second integral of \eqref{intzT}:
\begin{align}\label{DiffSat}
\sigma_T^{\rm diff}(x,Q^2) &\sim \frac{\sigma_0}{Q^2} \int_{1/Q^2}^{\infty} \frac{\rmd r^2}{r^4}\,{T^2(x,r)} \nonumber \\[0.2cm] 
&\sim \frac{\sigma_0}{Q^2} \left\{ \int_{1/Q^2}^{1/Q_s^2}\frac{\rmd r^2}{r^4} \left[r^2Q_s^2(x)\right]^{2\gamma_s} + \int_{1/Q_s^2}^{\infty} \frac{\rmd r^2}{r^4} \right\} \sim \sigma_0\,\frac{Q_s^2(x)}{Q^2}\,.
\end{align}
In evaluating Eq.~\eqref{DiffSat}, we have split the original integral into two pieces: one over the small dipoles with $r\ll 1/Q_s$, for which $T$ is evaluated in the single scattering approximation, and one for the larger dipoles with $r\gtrsim 1/Q_s$, for which $T\simeq 1$.
Importantly, both integrals are control by their common boundary at $r\sim 1/Q_s$—acting as the upper bound for the first integral and the lower bound for the second. They both contribute at the same parametric order to the final result in \eqref{DiffSat}. One can similarly check that, for a longitudinal photon, the dominant contribution is pinned at small sizes $r\sim 1/Q\ll 1/Q_s$ and is parametrically suppressed: $\sigma_{L}^{\rm diff}\sim \sigma_0(Q_s^2/Q^2)^{2\gamma_s}$. 
\begin{figure*}[t]
\begin{center}
\centerline{\includegraphics[width=0.9\textwidth]{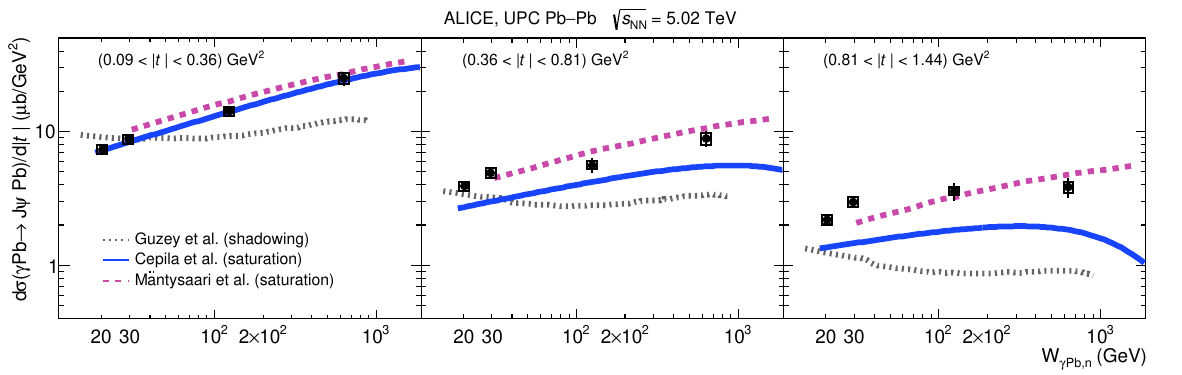}
} \caption{\sl  The LHC data for the cross-section for incoherent $J/\psi$ photonuclear production
 off $\text{Pb}$ nuclei~\cite{ALICE:2025cuw} are shown in three bins of $t$ (the
 squared transverse momentum transfer from the target) in terms of the COM energy $W_{\gamma\text{Pb,n}}$ and compared to several theoretical predictions: the leading-twist nuclear shadowing model in~\cite{Frankfurt:2011cs,Guzey:2018tlk} and the CGC-based  saturation models in~\cite{Mantysaari:2022sux,Mantysaari:2023xcu,Cepila:2023dxn}.
\label{fig:ALICE_incohdiff}}
\end{center}
\end{figure*}

We thus reach the important conclusion that hard DIS diffraction ($Q^2\gg Q_s^2$) at small $x$ is predominantly driven by aligned-jet configurations in the transverse photon wavefunction, which directly map the saturation scale. This has interesting implications for the structure of the final state:  the \textit{typical} $q\bar q$ pairs produced via hard diffractive scattering are inherently semi-hard, with transverse momenta of order $Q_s$ and a highly asymmetric longitudinal momentum sharing: $z(1-z)\sim Q_s^2/Q^2\ll 1$. Conversely, if one focuses on the diffractive production of a genuinely hard final state—such as a pair of jets with a relative momentum $P_\perp\gg Q_s$, or a heavy vector meson like the $\Upsilon$ ($b\bar b$) with a mass $M_V\gg Q_s$—the corresponding cross-section exhibits a power suppression of order $Q_s^2/P_\perp^2$ (or $Q_s^2/M_V^2$) due to the elastic nature of the scattering and the colour transparency of small configurations. 

The case of the $J/\psi$ meson ($c\bar c$) represents an interesting intermediate regime (see the left panel of Fig.~\ref{fig:DIFFR}): its mass $M_{J/\psi}\simeq 3.1$~GeV is not overwhelmingly larger than the nuclear saturation scale at LHC energies, rendering its cross-section marginally sensitive to saturation effects~\cite{Lappi:2010dd,Mantysaari:2016jaz,Mantysaari:2022sux,Mantysaari:2023xcu,Cepila:2023dxn,Mantysaari:2026vgx,Klein:2019qfb,Contreras:2026aqv}. This sensitivity is particularly important for \textit{incoherent} diffractive $J/\psi$ production~\cite{Lappi:2010dd}, which acts as a probe of spatial gluon density fluctuations in the transverse plane. The CGC predictions for the energy dependence of the respective cross-section~\cite{Mantysaari:2022sux,Mantysaari:2023xcu,Cepila:2023dxn} appear to be consistent with the ALICE measurement of incoherent $J/\psi$ photonuclear production
 in ultra-peripheral $\text{Pb}\text{--}\text{Pb}$ collisions~\cite{ALICE:2025cuw} (see Fig.~\ref{fig:ALICE_incohdiff}). This measurement provides compelling evidence of subnucleonic structures in the nuclear target down to spatial scales of \(0.2\)~fm, corresponding to a transverse momentum scale of \(\sim 1\)~GeV.

Thus far, our discussion has focused on the leading-order (LO) $q\bar q$ fluctuation of the virtual photon. The above LO estimates receive important real and virtual next-to-leading order (NLO) corrections associated with the emission of a soft gluon. Despite being suppressed by a factor of $\alpha_s$, these corrections are numerically sizeable, both because the QCD coupling is moderately weak  ($\alpha_s\sim 0.2\text{--}0.3$) and because they are enhanced by the large phase space open to gluon emissions. For instance, when $Q^2\gg Q_s^2$, these radiative corrections introduce terms of order $\alpha_s\ln(Q^2/Q_s^2)$, which reflect the DGLAP evolution of the PDFs in the target~\cite{Wusthoff:1997fz,GolecBiernat:1999qd,Kowalski:2008sa,Beuf:2022kyp,Iancu:2022lcw,Hauksson:2024bvv,Kaushik:2025roa}. By now, NLO corrections to the DIS impact factors within the dipole framework have been computed for both inclusive~\cite{Balitsky:2010ze,Balitsky:2012bs,Beuf:2011xd,Beuf:2016wdz,Beuf:2017bpd,Caucal:2021ent,Taels:2022tza,Bergabo:2022tcu,Iancu:2022gpw,Beuf:2021srj,Beuf:2021qqa,Beuf:2022ndu} and diffractive processes~\cite{Boussarie:2014lxa,Boussarie:2016ogo,Fucilla:2022wcg,Fucilla:2023mkl,Beuf:2024msh}. 

It is also interesting to notice that the leading-power contribution to the diffractive production of a pair of high-$p_T$ jets ($k_{1\perp}, k_{2\perp}\gg Q_s$) comes from special $q\bar q g$ configurations in which the two hard jets are accompanied by a semi-hard parton (either a gluon, or a quark) with transverse momentum $k_{3\perp}\sim Q_s$~\cite{Iancu:2021rup,Iancu:2022lcw,Hauksson:2024bvv,Rodriguez-Aguilar:2024efj,Caucal:2026xhj}, as shown in the right panel of Fig.~\ref{fig:DIFFR}. (See also~\cite{Iancu:2023lel,Shao:2024nor,Shao:2026doo} for phenomenological studies.) Such a ``2+1 jets'' configuration has a relatively large transverse size $R\sim 1/Q_s$ (the transverse separation between the semi-hard parton and the hard dijet) and therefore undergoes strong elastic scattering, thus avoiding the suppression due to the colour transparency of small $q\bar q$ dipoles. This mechanism is expected to dominate over the exclusive dijet production in the kinematics of $\text{Pb}\text{--}\text{Pb}$ UPCs at the LHC~\cite{CMS:2022lbi,Contreras:2026aqv} and also in diffractive processes at the EIC (at least, for sufficiently hard jets and/or open heavy quark production).

 \subsection{Geometric scaling and diffraction at HERA}
\label{sec:HERA}

\begin{figure*}[t]
\begin{center}
\centerline{\includegraphics[width=0.47\textwidth]{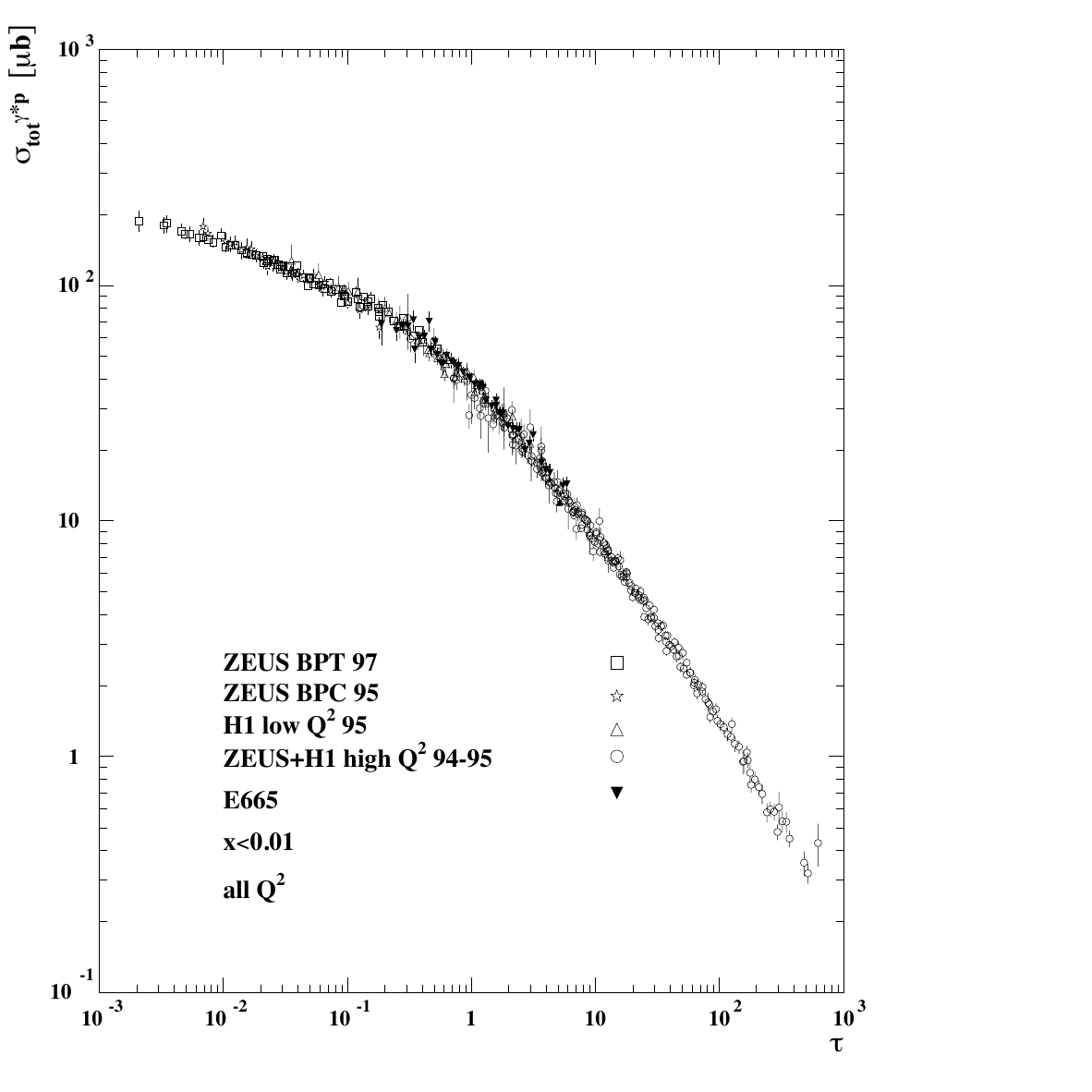}
\includegraphics[width=0.45\textwidth]{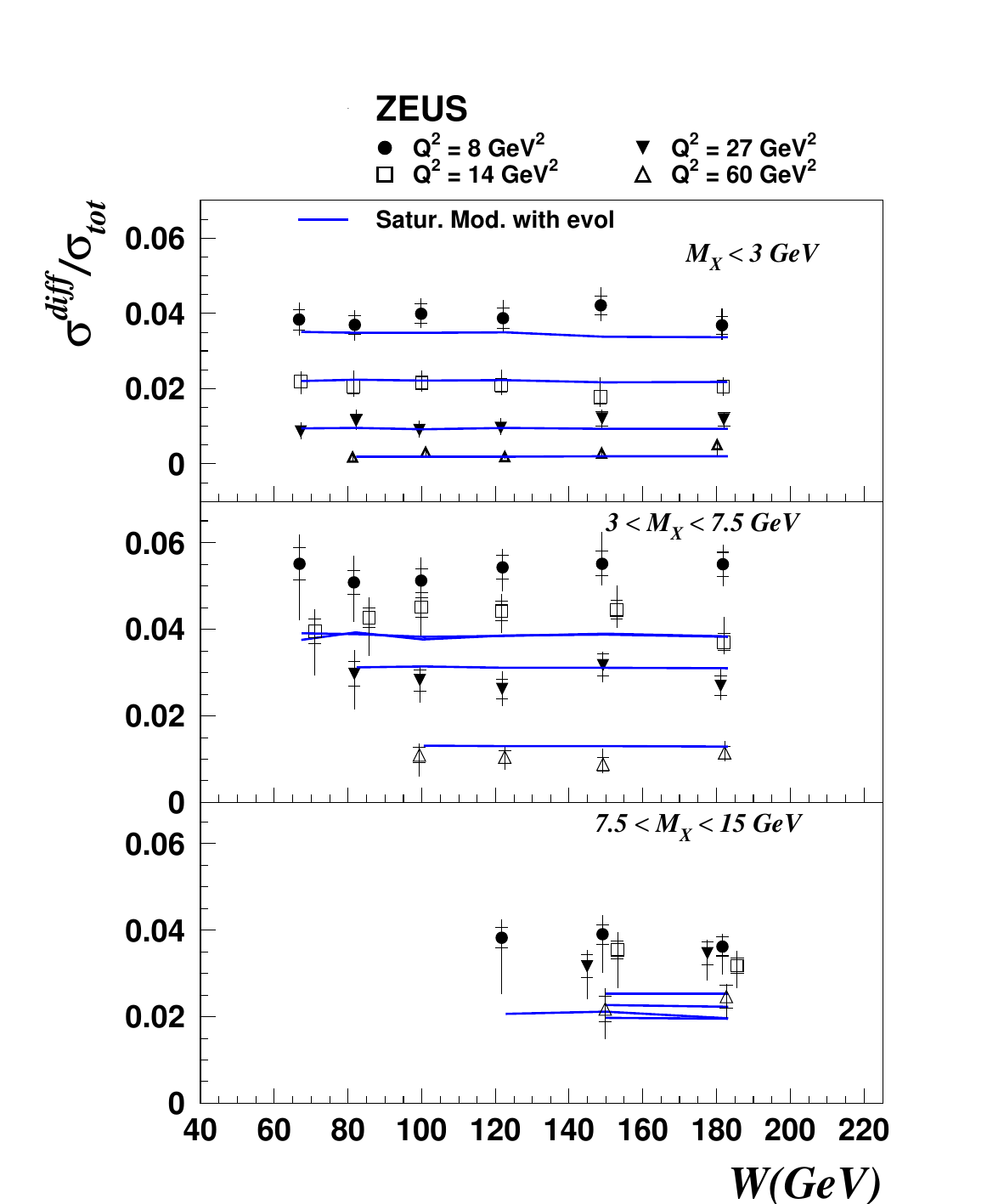}
} \caption{\sl Left: The DIS cross-section $\sigma_{\gamma^* p}(x,Q^2)$ 
at $x\le 0.01$ and $0.045\, {\rm GeV}^2 < Q^2 < 450\,{\rm GeV}^2$ plotted as a function of 
$\tau\equiv Q^2/Q_s^2(x)$~\cite{Stasto:2000er}. Right: 
The ratio $\sigma_{\rm diff}/\sigma_{\rm tot}$ between the diffractive and the total DIS cross-sections as a function of the COM energy $W=\sqrt{s}$ for several values of  $Q^2$: the points represent HERA data and the lines are the result of a calculation based on the dipole model supplemented with DGLAP evolution~\cite{Bartels:2002cj}. 
\label{fig:DISscaling}}
\end{center}
\end{figure*}

Based on this CGC framework for DIS at high energy, we can now provide a transparent physical interpretation for two remarkable regularities observed in the HERA data for electron-proton collisions at small $x$.  The first one refers to the observation of  geometric scaling in the total $\gamma^* p$ cross-section at $x\le 0.01$ over a wide range of virtualities ($0.045\,\text{GeV}^2 < Q^2 < 450\,\text{GeV}^2$)~\cite{Stasto:2000er}. When plotted as a function of the single dimensionless scaling variable $\tau\equiv Q^2/Q_s^2(x)$, with the parametrization $Q_s^2(x)\sim 1/x^{0.3}$ extracted from dipole-model fits to the same data~\cite{GolecBiernat:1998js,GolecBiernat:1999qd},  the small-$x$ data points for  $\sigma_{\rm tot}(x,Q^2)=\sigma_{T}+\sigma_{L}$ 
remarkably collapse onto a single universal curve, as displayed in the left panel of Fig.~\ref{fig:DISscaling}. This behaviour is naturally explained by the dipole picture together with the scaling form of the dipole amplitude, cf.  the second and third lines in Eq.~\eqref{TYrho}: via Eqs.~\eqref{DISdipole}--\eqref{sigmaD}, the geometric scaling property transmits from the dipole amplitude to the DIS cross-section, as confirmed by explicit derivations like \eqref{DISlowQ} and \eqref{DISscaling}.

At this point, it is useful to recall that the validity of geometric scaling for the dipole amplitude, $T(x,r)\simeq T(r^2Q_s^2(x))$, is highly non-trivial: this is not satisfied in the MV model, cf. Eq.~\eqref{SMV}, nor for the BK solution at very large values of $\rho=\ln(1/r^2\Lambda)$ (or very small dipole sizes $r$), well outside the scaling window (cf. the first line in Eq.~\eqref{TYrho}). As emphasised in the discussion following Eq.~\eqref{TGS}, this property is a consequence of the colour correlations introduced by the high-energy evolution. And indeed, 
the kinematic range of this scaling window widens progressively with increasing rapidity $Y$ (cf. Eq.~\eqref{gswindowrun}).  It it also interesting to notice that, for larger values $x>  0.01$, where non-linear gluon saturation effects have not yet set in, the HERA data show no signs of geometric scaling~\cite{Stasto:2000er}.

The other interesting experimental observation refers to the ratio $\sigma_{\rm diff}/\sigma_{\rm tot}$ between the diffractive and the total DIS cross-sections: when plotted as a function of the COM energy $W=\sqrt{s}$ for fixed $Q^2$, this ratio is roughly flat, as shown in the right plot in Fig.~\ref{fig:DISscaling}~\cite{Bartels:2002cj}. This behaviour not only is consistent with the saturation picture, as we shall shortly argue, but it also invalidates other scenarios which lack saturation and where unitarity is attributed to non-perturbative phenomena, like confinement.

To see this, consider the single scattering approximation (SSA) $T\simeq T_{2g}$, with $T_{2g}$ related to the gluon PDF as shown in Eq.~\eqref{T2gxG}: $T_{2g}(x, r)\sim \alpha_s r^2 xG(x, 1/r^2)/R_p^2$. The total cross-section $\sigma_{\rm tot}$ is linear in $T$ (cf. Eq.~\eqref{sigmaD}), so when computed in the SSA it grows with $1/x$ in the same way as the gluon PDF. The diffractive cross-section $\sigma_{\rm diff}$, on the other hand, is quadratic in $T$ (cf. Eq.~\eqref{sigmaDel}), hence one may expect it to grow like the gluon PDF {\it squared}. This argument is however incomplete, because of potential infrared divergences: when computing $\sigma_{\rm diff}$ in the SSA, the integral over $r^2$ develops a  quadratic divergence at large $r$, due to the fact that $T_{2g}^2(r)\propto r^4$.  This divergence is expected to be cut off by the unitarity corrections, which however work differently if there is saturation, or not.

 In the saturation scenario, we have already seen in the previous section that multiple scattering cuts off this divergence at the $x$-dependent scale $r\sim 1/Q_s(x)$, cf. Eq.~\eqref{DiffSat}. This mechanism slows down the growth  of $\sigma_{\rm diff}$ with $1/x$ as compared to the respective behaviour of the amplitude squared $T_{2g}^2(x,r)$. For $Q^2$ within the scaling window, one can estimate the ratio $\sigma_{\rm diff}/\sigma_{\rm tot}$ by using the results in  Eqs.~\eqref{DiffSat} and \eqref{DISscaling}. This yields  $\sigma_{\rm diff}/\sigma_{\rm tot}\sim [Q_s^{2}(x)]^{1-\gamma_s}\sim 1/x^{\delta}$, where $\delta\equiv (1-\gamma_s)\lambda_s\lesssim 0.1$ is a small exponent. This is indeed consistent with the almost flat behaviour seen in the HERA data.

In the non-perturbative scenario,  the upper limit on $r$ is set by confinement, $r_{\rm max}\sim R_p$, and is independent of $x$. Accordingly, the $x$-dependence of the dipole amplitude $T_{2g}$ directly transmits to the cross-section. Using again Eq.~\eqref{T2gxG}, one finds the following estimate 
\begin{equation}
\frac{\sigma_{\rm diff}}{\sigma_{\rm tot}}\,\sim \,\frac{ \alpha_s xG(x, Q^2)}{Q^2 R_p^2}\,,\end{equation}
which predicts a power increase $1/x^{0.3}$ with decreasing $x$.
Clearly, this growth is too fast to be consistent with the HERA data in Fig.~\ref{fig:DISscaling} (right).

To summarise, both the experimental observation of geometric scaling at HERA and the respective measurements of the energy dependence of the diffractive cross-section bring important evidence in favour of gluon saturation. What makes these observables truly special is the fact that they are sensitive to saturation even for relatively large photon virtualities $Q^2\gg Q_s^2(x)$. As a matter of fact, most of the small-$x$ data collected at HERA correspond to such a ``hard'' regime.

\subsection{Back-to-back dijets in $\gamma A$ collisions: TMD factorisation}
\label{sec:dijet}

In Sect.~\ref{sec:DIStot}, we established that the cross-section for diffractive scattering in DIS remains sensitive to gluon saturation even when the external scale is relatively hard ($Q^2 \gg Q_s^2$). We now present another classic example of this type, where the underlying scattering is predominantly \textit{inelastic}. Specifically, we consider the exclusive DIS process in which both the quark and the antiquark are explicitly measured in the final state (see the left panel of Fig.~\ref{DDcoll}). As before, we compute the wave function of the virtual photon in the transverse coordinate representation, enabling a streamlined treatment of multiple scatterings within the eikonal approximation. However, we must preserve the distinction between the transverse coordinates $\bx_\perp, \by_\perp$ of the $q\bar q$ pair in the direct amplitude (DA) and the corresponding coordinates $\bm{\overline{x}}_\perp, \bm{\overline{y}}_\perp$ in the complex conjugate amplitude (CCA). This separate tracking is mandatory since one eventually needs to compute the Fourier transforms from $\bx_\perp-\bm{\overline{x}}_\perp$ to $\bk_{1\perp}$, and similarly from $\by_\perp-\bm{\overline{y}}_\perp$ to $\bk_{2\perp}$, to fix the transverse momenta,  $\bk_{1\perp}$ and  $\bk_{1\perp}$, of the produced fermions.

Consider a transverse photon, for definiteness: after averaging over the photon polarisations and summing over the quark and antiquark helicities and colours, the dijet cross-section  takes the following form~\cite{Dominguez:2011wm}  (we omit the symbol $\perp$ on 2-dimensional vectors in the transverse plane):
\begin{align}\label{dijetcross}
\frac{d\sigma^{\gamma^*_{T} A\rightarrow q\bar qX}}{\rmd z_1\rmd^{2}\bm{k}_{1}\rmd z_2\rmd^{2}\bm{k}_{2}}&\,=
 2(2\pi)^2 N_c \alpha_{\rm em} e_f^2\,\delta(1-z_1-z_2)\,
 [z_1^2+ z_2^2]
\int\frac{\rmd^2\bx}{(2\pi)^2}\frac{\rmd^2\by}{(2\pi)^2}\frac{\rmd^2\bm{\overline{x}}}{(2\pi)^2}\frac{\rmd^2\bm{\overline{y}}}{(2\pi)^2}
\nonumber\\*[0.2cm]
&\times\,
\rme^{-i\bm{k}_{1}\cdot(\bm{x}-\overline{\bm{x}})-i\bm{k}_{2}\cdot(\bm{y}-\overline{\bm{y}})}\, 
\frac{\br\cdot \overline{\br}}{r\, \bar r}\,\bar{Q}^2 K_1(\bar{Q}r) K_1(\bar{Q}\bar r)\,
\mathcal{W}_{x}\left(\bm{x}, \bm{y}, \bm{\overline{y}}, \bm{\overline{x}}\right),
\end{align}
where $\br\equiv\bx-\by$, $\overline{\br}\equiv \overline{\bx}-\overline{\by}$,
$\bar Q^2\equiv z_1z_2Q^2$, the modified Bessel functions $K_1(\bar{Q}r) $ and $K_1(\bar{Q}\bar r)$ come from the transverse photon wavefunction, and the function $\mathcal{W}_x\left(\bm{x}, \bm{y}, \bm{\overline{y}}, \bm{\overline{x}}\right)$ encodes the Wilson lines describing the scattering of the quark and the antiquark in the eikonal approximation:
\begin{align}
\label{Wdef}
\mathcal{W}_x\left(\bm{x}, \bm{y}, \bm{\overline{y}}, \bm{\overline{x}}\right)
&\,\equiv\,\frac{1}{N_{c}}\,\left\langle \mathrm{tr}\big[V(\bm{x})
V^{\dagger}({\bm{y}})-1\big]\big[V(\bm{\overline{y}})
V^{\dagger}(\bm{\overline{x}})-1\big]\right\rangle_x \nonumber\\*[0.2cm]
&={Q}_x\left(\bm{x}, \bm{y}, \bm{\overline{y}}, \bm{\overline{x}}\right)-S_x\left(\bm{x}, \bm{y}\right)-S_x\left(\bm{\overline{y}}, \bm{\overline{x}}\right)+1,
\end{align}
where $S_x\left(\bm{x}, \bm{y}\right)$ is the dipole $S$-matrix introduced in Eq.~\eqref{Sdipole} and 
\begin{equation}
\label{quadrupole}
{Q}_x\,(\bm{x}, \bm{y}, \overline{\bm{y}}, \overline{\bm{x}})\,\equiv\,\frac{1}{N_{c}}\,\left\langle
\mathrm{tr}\big(V(\bm{x})\,V^{\dagger}(\bm{y})\,V(\overline{\bm{y}})\,
V^{\dagger}(\overline{\bm{x}})\big)\right\rangle_x,
\end{equation}
is the quadrupole $S$-matrix describing the elastic scattering of the
colourless systems made with the $q\bar q$ pair in the DA and its mirror pair in the CCA. 
The CGC average in \eqref{Wdef} must be evaluated at a value $x$ determined from the condition that the final 
$q\bar q$ pair be on shell. Energy-momentum conservation implies $xP^-+q^-=k_1^-+k_2^-$, with $k_i^-=k^2_{i\perp}/(2z_iq^+)$ and $q^-=-Q^2/2q^+$, hence
 \begin{equation}\label{xg}
 x = \frac{1}{2P\cdot q}\left(\frac{k^2_{1\perp}}{z_1} + \frac{k^2_{2\perp}}{z_2} + Q^2\right).
 \end{equation}
 If the quarks are not explicitly measured, integrating Eq.~\eqref{dijetcross} over the final-state phase space correctly recovers the inclusive total DIS cross-section \eqref{DISdipole}. Specifically, the integrations over $\bk_{1}$ and $\bk_{2}$ identify $\bx=\bm{\overline{x}}$ and respectively $\by=\bm{\overline{y}}$,  collapsing the colour structure to $\mathcal{W}_x = 2T_x(\bx, \by)$ in agreement with Eq.~\eqref{sigmaD}.\footnote{This mapping ignores the explicit momentum dependence of $x$ in Eq.~\eqref{xg}. This is legitimate since, at leading logarithmic accuracy, the fraction $x$ is well approximated by the standard Bjorken variable, $x \simeq \xbj = Q^2/(2P\cdot q)$.}
 
 As compared to the total cross-section \eqref{DISdipole}, the dipole sizes $r$ and $\bar r$ in Eq.~\eqref{dijetcross} are controlled not only by the photon virtuality (via the modified Bessel functions $K_1(\bar{Q}r) $ and $K_1(\bar{Q}\bar r)$  which enforce $r,\,\bar r\lesssim 1/\bar{Q}$), but also by the final momenta 
$\bm{k}_{1}$ and $\bm{k}_{2}$ (via the respective Fourier transforms). As we will now demonstrate, the dijet cross-section \eqref{dijetcross}  can be sensitive to saturation effects even in the hard regime where all the aforementioned external scales, $Q^2$, $k_{1\perp}^2$ and $k_{2\perp}^2$, are much larger than\footnote{Strictly speaking, the relevant scale for this problem is the adjoint saturation momentum $\tilde Q_s$ introduced in Sect.~\ref{sec:sat}. We retain the generic notation $Q_s$ for physical discussions, but the proper adjoint value is employed for numerical evaluations.}  $Q_s^2(x,A)$. This property is highly valuable since experimental jet measurements are inherently hard.

Specifically, we consider the ``correlation limit'' where the individual transverse momenta  of the produced fermions are much larger than their imbalance: $k_{1\perp}\simeq  k_{2\perp}\gg K_\perp\equiv |\bm{k}_{1}+\bm{k}_{2}|$,  with $K_\perp$ of order $Q_s(x)$ (since controlled be multiple scattering off the saturated gluons).  It is then convenient to make the following change of variables:
\begin{align}
	\label{pandk}
	\br\equiv \bx-\by\,,
	\qquad \bb\equiv z_1\bx+z_2\by\,,	\qquad 
	\bP \equiv z_2\bk_1 - 
	z_1 \bk_2\,,
	\qquad
	\bK \equiv \bk_1 + \bk_2\,,
\end{align}
so that $k_{1\perp}\simeq  k_{2\perp}\simeq P_\perp\gg K_\perp \sim Q_s(x)$. In this regime, the final quark and antiquark propagate nearly back-to-back in the transverse plane: the relative azimuthal angle $\Delta\Phi$  between $\bk_1$ and $\bk_2$ is peaked at $\Delta\Phi=\pi$ with a narrow width $|\Delta\Phi-\pi|\sim Q_s/P_\perp$.  

The Fourier phases in Eq.~\eqref{dijetcross} can be rewritten as
$\bk_1 \cdot \bx + \bk_2 \cdot \by=\bP \cdot \br +\bK \cdot \bb$ and similarly for the CCA. This shows 
that the dijet relative momentum $\bP$ and the dipole size $\br$ (or $\bar\br$) are conjugated with each other, so when $P_\perp$ is large, the dipoles are small: $r, \,\bar r \sim 1/P_\perp\ll 1/Q_s$. One can then expand the Wilson lines in \eqref{Wdef} to deduce (with the notation $V_{\bx} \equiv V({\bx})$) 
\begin{align}
	\label{Vqq}
	V_{\bx} V_{\by}^\dagger - 1\simeq -r^i V_{\bb}\partial^i V_{\bb}^\dagger\quad\Longrightarrow\quad
	\mathcal{W}_x\simeq -r^i\bar r^j\left\langle\big(V_{\bb}\partial^i V_{\bb}^\dagger\big) \big(V_{\bbb} \del^j V_{\bbb}^{\dagger}\big)\right\rangle_{x}.
	\end{align}
Accordingly, the dependence of the integrand in   \eqref{dijetcross} upon the dipole sizes $\br$ and $\bbr$ {\it factorises} from its dependence upon the central coordinates $\bb$ and $\bbb$. After the Fourier transforms, this implies a similar factorisation for the dijet cross-section in terms of  $\bP$ and $\bK$~\cite{Dominguez:2010xd,Dominguez:2011wm}:
\begin{align}
	\label{sigma0}
	\frac{\dif \sigma^{\gamma_{T}^* A \to q\bar{q}X}}
	{\rmd z_1
  	\rmd z_2\dif^2\bP\dif^2\bK } &\,=  {\alpha_{em}\alpha_s}
	e_f^2\delta(1-z_1-z_2)\left(z_1^{2} + 
	z_2^{2}\right) \frac{P_{\perp}^4 + \bar{Q}^4}
	{(P_{\perp}^2 + \bar{Q}^2)^4}\nonumber\\*[0.2cm]
	&\, \times \left\{\,xf_g(x, K_\perp) 
	 -  \cos(2\phi)
\, \frac{2 P_{\perp}^2 \bar{Q}^2}{P_{\perp}^4 + \bar{Q}^4}   \,     xh_g(x,K_\perp)\right\},
\end{align}
where $\phi$ is the azimuthal angle between $\bP$ and $\bK$ (distinct from the relative jet angle $\Delta\Phi$) and $xf_g(x,K_\perp)$ and $xh_g(x,K_\perp)$ are the two independent scalar components of the tensorial Weizsäcker-Williams (\text{WW}) transverse-momentum dependent (\text{TMD}) gluon distribution~\cite{Dominguez:2011wm,Metz:2011wb,Dominguez:2011br}:
 \begin{align}
 \hspace*{-0.2cm}
 	\label{dGWW} \mcal{F}_{WW}^{ij}(x, \bK)
	&\ \equiv  
  \int\frac{\rmd^2{\bm{b}}\, \rmd^2 \bbb}{(2\pi)^4} \ 
   \rme^{-i\bK\cdot(\bb-\bbb)}\   \frac{-2}{\alpha_s}
\left\langle {\rm tr} \left[ 
V_{\bb}\big (\del^i V_{\bb}^{\dagger}\big)\,
     V_{\bbb} \big(\del^j V_{\bbb}^{\dagger}\big)\right]\right\rangle_{x}
     \nonumber\\*[0.2cm]
	&\ \equiv \,\frac{\delta^{ij}}{2}\,
xf_g(x, K_\perp) +
        \left(\frac{K_{\perp}^i K_{\perp}^j}{K_{\perp}^2}
        -\frac{\delta^{ij}}{2} \right)
        xh_g(x,K_\perp).
   \end{align}
Comparing the first line of Eq.~\eqref{dGWW} with Eq.~\eqref{xfg} confirms that $xf_g(x, K_\perp) = \mcal{F}_{\text{WW}}^{ii}$ is identical to the UGD introduced in Eq.~\eqref{unint}. In this context, $xf_g(x, K_\perp)$ is identified as the unpolarised WW gluon TMD, while $xh_g(x, K_\perp)$ represents the distribution of linearly polarised gluons inside an unpolarised target~\cite{Metz:2011wb,Dominguez:2011br}. 
Two remarks are in order: \texttt{(i)} these two distributions differ significantly within the non-linear saturation domain ($K_\perp \lesssim Q_s$), but merge asymptotically at large momentum ($K_\perp \gg Q_s$) towards the power-law tail \eqref{noccup-high}; \texttt{(ii)} the contribution of linearly polarised gluons to the dijet cross-section \eqref{sigma0} vanishes after integrating over the azimuthal angle $\phi$, and drops out completely in the photo-production limit ($Q^2 \to 0$).
      
Eq.~\eqref{sigma0} exhibits {\it TMD factorisation}, separately for the unpolarised and the linearly-polarised contributions.  Each contribution factorises  into a hard matrix element depending on the kinematics $(z_i, P_\perp, Q^2)$ of the dijet and a specific WW gluon TMD describing small-$x$ gluons in the target. Importantly, while the relative jet momentum is hard ($P_\perp \gg Q_s$), the momentum imbalance $K_\perp$ probes the saturation scale.  Hence, a differential measurement of the dijet distribution  at the future EIC will allow for direct studies of gluon saturation: the broadening of the back-to-back peak at $\Delta \Phi=\pi$ provides a direct measure of the saturation momentum~\cite{Marquet:2007vb,Zheng:2014vka,Kolbe:2020tlq,Zhao:2021kae} (see also Refs~\cite{Albacete:2010pg,Lappi:2012nh,Albacete:2014fwa,Albacete:2018ruq} for related studies for $pA$ collisions), whereas the $\cos(2\phi)$ asymmetry cleanly isolates the distribution of linearly polarised gluons~\cite{Metz:2011wb,Dumitru:2015gaa,Dumitru:2018kuw,Marquet:2017xwy,Mantysaari:2019hkq,Hatta:2020bgy,Rodriguez-Aguilar:2023ihz}.

\begin{figure}[t]
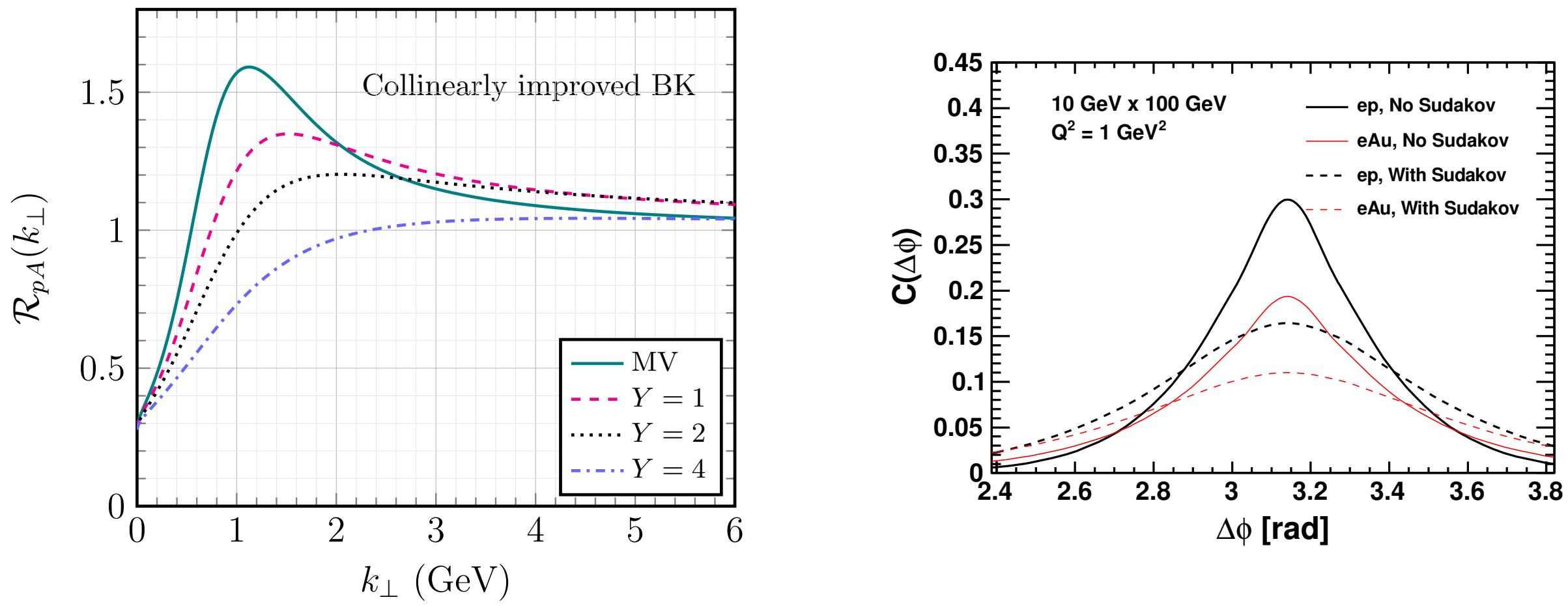
 
  \centering
  \begin{minipage}[t]{0.45\textwidth}
    \mbox{ }\\
    \vspace*{1.cm}\includegraphics[page=,width=\textwidth]{PDF/RpA_Occupation_BK.pdf}
  \end{minipage}
  \hfill%
  \begin{minipage}[t]{0.48\textwidth}
    \mbox{ }\\
      \includegraphics[page=,width=\textwidth]{PDF/BowenElke-Sudakov.pdf}
        \end{minipage}
        \caption{\sl  Left: The $R_{pA}$ ratio for the gluon occupation number,  Eq.~\eqref{RpAWW}. For the MV model, we use  $\tilde Q^2_A=4 \tilde Q^2_p=0.68$~GeV$^2$ (yielding $ \tilde Q_s^2(A)\simeq 2.3$~GeV$^2$). The high-energy evolution is computed by using the collinearly improved BK equation with running  coupling.  The Cronin peak predicted by the MV model is gradually washed out by the evolution with $Y$.
 Right: The dijet distribution in the azimuthal angle $\Delta \Phi$ between $\bk_1$ and $\bk_2$ for both $ep$ and $eA$ collisions, in the MV model (the continuous lines) and after adding the Sudakov effect (the dotted lines)~\cite{Zheng:2014vka}. One can see the additional broadening due to final-state radiation.
\label{fig:dijet}}
\end{figure}

One suggestive indicator of the saturation effects in the nuclear target is the nucleus-to-proton ratio, $\mcal{R}_{pA}(k_\perp)$, for the dijet cross-section averaged over the azimuthal angle $\phi$. Due to TMD factorization, this observable reduces  to the properly normalised ratio of the respective gluon occupation numbers introduced in Eq.~\eqref{nxk}:
\begin{equation}\label{RpAWW}
\mcal{R}_{pA}(k_\perp) \equiv \frac{1}{A^{1/3}}\,\frac{n_A(x, k_\perp^2)}{n_p(x, k_\perp^2)}\,.
\end{equation}
This ratio is normalised such that it approaches unity at large $k_\perp \gg \tilde Q_s(A, x)$. Within the classical MV model, the occupation numbers are evaluated by applying Eq.~\eqref{xfgMV} to both the nucleus and the proton, with $\tilde Q^2_A = A^{1/3} \tilde Q^2_p$. The high-energy evolution with increasing rapidity $Y \equiv \ln(x_0/x)$ is subsequently computed using the Gaussian approximation to the JIMWLK equation, as illustrated in the right panel of Fig.~\ref{fig:WW}.  This procedure leads to the numerical results shown in the left panel of Fig.~\ref{fig:dijet}. They reveal a close similarity with the corresponding results for quark production in $pA$ collisions, as visible in the right panel of Fig.~\ref{fig:pT}. This qualitative resemblance is physically natural, as both observables probe the same underlying phenomenon: transverse momentum broadening driven by multiple scatterings off the saturated gluons.

It is important to emphasise that these leading-order (LO) results can be severely altered by higher-order radiative corrections that are numerically sizeable. The most profound modification stems from the \textit{Sudakov effect}, which physically describes a shift in the dijet transverse momentum imbalance towards larger values $K_\perp \gg Q_s$. This transverse broadening is triggered by the recoil from soft, collinear gluons radiated by the final-state quark and antiquark legs~\cite{Mueller:2013wwa,Xiao:2017yya,Taels:2022tza,Caucal:2022ulg,Caucal:2024bae}. Because the Sudakov effect is numerically large, it can dominate the overall broadening of the back-to-back peak, potentially masking the genuine signatures of gluon saturation~\cite{Zheng:2014vka,Albacete:2018ruq,Stasto:2018rci,vanHameren:2016ftb,vanHameren:2021sqc,vanHameren:2023oiq,Cassar:2025vdp,Caucal:2025zkl}. This interplay is illustrated in the right panel of Fig.~\ref{fig:dijet} (adapted from Ref.~\cite{Zheng:2014vka}), which compares the azimuthal dijet distribution $\Delta\Phi$ computed with and without Sudakov resummations.

We conclude this section with a brief overview of recent theoretical developments regarding TMD factorisation at small $x$ (see the recent review~\cite{Caucal:2026yzl} for more details and references). Historically, this factorisation scheme was established for selected processes—such as Semi-Inclusive Deep Inelastic Scattering (SIDIS) and the Drell-Yan process—at moderate values of $x \gg 0.01$, where collinear factorisation serves as the standard baseline for QCD hard scatterings~\cite{Collins:2011zzd,Boussarie:2023izj}. The emergence of a TMD structure from the CGC effective theory at small $x$ is highly non-trivial. Indeed, in general, the CGC results for the cross-sections do not factorise in transverse space; rather, they are expressed as multi-dimensional convolutions, as visible in Eq.~\eqref{dijetcross}. 

TMD factorisation emerges from the CGC only after additional kinematic approximations that exploit the wide separation of transverse scales (recall the multipole expansion leading from Eq.~\eqref{dijetcross} to Eq.~\eqref{sigma0}). To leading order in $\alpha_s$, this factorisation has been formally proven for several processes characterised by two widely disparate transverse momentum scales. These include the semi-inclusive~\cite{Marquet:2009ca,Caucal:2026xhj} and diffractive~\cite{Hatta:2022lzj,Hauksson:2024bvv} production of a single forward jet in $e\text{--}A$ DIS; inclusive dijets and isolated photon-jet correlations in both $e\text{--}A$ and $p\text{--}A$ collisions~\cite{Dominguez:2010xd,Dominguez:2011wm,Metz:2011wb,Dominguez:2011br,Marquet:2017xwy,Altinoluk:2024tyx,Caucal:2025xxh,Caucal:2025qjg}; and diffractive $2+1$ jet production in $e\text{--}A$ and ultra-peripheral heavy-ion ($A\text{--}A$) collisions~\cite{Iancu:2021rup,Iancu:2022lcw,Iancu:2023lel,Hauksson:2024bvv,Caucal:2026xhj,Hatta:2022lzj}. 
The underlying TMD distributions in these examples describe small-$x$ partons (gluons or sea quarks) within the dense target. For diffractive channels, the corresponding diffractive TMDs describe partons extracted from the "Pomeron"—the colourless gluonic excitation exchanged in the elastic scattering between the projectile and the nuclear target~\cite{Iancu:2021rup,Iancu:2022lcw,Iancu:2023lel,Hauksson:2024bvv,Caucal:2026xhj,Hatta:2022lzj} (see Fig.~\ref{fig:DIFFR}).

Due to gluon saturation, the TMD distributions derived from the CGC naturally incorporate the intrinsic semi-hard scale $Q_s(A,x)$, which regularises the infrared domain and permits perturbative expansions even at tree level (cf. Sect.~\ref{sec:sat}). Furthermore, they acquire an explicit dependence on the external hard scale (such as the relative dijet momentum $P_\perp$ or the photon virtuality $Q^2$) through the Sudakov effects shown in Fig.~\ref{fig:dijet}. As is standard at moderate $x$~\cite{Collins:2011zzd,Boussarie:2023izj}, these Sudakov effects can be resummed to all orders by solving the Collins-Soper-Sterman (CSS) equation~\cite{Collins:1981uk,Collins:1981uw,Collins:1984kg}. 

At small $x$, the CSS equation has been derived from one-loop calculations within the CGC framework~\cite{Caucal:2024bae,Caucal:2025mth,Caucal:2024vbv}. As compared to its  
moderate-$x$ counterpart, the phase space for soft radiation is dynamically bounded by the saturation momentum. This non-linear constraint acts as an infrared cutoff, significantly mitigating or entirely removing the sensitivity of the CSS solutions to the non-perturbative physics of colour confinement~\cite{Iancu:2025jsu}. In addition to CSS scale evolution, the small-$x$ TMDs also satisfy the non-linear JIMWLK evolution with decreasing $x$—see Refs.~\cite{Dominguez:2011gc,Caucal:2025mth} for the functional equations satisfied by the Weizsäcker-Williams gluon distribution, and Refs.~\cite{Marquet:2016cgx,Marquet:2017xwy} for numerical solutions constructed via the Langevin-JIMWLK method.

\section{Summary and outlook}
\label{sec:outlook}

Throughout this review, we have traveled across the theoretical and phenomenological landscape of the Colour Glass Condensate (CGC), exploring how the growth of the gluon occupation number with increasing energy and/or increasing the nuclear mass number $A$ naturally terminates at a universal, non-linear boundary—a phenomenon known as gluon saturation. Grounded in the separation of timescales via high-energy kinematics, the CGC effective field theory provides a rigorous, first-principles description of this dense bosonic medium in terms of strong classical colour fields, eikonal Wilson lines, and non-linear equations for the evolution of scattering amplitudes with increasing energy, or decreasing $x$.

The non-linear effects encoded in the B-JIMWLK or BK evolution equations dynamically predict the emergence of an intrinsic, energy-dependent transverse momentum scale—the saturation scale $Q_s(x,A)$—which is semi-hard for sufficiently small $x$ and/or large enough $A$, thus efficiently screening the non-perturbative physics of confinement and enabling controlled calculations within perturbative QCD. The CGC framework allows for a systematic reorganisation of QCD perturbation theory in which strong classical colour fields representing the saturated gluons offer a background for the propagation of quantum fluctuations, whose effects can be computed order-by-order in the expansion in powers of $\alpha_s$, but to all orders in the background fields.

A two-decade long, community-wide effort to push both the non-linear evolution equations and the partonic cross-sections  to next-to-leading order (NLO) accuracy, supplemented by essential collinear and running-coupling resummations, has transformed the CGC from a semi-quantitative framework into an accurate predictive tool. Besides bringing stability to CGC predictions at NLO, these collinear resummations are also mandatory to disentangle the large but standard radiative logarithms associated with linear evolutions (DGLAP and CSS) from the non-linear signatures of saturation.

On the experimental front, the physics of gluon saturation is entering a decisive era, transitioning from indirect indications towards a precise, multidimensional, tomographic mapping, which should pave the way for its definitive confirmation. This paradigm shift is driven by two major upcoming pillars:

\begin{enumerate}
\item \textbf{The Forward Upgrades at the LHC (Run 4):}
 The imminent installation of the Forward Calorimeter (FoCal) upgrade~\cite{ALICE:2024jtt} within the ALICE experiment (scheduled for the late 2020s) will unlock unprecedented access to forward particle production. By covering a broad forward pseudorapidity window ($3.2 < \eta < 5.8$), FoCal will sample gluon distributions down to uncharted kinematic limits ($x \sim 10^{-6}$) in $pp$, $p\text{--}\text{Pb}$, and ultra-peripheral $\text{Pb}\text{--}\text{Pb}$ collisions (UPCs). In UPCs, the exclusive photoproduction of vector mesons (like the $J/\psi$) will provide precise maps of the spatial structure of gluons. Furthermore, high-precision measurements of isolated forward direct photons, forward di-hadrons, and photon-jet azimuthal de-correlations will directly map the saturated, transverse-momentum-dependent parton distributions and test the core predictions of the CGC effective theory without final-state plasma ambiguities. 
 
 In a complementary fashion, LHCb—which operates natively as a forward spectrometer—is uniquely suited to explore the nuclear parton structure at small $x$. Benefiting from major upgrades in Run 4, LHCb will deliver dedicated studies of gluon saturation in $p\text{--}\text{Pb}$ collisions down to zero transverse momentum. This will be achieved notably via the production of heavy-flavour quark pairs (open charm/beauty) and the study of azimuthal correlations between pairs of $D$ mesons ($D\text{--}\bar{D}$). A modification of these correlations compared to $pp$ baselines should help us disentangling traditional shadowing models from gluon saturation.

\item \textbf{The Electron-Ion Collider (EIC):} In parallel, the construction of the future EIC at Brookhaven National Laboratory (BNL)~\cite{AbdulKhalek:2022hcn} will provide an ultra-clean, high-luminosity leptonic microscope. While operating at more modest center-of-mass energies than the LHC, the EIC versatile capability to collide polarised electrons with a vast selection of nuclear targets, including heavy nuclei like lead, will permit precise multidimensional scans of the gluon distribution in the high density regime. The physics program with heavy ions will be notably driven through exclusive vector meson photoproduction, diffractive structure functions, and the transverse momentum dependent (TMD) unpolarised and linearly polarised Weizsäcker-Williams gluon distributions introduced in Sect.~\ref{sec:dijet}.
\end{enumerate}

In conclusion, the close synergy between state-of-the-art NLO CGC calculations and upcoming experimental data from Run 4 at the LHC and the EIC is poised to deliver a definitive answer to the long-standing quest for gluon saturation. By matching the theoretical precision of QCD resummations with the exquisite performance of modern detectors, the nuclear physics community is on the verge of establishing the Colour Glass Condensate as a fundamental state of matter, reshaping our understanding of the subnucleonic structure of hadrons and nuclei at high energies.

\section*{Acknowledgements}
I am grateful to Shu-Yi Wei for producing all the original numerical results and the associated plots that I used to illustrate various points of physics. Figures representing Feynman graphs were created with JaxoDraw~\cite{Binosi:2003yf}.

\bigskip
\bibliographystyle{JHEP}
\bibliography{refs}
\end{document}